\documentclass[11pt,headings=big,numbers=noenddot,DIV=14,a4paper]{article}%

\pdfoutput=1

\usepackage[margin=1 in]{geometry}
\usepackage{comment}
\usepackage{graphicx}  
\usepackage{dcolumn} 
\usepackage{bm,relsize}   
\usepackage{amsfonts,amsmath,amssymb,amsthm,mathtools}
\usepackage{slashed}
\usepackage{placeins}
\usepackage{adjustbox}
\usepackage[normalem]{ulem}
\usepackage{lscape}
\usepackage{multirow}
\usepackage{lipsum, color}
\usepackage[usenames,dvipsnames,svgnames,table]{xcolor}
\usepackage[linktoc=page,bookmarks=false,colorlinks=true,linkbordercolor=RoyalBlue,citebordercolor=ForestGreen,urlbordercolor=CornflowerBlue]{hyperref}
\hypersetup{
    colorlinks=true,
    linkcolor=ForestGreen,
    filecolor=ForestGreen,      
    urlcolor=ForestGreen,
    citecolor=ForestGreen,
}

\usepackage[sort&compress,numbers,merge]{natbib}
\usepackage{breakcites}

\usepackage{tikz-feynman}

\def\beq{\begin{equation}}
\def\eeq{\end{equation}}
\def\bea{\begin{eqnarray}}
\def\eea{\end{eqnarray}}

\def\RH			{{\mathsmaller{\rm RH}}}
\def\IR			{{\mathsmaller{\rm IR}}}

\def\MDM		{m_{\mathsmaller{\rm DM}}}
\def\MPl		{M_{\mathsmaller{\rm Pl}}}
\def\gSM		{g_{\mathsmaller{\rm SM}}}

\newcommand{\dilaton}{\chi}
\newcommand{\dilatonvev}{f}

\newcommand{\Tnuc}{T_\text{nuc}}
\newcommand{\Tstart}{T_{\rm start}}

\def\DIS			{{\mathsmaller{\text{DIS}}}}

\def\LO			{{\mathsmaller{\text{LO}}}}
\def\NLO			{{\mathsmaller{\text{NLO}}}}

\def\SM			{{\mathsmaller{\text{SM}}}}
\def\SC			{{\mathsmaller{\text{SC}}}}
\def\TC			{{\mathsmaller{\text{TC}}}}
\def\RH			{{\mathsmaller{\text{RH}}}}

\def\mDM		{m_{\mathsmaller{\text{DM}}}}

\def\MPl		{M_{\mathsmaller{\text{Pl}}}}

\def\Tstart		{T_{\mathsmaller{\text{start}}}}

\def\TRH		{T_{\mathsmaller{\text{RH}}}}

\def\gSM		{g_{\mathsmaller{\text{SM}}}}

\def\gwp		{\gamma_{\mathsmaller{\text{wp}}}}
\def\gwc		{\gamma_{\mathsmaller{\text{wc}}}}
\def\gcp		{\gamma_{\mathsmaller{\text{cp}}}}
\def\ECM		{E_{\mathsmaller{\text{CM}}}}

\begin{document}

\begin{flushright}
\footnotesize
DESY 20-122\\
ULB-TH/20-08
\end{flushright}
\color{black}

\begin{center}

{\LARGE \bf
String Fragmentation in Supercooled Confinement\\ \vspace{.2 cm}
and Implications for Dark Matter
}

\medskip
\bigskip\color{black}\vspace{0.5cm}

{
{\large Iason Baldes},$^a$
{\large Yann Gouttenoire},$^{b,c}$
{\large Filippo Sala}$^{b,c}$
}
\\[7mm]

{\it \small $^a$ Service de Physique Th\'eorique, Universit\'e Libre de Bruxelles, \\ Boulevard du Triomphe, CP225, B-1050 Brussels, Belgium}\\
{\it \small $^b$ DESY, Notkestra{\ss}e 85, D-22607 Hamburg, Germany}\\
{\it \small $^c$ LPTHE,  CNRS \& Sorbonne Universit\'e, 4 Place Jussieu, F-75252, Paris, France}\\
\end{center}

\bigskip

\centerline{\bf Abstract}
\begin{quote}
\color{black}

A strongly-coupled sector can feature a supercooled confinement transition in the early universe.
We point out that, when fundamental quanta of the strong sector are swept into expanding bubbles of the confined phase, the distance between them is large compared to the confinement scale.
We suggest a modelling of the subsequent dynamics and find that the flux linking the fundamental quanta deforms and stretches towards the wall, producing an enhanced number of composite states upon string fragmentation.
The composite states are highly boosted in the plasma frame, which leads to additional particle production through the subsequent deep inelastic scattering. We study the consequences for the abundance and energetics of particles in the universe and for bubble-wall Lorentz factors.
This opens several new avenues of investigation, which we begin to explore here, showing that the composite dark matter relic density is affected by many orders of magnitude.
\end{quote}

\clearpage
\noindent\makebox[\linewidth]{\rule{\textwidth}{1pt}} 
\tableofcontents
\noindent\makebox[\linewidth]{\rule{\textwidth}{1pt}}

\section{Introduction}

The possible existence of new confining sectors is motivated by most major failures of our understanding of Nature at a fundamental level.
First, the stability of particle Dark Matter can be elegantly achieved as an accident if it is a composite state of a new strongly-coupled sector, similarly to proton stability in QCD, see e.g.~\cite{Antipin:2015xia}.
The hierarchy problem of the Fermi scale is solved via dimensional transmutation by new confining gauge theories, whose currently most appealing incarnation is that of composite Higgs models~\cite{Contino:2010rs,Panico:2015jxa}. Analogous composite pictures can UV-complete~\cite{Geller:2014kta,Barbieri:2015lqa,Low:2015nqa} twin-Higgs scenarios~\cite{Chacko:2005pe}, and so ameliorate also the little hierarchy problem. A rationale to understand the SM hierarchies of masses and CKM mixing angles is provided by partial compositeness of the SM fermions~\cite{Kaplan:1991dc}. Finally, new confining sectors play crucial roles in addressing the strong CP problem~\cite{Rubakov:1997vp,Redi:2016esr}, the baryon asymmetry~\cite{Konstandin:2011ds,Servant:2014bla}, etc.

Given their ubiquity, it makes sense to look for predictions of confining sectors that do not depend on the specific way they address a given SM issue.
Cosmology naturally offers such a playground, in association with the confinement phase transition (PT) in the early universe.
The low-density QCD phase transition would for example be strongly first-order if the strange or more quarks had smaller masses~\cite{Pisarski:1983ms}, with associated signals in gravitational waves~\cite{Witten:1984rs,Helmboldt:2019pan}. New confining sectors could also well feature a similar PT.
In addition, the confinement transition could be supercooled, a property that for example arises naturally in 5-dimensional (5D) duals of 4D confining theories~\cite{Creminelli:2001th,Randall:2006py,Nardini:2007me}.

Generically, supercooling denotes a PT in which bubble percolation occurs significantly below the critical temperature.
Here we are interested in the case where a cosmological PT becomes sufficiently delayed so that the radiation energy density becomes subdominant to the vacuum energy.
The universe then experiences a stage of inflation until the PT completes~\cite{Kolb:1979bt}. This implies a dilution of any pre-existing relic, such as dark matter (DM), the baryon or other asymmetries, topological defects, and gravitational waves, see e.g.~\cite{Hambye:2018qjv,Barreiro:1996dx,Easther:2008sx}.

In this paper we point out an effect that, to our knowledge, had been so far missed: when the fundamental quanta of the strong sector enter the expanding bubbles of the confined phase, their relevant distance can be much larger than the inverse of the confinement scale, thus realising a situation whose closest known analogues are perhaps QCD jets in particle colliders or cosmic ray showers.
We anticipate that our attempt to model this phenomenon implies an additional production mechanism of any composite resonance --- string fragmentation followed by deep inelastic scattering --- which introduces a mismatch between the dilution of composite and other relics. This opens new model building and phenomenological avenues, which we begin exploring here in a model independent manner for the case of composite DM. The application of our findings to a specific model, namely composite dark matter with dilaton mediated interactions, will appear elsewhere~\cite{2ndpaper}.

\section{Synopsis}
\label{sec:synopsis}
Due to the numerous effects which will be discussed in the following sections, it is perhaps useful for the reader that we
summarise the overall picture in a few paragraphs. We begin in the deconfined phase in which the techniquanta $\TC$ of the new strong sector (which we will call quarks and gluons) are in thermal equilibrium. Their number density normalised to entropy takes a familiar form	
	\begin{equation}
	\label{eq:quark_density}
	Y_\TC^{\rm eq} = \frac{45\, \zeta(3) \, g_\TC}{2\pi^4 g_s},
	\end{equation}
where $g_\TC$ ($g_{s}$) are the degrees of freedom of the quarks and gluons (entropic bath) respectively.
Next a period of supercooling occurs, in which the universe finds itself in a late period of thermal inflation, which is terminated by bubble nucleation. As is known from previous studies, such a phase will dilute the number density of primordial particles. The dilution factor is given by
	\begin{equation}
	D^{\rm SC} = \left( \frac{T_{\rm nuc}}{T_{\rm start}} \right)^3 \frac{T_{\rm RH}}{T_{\rm start}},
	\end{equation}
where $T_{\rm nuc}$ is the nucleation temperature, $T_{\rm start} \propto f$ is the temperature at which the thermal inflation started, $T_{\rm RH}$ is the temperature after reheating, and $f$ is the energy scale of confinement. We assume reheating to occur within one Hubble time, so that $T_{\rm RH} \propto f$.
The supercooled number density of quarks and gluons then becomes
	\begin{equation}
	Y_\TC^{\SC} =  D^{\rm SC}~Y_\TC^{\rm eq} \propto \left( \frac{T_{\rm nuc}}{f} \right)^3.
	\label{eq:SCintro}
	\end{equation}
	For completeness, the details entering Eq.~(\ref{eq:SCintro}) will be rederived in Sec.~\ref{sec:supercool_relics}. 
	
	When the fundamental techniquanta are swept into the expanding bubbles, they experience a confining force. Because $f \gg T_{\rm nuc}$ in the supercooled transition, the distance between them is large compared to the size of the composite states  $\psi$ (which we will equivalently call `hadrons').
The field lines attached to a quark or gluon then find it energetically more convenient to form a flux tube oriented towards the bubble wall, rather than directly to the closest neighbouring techniquantum, which is in general much further than the wall (see Fig.~\ref{fig:wall_diagram}).
The string or flux tube connecting the quark or the gluon and the wall then fragments, producing a number of hadrons inside the wall.
Additionally, because of charge conservation, techniquanta must be ejected outside the wall to compensate (see Fig.~\ref{fig:string_breaking}).
The process is conceptually analogous to the production of a pair of QCD partons at colliders, and we model it as such. The details are explained in~Sec.~\ref{sec:inside_bubble}.
The result is an increase of the yield of composite particles, compared to the naive estimate following directly from Eq.~\eqref{eq:SCintro}, by a string fragmentation factor $K^{\rm string}$,
	\begin{equation}
	Y_\psi^{\rm \SC+ string } =  K^{\rm string}  D^{\rm SC} ~Y_\TC^{\rm eq}
	\propto \left( \frac{ \Tnuc}{ f } \right)^3 \times  \text{logs}{\left(\frac{ \gwp \Tnuc }{ f }\right)},
	\end{equation}
where $\gwp > f/\Tnuc \gg 1$ is the Lorentz factor of the bubble wall at the time the quarks enter.

The Lorentz factor is estimated in Sec.~\ref{sec:wall_speed}.
In Sec.~\ref{sec:our_picture_relevant} we show that our picture can be relevant already for $\Tnuc/\Tstart\lesssim 1$.
The quarks ejected from the bubbles are treated in detail in Sec.~\ref{sec:outside_bubble}. 
We find they enter neighbouring bubbles and confine there into hadrons.
Acting as a cosmological catapult, string fragmentation at the wall boundary gives a large boost factor to the newly formed hadrons, such that their momenta in the plasma frame can be $\gg f$.

The composite states and their decay products can next undergo scatterings with other particles they encounter, e.g~with particles of the preheated `soup' after the bubbles collide.
Since the associated center-of-mass energy can be much larger than $f$, the resulting deep inelastic scatterings (DIS) increase the number of hadrons.
We explore this in detail in Sec.~\ref{sec:DIS}. The resulting effect on the yield can be encapsulated in a factor $K^{\rm DIS}$, and reads
	\begin{equation}
	Y_\psi^{\rm \SC+string+\DIS }
	=  K^{\rm DIS} D^{\rm SC}~Y_\TC^{\rm eq}
	\propto \left( \frac{ \Tnuc}{ f } \right)^3 \gwp
	\xmapsto{\rm if~runaway} \left( \frac{ \Tnuc}{ f } \right)^4 \frac{M_\text{Pl}}{m_*}\,,
	\label{eq:Yintro_SC+str+DIS}
	\end{equation}
	where $M_\text{Pl}$ is the Planck mass and $m_* = g_*f$ is the mass scale of hadrons. The last proportionality holds in the regime of runaway bubble walls, relevant for composite DM.

Finally the late-time abundance of the long-lived and stable hadrons, if any, evolves depending on their inelastic cross section in the thermal bath $\langle \sigma v_{\mathsmaller{\rm rel}}  \rangle$, and on $Y_\psi^{\rm \SC+string+\DIS }$ as an initial condition at $\TRH$. We compute it in Sec.~\ref{sec:DMabundance} by solving the associated Boltzmann equations.

By combining all the above effects we arrive at an estimate of the final relic abundance of the composite states. 
Our findings impact their abundance by several orders of magnitude, as can be seen in Fig.~\ref{fig:compositeDM_generic} for the concrete case where the relic is identified with DM.
The formalism leading to this estimate can readily be adapted for other purposes. For example, if $\psi$ instead decays out-of-equilibrium, it could source the baryon asymmetry. The estimate of $Y_\psi^{\rm \SC+string+\DIS}$ would then act as the first necessary step for the determination of the baryonic yield.

\section{Supercooling before Confinement}
\label{sec:supercool_relics}
\subsection{Strongly coupled CFT}

Although striving to remain as model independent as possible in our discussion, we shall be making a minimal assumption that the confined phase of the strongly coupled 
theory can be described as an EFT with a light scalar $\chi$, e.g.~a dilaton.
The scalar VEV, $\langle \chi \rangle$, then parametrizes the local value of the strong scale. It can be thought of as a scalar condensate of the strong sector, such as a glueball- or pion-like state. The scalar VEV at the minimum of its zero-temperature potential is identified with $\langle \chi \rangle = f$, where $f$ is the confinement energy scale, while $\langle \chi \rangle = 0$ at large enough temperatures.
In order to have strong supercooling, we require the approximate (e.g.~conformal) symmetry to be close to unbroken, thus justifying the lightness of the associated pseudo-Nambu-Goldstone boson (e.g.~the dilaton~\cite{Bardeen:1985sm}). That supercooling occurs with a light dilaton is known from a number of previous studies~\cite{Creminelli:2001th, Randall:2006py, Nardini:2007me}, see~\cite{Paterson:1980fc, Bruggisser:2018mus, Bruggisser:2018mrt, Baratella:2018pxi, Agashe:2019lhy, DelleRose:2019pgi, vonHarling:2019gme,Bloch:2019bvc,Azatov:2020nbe} for studies in a confining sector and 
\cite{Hassanain:2007js, Konstandin:2010cd, Konstandin:2011dr, vonHarling:2017yew,Fujikura:2019oyi,Bunk:2017fic, Dillon:2017ctw, Megias:2018sxv,Megias:2020vek, Agashe:2020lfz} for studies of holographic dual 5D warped extra dimension models.

\subsection{Thermal history}
\label{sec:thermal_history}
The vacuum energy before the phase transition is given by
	\begin{equation}
	\label{eq:lambda_vac}
	\Lambda_{\rm vac}^{4} \equiv c_{\rm vac} \, f^{4},
	\end{equation}
with some model dependent $c_{\rm vac} \sim \mathcal{O}(0.01)$ constant. The radiation density is given by
	\begin{equation}
	\rho_{\rm rad} = \frac{g_{R}\pi^{2}}{30}T^{4},
	\end{equation}
where $g_{R}$ counts the effective degrees of freedom of the radiation bath. We define $g_{R} \equiv g_{Ri} \; (g_{Rf})$ in the deconfined (confined) phase. Now consider the case of strong supercooling. The universe will enter a vacuum-dominated phase at a temperature
	\begin{equation}
	T_{\rm start} = \left( \frac{ 30\, c_{\rm vac} }{  g_{Ri} \, \pi^{2}} \right)^{\! 1/4}f,
	\label{eq:Tstart}
	\end{equation}
provided the phase transition has not yet taken place beforehand. The vacuum domination signals a period of late-time inflation. The phase transition takes place at the nucleation temperature, $T_{\rm nuc}$, when the bubble nucleation rate becomes comparable to the Hubble factor.
Following the phase transition, the dilaton undergoes oscillations and decay, reheating the universe to a temperature
	\begin{equation}
	T_\RH = \left( \frac{g_{Ri}}{g_{Rf}} \right)^{1/4}T_{\rm start},
	\end{equation}     
At this point the universe is again radiation dominated. We have assumed the decay to occur much faster than the expansion rate of the universe such that we can neglect a matter-dominated phase~\cite{Hambye:2018qjv}.

\subsection{Dilution of the degrees of freedom}
\label{sec:dilution}
Now consider some fundamental techniquanta of the strong sector, e.g. techniquarks or technigluons (for simplicity we always refer to them as quarks and gluons). Prior to the phase transition the number density of techniquanta follows a thermal distribution for massless particles
	\begin{equation}
	n_\TC^{\rm eq} = g_\TC \frac{\zeta(3)}{\pi^{2}}T^{3},
	\end{equation}
where $g_\TC$ denotes the degrees of freedom of the quanta under consideration.
The entropy density is given by
	\begin{equation}
	s = \frac{2\pi^{2}g_s}{45}T^{3},
	\end{equation}
where $g_s$ are the total entropic degrees of freedom.\footnote{In a picture with $N_f$ flavours of quarks in fundamental representations of an $SU(N)$ confining gauge group, one has $g_q = 2 N_f N$, $g_g = 2 (N^2-1)$, $g_\TC = g_g + 3g_q/4$, $g_s = g_g + 7g_q/8$.
}
The number density normalized to entropy before the phase transition,
	\begin{equation}
	Y_\TC^{\rm eq} = \frac{45 \zeta(3) \, g_\TC}{2\pi^4 g_s},
	\label{eq:Yeqi}
	\end{equation}
remains constant up to the point when the phase transition takes place. 
The entropy density then increases during reheating giving 
	\begin{equation}
	\label{eq:qyield}
	Y_\TC^{\SC} = D^{\rm SC} ~Y_\TC^{\rm eq},
	\end{equation}  
when we find ourselves back in the radiation-dominated phase.
The dilution factor from the additional expansion during the vacuum-dominated phase can be derived by finding the increase in entropy between $\Tnuc$ and $T_{\rm RH}$. It reads
	\begin{equation}
	\label{eq:DSC}
	D^{\rm SC}
	\equiv \left( \frac{T_{\rm nuc}}{T_{\rm start}} \right)^3 \left( \frac{T_{\rm RH}}{T_{\rm start}} \right)
	\simeq \frac{g_{Ri}}{c_\text{vac}^{3/4} g_{Rf}^{1/4} } \, \Bigg(\frac{\Tnuc}{f}\Bigg)^3\,.
	\end{equation}
If the quarks and gluons were non-interacting following the phase transition, the yield today would be given by the above formula. (In the presence of interactions the above would be taken as an initial condition at $T_{\rm RH}$ for the Boltzmann equations describing the effects of number changing interactions between reheating and today.) The picture would then be analogous to that studied, in a theory without confinement, in~\cite{Hambye:2018qjv}.
The picture is completely changed, however, for supercooled confining phase transitions, which we elucidate next.

\section{Confinement and String Fragmentation}
\label{sec:inside_bubble}

\subsection{Where does confinement happen?}
\label{sec:when_confinement}

\begin{figure}[t]
\centering
\hspace{-1cm}
\raisebox{0cm}{\makebox{\includegraphics[height=0.35\textwidth, scale=1]{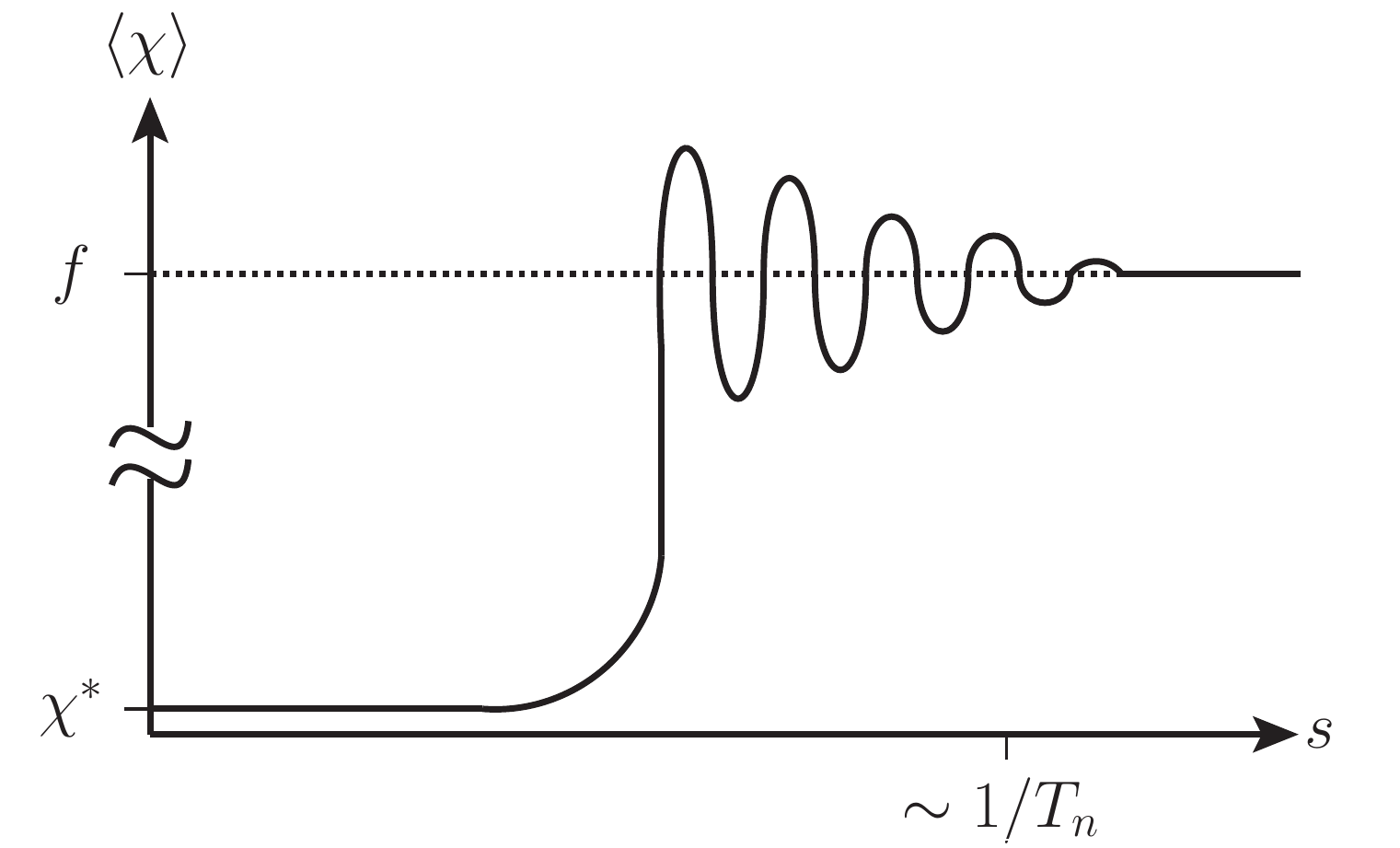}}}
\caption{\it \small A typical wall profile found in close-to-conformal potentials. After nucleating by tunneling to the exit point, $\chi^{\ast} \ll f$, the field rolls down and undergoes damped oscillations around the minimum of its potential. The typical wall thickness is $L_w \sim 1/\Tnuc$.}
\label{fig:wall_profile} 
\end{figure}

\paragraph{Bubble wall profile.}
The expanding bubble is approximately described by the Klein-Gordon equation~\cite{Jinno:2019bxw}
	\begin{equation}
	\label{eq:KG}
	\frac{ d^{2} \chi }{ d s^{2} } +  \frac{3}{s}\frac{d \chi}{ ds}  + \frac{d V }{ d\chi}   = 0,
	\end{equation}
where $s^{2} =t^2- r^{2}$ is the light-cone coordinate and $V$ is the scalar potential.
A sketch of a typical bubble profile for close-to-conformal potentials is shown in Fig.~\ref{fig:wall_profile}.
The key point here is that the wall thickness is 
	\beq
	L_{\rm w} \lesssim \frac{1}{\Tnuc},
	\label{eq:Lw}
	\eeq
as shown by numerical computations and analytical estimates, see App.~\ref{app:wall_profile} for a calculation in an explicit example.

\paragraph{Confinement time scale.}
The techniquanta (quarks and gluons) constitute a plasma with temperature of order $\Tnuc$ before entering the bubble. Once they enter the bubbles, they could in principle either confine in a region close to the bubble wall where $\langle \chi \rangle \ll f$,
or approach as free particles the region where $\chi$ has reached its zero-temperature expectation value $\langle \chi \rangle = f$.
To determine this, let us define a `confinement rate' and a `confinement length' as
	\beq
	\Gamma_\text{conf} = L_\text{conf}^{-1} = n_\TC\, v_\TC\, \sigma_\text{conf},
	\label{eq:confinement_rate}
	\eeq
where $n_\TC$ and $v_\TC$ are, respectively, the number density and the relative M{\o}ller velocity of the techniquanta $v_\TC \equiv \left[|\bold{v}_1-\bold{v}_2|^2 - |\bold{v}_1\times\bold{v}_2|^2  \right]^{1/2}$~\cite{moller1945general}, and $\sigma_\text{conf}$ is a `confining cross section'. 
We want to compare $L_\text{conf}$  with the length of the bubble wall, defined as the distance over which $\chi$ varies from its value at the exit point, $\langle \chi \rangle = \chi^* \ll f$, to $\langle \chi \rangle =f$.
Of course we need to perform this comparison in the same Lorentz frame, so we emphasise our definition of $L_w$ as the bubble-wall length in the bubble-wall frame, and $L_{\rm p} = L_{\rm w}/\gwp$ as the bubble-wall length in the frame of the center of the bubble, which coincides with the plasma frame, and where $\gwp$ is the boost factor between the two frames.
Let us now move to the confinement timescale of Eq.~(\ref{eq:confinement_rate}).
Since we expect confinement to happen `as soon as possible', we assume the related cross section to be close to the unitarity limit \cite{Griest:1989wd},
	\beq
	\sigma_\text{conf} \sim \frac{4\pi}{\Tnuc^2}\,.
	\label{eq:confinement_cross_section}
	\eeq
Since $n_\TC^2 v_\TC$ is Lorentz invariant~\cite{moller1945general, Gondolo:1990dk},
one then has that $n_\TC \,v_\TC$ transforms under boosts as $n_\TC^{-1}$.
The boost to apply in this case is $\gwp$, because by definition the string forms after confinement, so we can treat the plasma frame as the center-of-mass frame of the techniquanta.
Combining this with the Lorentz invariance of the cross section, we obtain
	\beq
	\label{eq:conf_rate}
	\Gamma_{\text{conf},w}
	= n_{\TC,\, \rm w}\,v_{\TC,\, \rm w}\,\sigma_\text{conf}
	=  \frac{ n_{\TC,\, \rm p}\,v_{\TC, \rm \,p} }{ \gwp }\,\sigma_\text{conf}
	\sim  \frac{ 4 \pi \,\Tnuc }{ \gwp },
	\eeq
where in the last equality we have used that the average relative speed and density of the techniquanta in the plasma frame satisfy, respectively, $v_{\TC,\,p} \simeq 1$ and $n_{\TC,\,p} \sim \Tnuc^3$, because they are relativistic.
This in turn implies
	\beq
	\label{eq:Lconf}
	L_{\text{conf, w}}
	\sim \frac{\gwp}{4\pi}L_{\rm w}.
	\eeq
	
\paragraph{Confinement takes place deep inside the bubble.}
For the regimes of supercooling we are interested in, the phase transition is of detonation type and the Lorentz factor $\gwp$ is orders of magnitude larger than unity.
Therefore, $L_{\text{conf},\,w} \gg L_{\rm w}$ such that confinement does not happen in the outermost bubble region where $\langle \chi \rangle \ll f$.
This conclusion is solid in the sense that it would be strengthened by using a confinement cross section smaller than what assumed in Eq.~(\ref{eq:confinement_cross_section}), which is at the upper end of what is allowed by unitarity. The end effect of the above discussion, is that for practical purposes, we can consider the wall profile to be a step-like function between the deconfined phase, $\langle \chi \rangle=0$, and confined phase, $\langle \chi \rangle=f$. Furthermore, as we shall discuss below, the quarks will not confine directly in pairs but rather form fluxtubes pointing toward the bubble wall as they penetrate the $\langle \chi \rangle=f$ region of the bubble.

\paragraph{The ballistic approximation is valid.}
Equation~\eqref{eq:Lconf}, together with the large wall-Lorentz-factors encountered in this study, implies that we can safely neglect the interactions between neighbouring techniquanta during the time when they cross the bubble wall. This is the so-called ballistic regime, see e.g.~\cite{Mancha:2020fzw}, which will be useful for deriving the friction pressure in Sec.~\ref{sec:wall_speed}.

\subsection{Fluxtubes attach to the wall following supercooling}
\label{sec:string_breaking}

\paragraph{A hierarchy of scale.}
Upon entering the region $\langle \chi \rangle = f$ of expanding bubbles, the techniquanta experience a confinement potential much stronger than in the region close to the wall.
This can be easily understood by taking the long-distance potential of the Cornell form~\cite{Eichten:1974af, Thorn:1979gv, Greensite:1988tj, Poulis:1996iv, Ko:1999yx, Greensite:2001nx, Bardakci:2002xi, Greensite:2003xf, Greensite:2003bk,Trawinski:2014msa}
	\begin{equation}
	\label{eq:stringenergy}
	E_\TC = c_\TC \,f^{2} \, d_{c},
	\end{equation}
where $d_c$ is the techniquanta seperation in their `center of interaction frame' (or equivalently `string center of mass frame')\footnote{Lattice simulations find that the QCD potential at $d_c \gtrsim$~fm saturates to a constant, a behavior which is interpreted in terms of pair creation of quarks from the vacuum, see e.g. the recent~\cite{Bulava:2019iut}. Therefore this realises an outcome that, for our purposes, coincides with having $E_\TC \propto d_c$ to larger distances. Lattice simulations with quarks only as external sources~\cite{Bali:2000gf}, so without sea quarks (`quenched'), find that the linear regime of the QCD Cornell potential extends up to the maximal distances probed, namely $d_c\simeq 3$~fm in the results reported in~\cite{Bali:2000gf}.
}, and $c_\TC$ is an adimensional constant\footnote{$c_\TC$ does not hide any `coupling dimension', indeed in units where $\hbar \neq 1$, $[f] = (\text{energy}/\text{distance})^{\!\frac{1}{2}}$.
}, $c_{q\bar{q}} \simeq 10$ in QCD~\cite{Trawinski:2014msa}.
A crucial point regarding the string energy in this context, besides the fact it grows proportionally to $\chi^2$, is that the inter-quanta distance is large compared to the natural confinement scale, i.e. $d_{c} \gg f^{-1}$, due to the supercooling.
Indeed the distance between quanta outside the wall, in the plasma and wall frames respectively, scales as $d_p \sim \Tnuc^{-1}$ and $d_w \sim \gwp^{-1/3} \Tnuc^{-1}$. Since $\gwp \ll (f/\Tnuc)^3$ (see Sec.~\ref{sec:wall_speed}) and $d_c \geq d_w$ (because $d_c = d_p$ outside the wall, and because the quarks and gluons cannot be accelerated upon entering so $d_w$ is Lorentz contracted with respect to $d_p$), one ends up with $d_c \gg f^{-1}$.
What happens then to the techniquanta and to the fields connecting them?

\begin{figure}[t]
\begin{minipage}[t]{0.6\linewidth}
    \centering
    \includegraphics[height=7 cm]{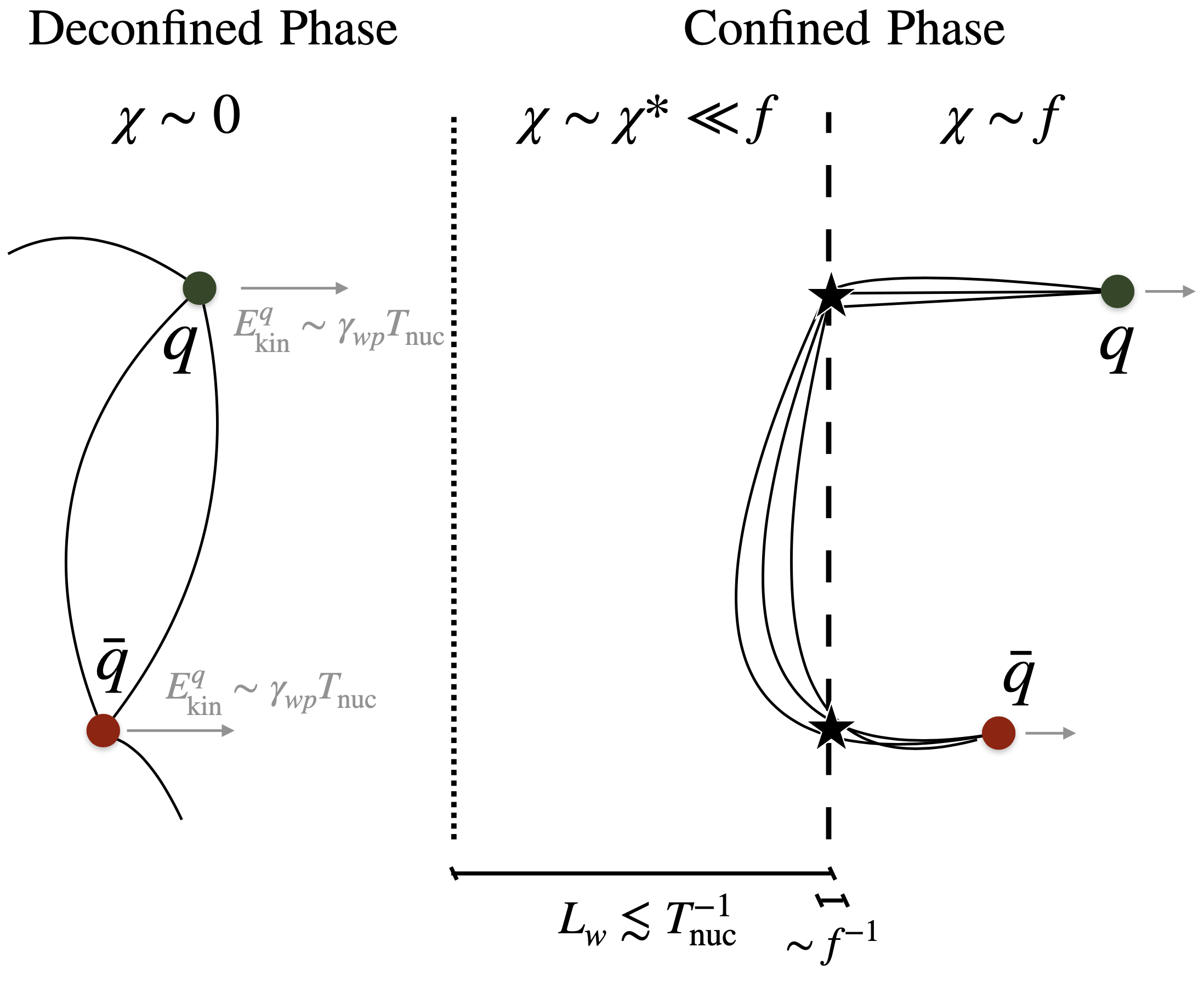}
    \caption{\it \small  Quarks entering the bubble as seen in the frame of the bubble wall, together with the associated field lines and quantities defined in the text. The rest energy of the string is minimized if the fluxtubes in the region $\chi = f$ point to the bubble wall, rather than if they point to the closest color charge.}
    \label{fig:wall_diagram}
\end{minipage}
\hspace{0.02\linewidth}
\begin{minipage}[t]{0.38\linewidth} 
    \centering
    \includegraphics[height=7 cm]{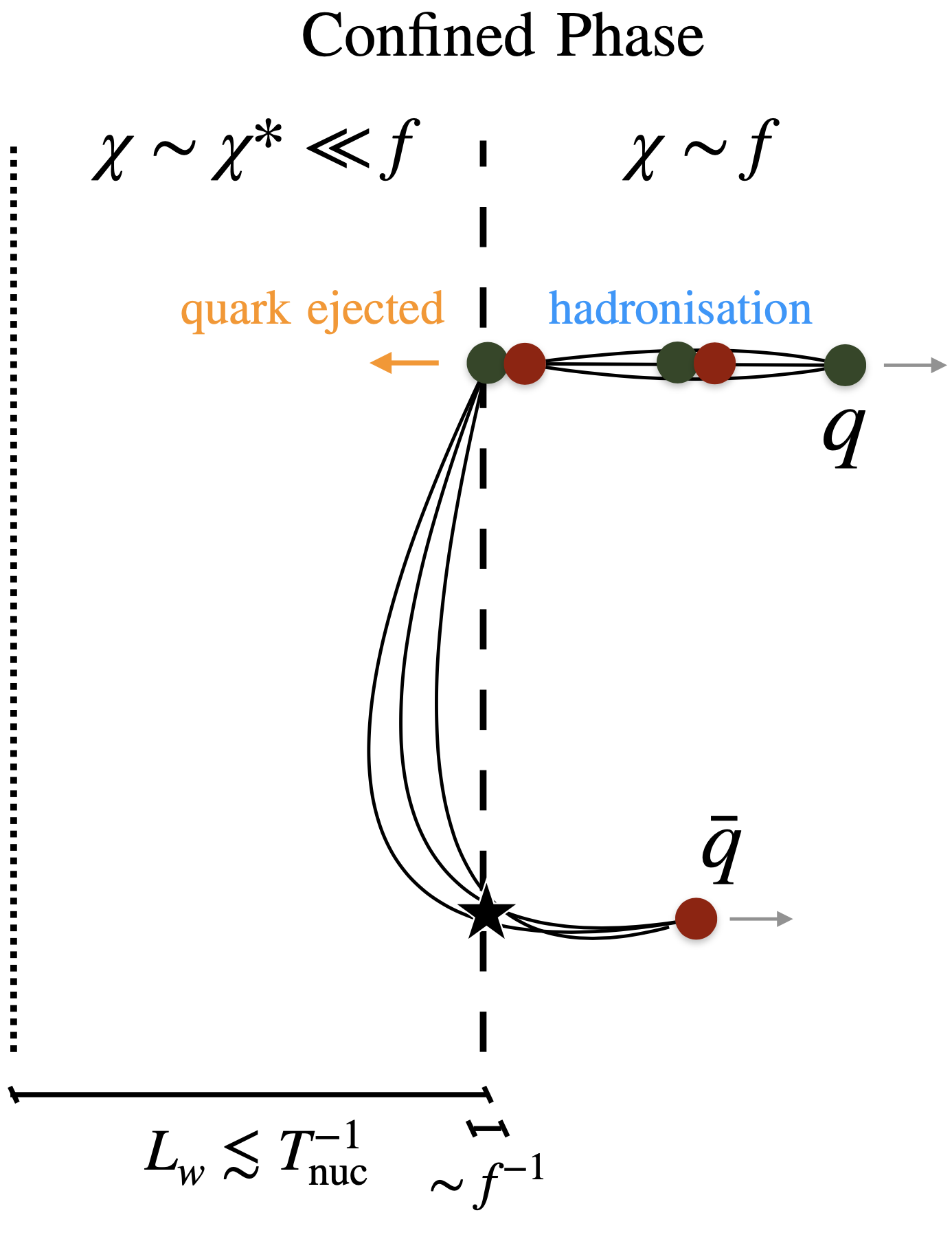}
    \caption{\it \small The string inside the wall breaks, producing hadrons (Sec.~\ref{sec:fragmentation_multiplicity_energy}), and a quark is ejected from the wall (Sec.~\ref{sec:ejected}).}
    \label{fig:string_breaking}
\end{minipage}        
\end{figure}  

\paragraph{Flux tubes minimize their energy.}

In a picture without hierarchy of scales, the fields would compress in fluxtubes connecting different charges, `isolated' in pairs or groups to form color-singlets. Here, we argue that the fluxtubes have another option, which is energetically preferable: that of orienting themselves towards the direction of minimal energy, i.e. as perpendicular as possible to the bubble-wall\footnote{We wish to express our gratitude to Benedict von Harling, Oleksii Matsedonskyi, and Philip Soerensen, for discussions which lead us to develop the picture we employ in this paper.}, and to keep a `looser' connection in the outer region where $\chi \ll f$.
Indeed, a straight-line connection between techniquanta would result in a much longer portion of fluxtubes in the region $\langle \chi \rangle = f$, with respect to our picture of fluxtubes perpendicular to the wall.
Via Eq.~\eqref{eq:stringenergy}, this would in turn imply a much higher cost in energy, disfavoring that option.
We stress that, in our picture, the fluxtubes are still connecting techniquanta in such a way to form an overall color singlet, just these fluxtubes minimise their length in the region $\chi = f$, and partly live in a region $\chi \simeq \chi^* \ll f$.
This picture is visualised in Fig.~\ref{fig:wall_diagram}. Note the nearest neighbour quark from the plasma may also be located outside the bubble.

\paragraph{Condensed matter analogy.}
An interesting analogue to the picture above is the vortex string of magnetic flux in the Landau-Ginzburg model of superconductivity. To match onto confinement dynamics a dual superconductor is pictured, in which the external colour-electric field --- rather than the magnetic field --- is expelled by the Meissner effect~\cite{Ripka:2003vv}. Here the bubble of confining phase corresponds to the superconductor from which the colour-electric field is expelled. Quarks entering the bubble then map onto magnetic monopoles being fired into a regular superconductor.

\subsection{String energy and boost factors}
\label{sec:boosts}

To possibly be quantitative on the implications of the picture we just outlined, we first need to determine the string energy and the Lorentz boosts among the frames of the plasma, wall and center-of-mass of the string.

\paragraph{String end-points.}
Let us define as $\TC_i$ the quark or gluon that constitutes an endpoint, inside the bubble, of a fluxtube pointing towards the wall, and $\bigstar$ the end-point of the fluxtube on the wall. The energy of the incoming techniquantum in the wall frame is $E_{i,\text{w}} = 3 \gwp \Tnuc$, where for simplicity we have averaged over their angle with respect to the wall. We assume $\bigstar$ to be at rest or almost, and to carry some $\mathcal{O}(1)$ fraction of the inertia of the string. Hence the respective four-momenta are
\beq
p_{i,\text{w}} = 
\begin{pmatrix}
 3\,\gwp \Tnuc \\
\sqrt{9\, \gwp^2 \Tnuc^2 - m_i^2}
\end{pmatrix},
\quad
p_{\bigstar,w} = 
\begin{pmatrix}
m_f \\
\epsilon f
\end{pmatrix},
\qquad \epsilon \ll 1,
\qquad m_i \simeq m_f = q f, \quad q\leq \frac{1}{2}\,.
\label{eq:momenta_star}
\eeq

\paragraph{String center-of-mass.}
Then we define the center-of-mass of the string as the one of $\TC_i$ and $\bigstar$, and find
\beq
\ECM = |p_{\bigstar,w} + p_{i,\text{w}}|
\simeq \sqrt{3\, \gwp \, \Tnuc\, f}\,,
\label{eq:ECMstring}
\eeq
where the second expression is valid up to relative orders $(\gwp f/\Tnuc)^{-1} \ll 1$.
By employing a Lorentz boost between the wall and center-of-mass frames, and imposing $\vec{p}_{i,c} = - \vec{p}_{\bigstar,c}$, we find
\beq
\gwc \simeq \sqrt{3\,\gwp\frac{\Tnuc}{f}}\,.
\label{eq:gwc}
\eeq
On the right-hand side of the equations above we have omitted a factor of $\sqrt{2 (q-\epsilon)}$, in~(\ref{eq:ECMstring}), and of $1/\sqrt{2 (q-\epsilon)}$, in~(\ref{eq:gwc}), because for simplicity we take these to be $\approx 1$ from now on (as per the benchmark $q=1/2$, $\epsilon=0$).
Finally we determine the boost between the center-of-mass frame of the string and the plasma frame as
\beq
\gcp \simeq\frac{\gwp}{2\gwc} = \frac{1}{2} \sqrt{\frac{\gwp}{3} \frac{f}{\Tnuc}},
\label{eq:gcp}
\eeq
which is valid up to a relative order $(\gwp f/\Tnuc)^{-1} \ll 1$.

\subsection{Hadrons from string fragmentation: multiplicity and energy}
\label{sec:fragmentation_multiplicity_energy}

The fluxtubes connecting a quark or gluon to the wall will fragment and form hadrons, singlet under the new confining gauge group. We would now like to determine:
\begin{itemize}
\item The number of hadrons formed per fluxtube.
\item The momenta of said hadrons.
\end{itemize}
\paragraph{Collider analogy.}
We start by noticing that the process of formation of a fluxtube, in our picture, is analogous to two color charges in an overall-singlet state, $\TC_i$ and $\bigstar$, moving apart with a certain energy $\ECM$, where $\ECM =  \sqrt{3\,\gwp\, \Tnuc\, f}$ in the modelling of Sec.~\ref{sec:boosts}.
This physical process appears entirely analogous to what would happen in a collider that produces a pair of techniquanta of the new confining force, starting from an initial singlet state. In light of this observation, we then decide to model the process
by analogy with a very well-studied process observed in Nature, that of QCD-quark pair production at electron-positron colliders, where the analogy lies also in the fact that the initial state electron-positron pairs is in a color singlet state.
Needless to say, a BSM confining sector needs not behave as QCD in terms of number and momenta of hadrons produced per scattering, see e.g.~\cite{Strassler:2008bv}.
However, QCD constitutes a well studied and tested theory, so that we find it reasonable to use it as our benchmark. Moreover, we anticipate from Sec.~\ref{sec:DIS} that our final result for the cosmological abundance of hadrons, in the assumption of efficient-enough interactions between them and the SM, will only depend on the initial available energy $\ECM$. This suggests that, within that assumption, our final findings hold for confining sectors that distribute this energy over a number of hadrons different from~QCD.

\begin{figure}[t]
\centering
\begin{adjustbox}{max width=1.2\linewidth,center}
\includegraphics[width= 0.5\textwidth]{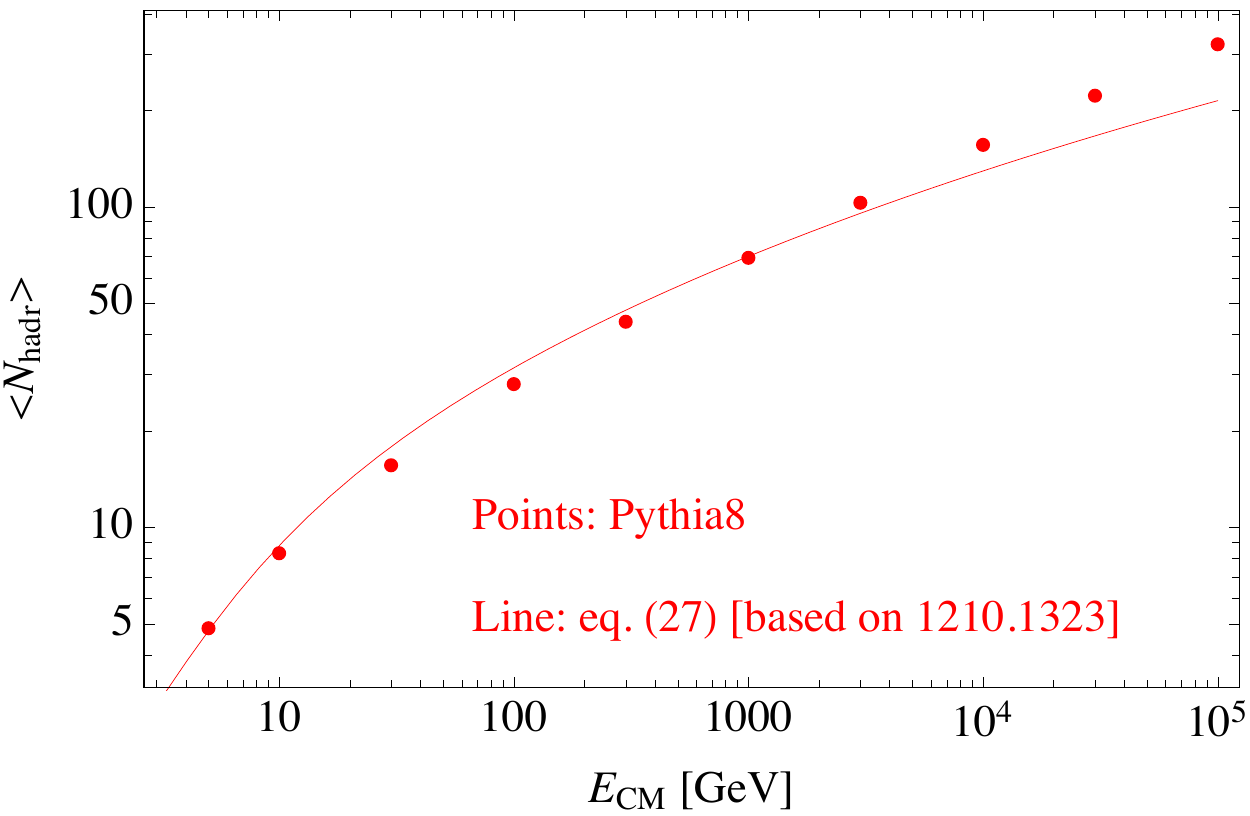}
\;
\includegraphics[width= 0.5\textwidth]{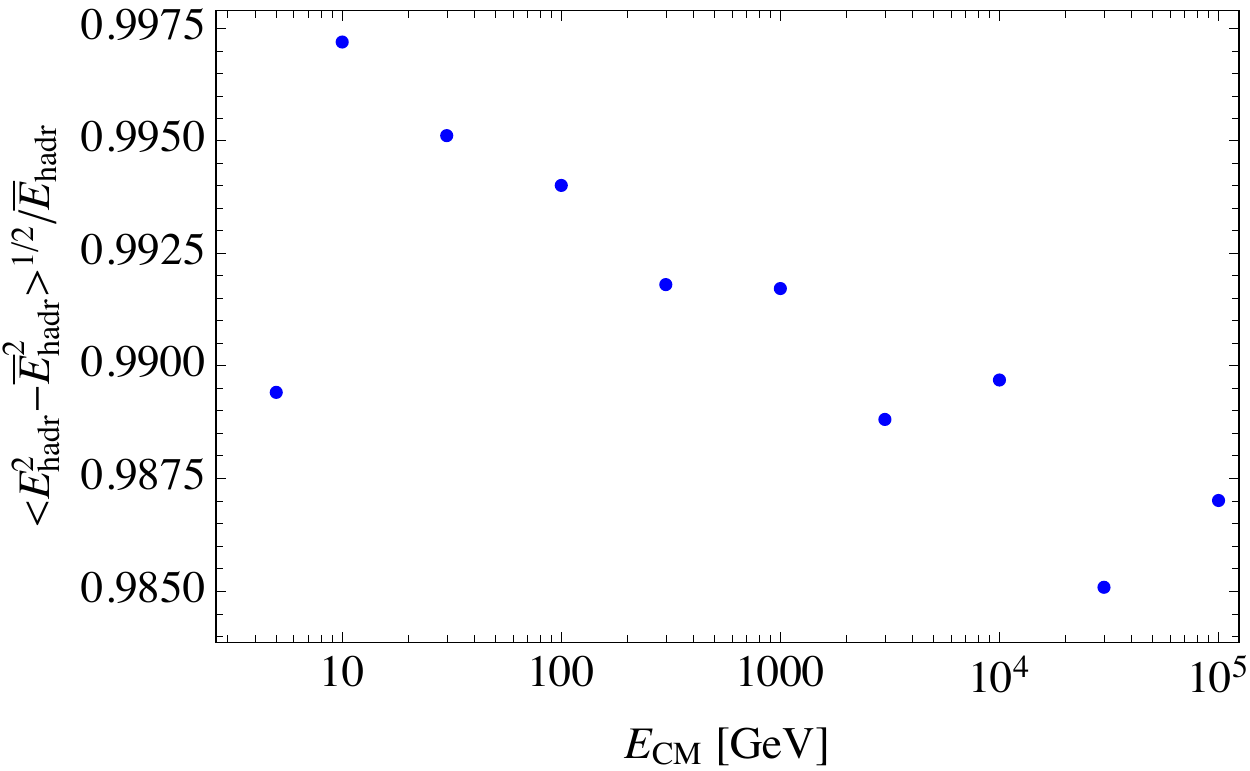}
\end{adjustbox}
\caption{\it \small Left: average hadron multiplicity per single QCD scattering $e^-e^+ \to q\bar{q}$.
Right: Root square mean of the hadron energy per single QCD scattering $e^-e^+ \to q\bar{q}$, where $\bar{E}_\text{hadr} = \ECM/\langle N_\text{hadr}\rangle$ is the average hadron energy per scattering.
Dots in both plots are extracted via MadAnalysis v1.8.34~\cite{Conte:2012fm} by simulations with MadGraph v2.7.0~\cite{Alwall:2014hca} plus Pythia v8.2~\cite{Sjostrand:2014zea}, the line in the left-hand plot displays Eq.~(\ref{eq:Npsi_string}).
Results are expressed as a function of the center-of-mass energy of the scattering in GeV, to export them to our cosmological picture we simply substitute $\mathrm{GeV}\simeq 4\pi f_\pi$ by $m_* = g_* f$, and use $\ECM = \sqrt{3\, \gwp \, \Tnuc\, f}$.}
\label{fig:pythia} 
\end{figure}

\paragraph{Numerical simulations.}
We use Pythia v8.2~\cite{Sjostrand:2014zea} interfaced to MadGraph v2.7.0~\cite{Alwall:2014hca} to simulate the process $e^-e^+ \to q\bar{q}$ for different center-of-mass energies, and MadAnalysis v1.8.34~\cite{Conte:2012fm} to extract from these simulations both the total number of hadrons produced per scattering and their energy distribution. We thus recover known QCD results and display them in Fig.~\ref{fig:pythia}.
We translate them to our picture by replacing the units of a GeV$\simeq 4 \pi f_\pi$ used by Pythia, with the generic mass of a composite state $m_* = g_* f$, where $1\leq g_* \leq 4\pi $ is some strong effective coupling.
These results can be summarised as follows:
\begin{itemize}
\item The number of hadrons produced per fluxtube grows logarithmically in $\ECM$.
\item The distribution of hadron energies is such that its root square mean coincides, to a percent level accuracy, with the average energy per hadron
\beq
\bar{E}_\text{hadr} = \frac{\ECM}{\langle N_\text{hadr}\rangle}\,.
\label{eq:energy_per_hadron}
\eeq
This will support, in Sec.~\ref{sec:DIS_hadron_abundance}, our simplifying assumption that all hadrons produced by the string fragmentation carry an energy of order $\bar{E}_\text{hadr}$.
\end{itemize}

\medskip
\paragraph{Results from the literature.}
The multiplicity of QCD hadrons from various scattering processes has been the object of experimental and theoretical investigation, since the late 1960s~\cite{Feynman:1969ej}.
We now leverage such studies both to check the results of our simulation and to obtain analytical control over them.
Collider studies have typically focused on the multiplicity of charged QCD resonances per scattering, $\langle n_\text{ch}\rangle $. In particular, works such as~\cite{GrosseOetringhaus:2009kz,Kumar:2012hx} have carried out the exercise of collecting the most significant measurements of $\langle n_\text{ch}\rangle$ and `filling' the missing phase space --- not covered by detectors --- with the output of MC programs, thus obtaining a full-phase-space quantity. We take as our starting point the result provided in~\cite{Kumar:2012hx} from $pp$ collisions, which reads
\beq
\langle n_\text{ch}\rangle (\ECM) = a + b \log \frac{\ECM}{m_*} + c \log^2 \frac{\ECM}{m_*} + d \log^3 \frac{\ECM}{m_*},
\label{eq:ncharged_s}
\eeq
with $(a,b,c,d) = (0.95,0.37,0.43,0.04)$. Here, as already explained, we substituted the normalisation of a GeV with $m_* = g_* f$. 
\paragraph{Our modelling.}
To obtain the total number of hadrons from $e^+e^-$ collisions we proceed as follows.
First, most hadrons coming out from hard scatterings consist in the lightest ones, i.e. the pions.
Second, the total number of pions produced is very well approximated by $3 \langle n_\text{ch}\rangle /2$, because of isospin conservation. By the first argument, this coincides with very good approximation to the total number of hadrons produced. 
Third, the multiplicity of composite states from $e^+e^-$ collisions has been found to roughly match the one from $pp$ collisions, upon increasing the $e^+e^-$ energy by a factor of 2, see e.g.~Sec.~2.2 in~\cite{Rosin:2006av}\footnote{\label{foot:Npsi_ee_vs_pp}This is qualitatively understood by the fact that, in purely leptonic initial states, there is more energy available to produced hadrons, while in the case with protons in the initial state much energy is carried over by the initial hadron remnant.
}.
We then model the total number of composite states produced, per string fragmentation, as
\beq
N_\psi^\text{string}(\ECM)
\simeq
\frac{3}{2} \langle n_\text{ch}\rangle (2 \ECM) \exp(-3 m_*/\ECM) + 1\,,
\label{eq:Npsi_string}
\eeq
where we have multiplied by an exponential and added one to smoothen $N_\psi^\text{string}(\ECM)$ to 1 as  $\ECM \to m_*$, because this physical regime was not taken into account in\cite{Kumar:2012hx}.
In the left-hand panel of Fig.~\ref{fig:pythia} one sees that Eq.~(\ref{eq:Npsi_string}) reproduces the results of our Pythia simulation for $\ECM$ smaller than a few TeV rather well. This was to be expected since Eq.~(\ref{eq:ncharged_s}) was determined in~\cite{Kumar:2012hx} from fits to data up to that energy. It is not the purpose of this paper to improve on this fit, as stated above, we simply use the above results as a check of our Pythia simulation.

\subsection{Enhancement of number density from string fragmentation}
\label{sec:string_summary}

\paragraph{Production of composite states.}
Prior to (p)reheating, we then have a yield of composite states given by the yield of strings, which can be estimated from Eq.~(\ref{eq:qyield}), multiplied by the number of composite states per string
\begin{equation}
K^{\rm string} = 
\left\{  \begin{array}{ll}
             \frac{\frac{3}{4}g_q N_\psi^\text{string}(\ECM) + g_g (N_\psi^\text{string}(\ECM)-1)}{g_\TC} & \quad \text{heavy~composite~state}, \\
               N_\psi^\text{string}(\ECM) & \quad \text{light~composite~state}\,,
                \end{array}
              \right.
\label{eq:Kstring}
\end{equation}
where $N_\psi^\text{string}(\ECM)$ is given by Eq.~\eqref{eq:Npsi_string} and $\ECM = \sqrt{3\, \gwp \, \Tnuc\, f}$ in Eq.~\eqref{eq:ECMstring}.
We have distinguished the cases where the composite state of interest is heavier or lighter than the glueballs (e.g. the analogous of a proton or a pion in QCD). In the former case, the $-1$ we added to the factor multiplying $g_g$ accounts for the fact that, if the final composite states produced by string fragmentation do not undergo other additional interactions, then glueballs decay to the light composite states and do not contribute to the final yield of any heavy composite state of quarks. 
The yield of composite states $\psi$ then reads
\begin{align}
	Y_{\psi}^{\rm \SC + string} & =  \,Y_\TC^{\rm eq} ~ D^{\rm SC}  ~ K^{\rm string}
	\propto \left( \frac{ \Tnuc}{ f } \right)^3 \times  \text{logs}{\left( \frac{ \gwp\Tnuc }{ f } \right)}.
	\label{eq:Ystring} 
	\end{align}
	The appearance of $Y_\TC^{\rm eq}$ in Eq.~(\ref{eq:Ystring}) accounts for string formation from both quarks and gluons. Hence, not only is the number of $\psi$'s enhanced by the string fragmentation, relative to the case with no confinement, but also by the possibility of gluons to form strings. $K^{\rm string}$ and $Y_{\psi}^{\rm \SC + string} $ are plotted in Fig.~\ref{fig:contributions}.

\paragraph{Hadrons are highly boosted in the plasma frame.}
The hadrons formed after string fragmentation schematically consist of two equally abundant groups.  
Hadrons in the first group, which for later convenience we call `Population A', move towards the bubble wall with an average energy
\beq
E_\text{A,p} \simeq 2 \gcp \frac{ \ECM}{N_\psi^\text{string}(\ECM)} \simeq \frac{\gwp f}{N_\psi^\text{string}(\ECM)},
\label{eq:EAp}
\eeq
 where we have boosted the energy per hadron of Eq.~(\ref{eq:energy_per_hadron}) to the plasma frame with the $\gcp$ of Eq.~(\ref{eq:gcp}), and also used Eqs.~(\ref{eq:ECMstring}) and (\ref{eq:Kstring}).
  We conclude from Eq.~\eqref{eq:EAp} that the newly formed hadrons have large momenta in the plasma frame. The formation of a gluon string between the incoming techniquanta and the wall acts as a cosmological catapult which propels the string fragments in the direction the wall is moving.
Hadrons in the second group move, in the wall frame, towards the bubble wall center, and their energy in the plasma frame is negligible compared to~(\ref{eq:EAp}).
Note that if only one hadron is produced on average per every string, then it would roughly be at rest in the center-of-mass frame of the string, with an energy (mass) of order $\ECM$. In the plasma frame, its energy would then read $E_\text{p} \simeq \gcp \ECM \simeq \gwp f/2$. As we will see in~Sec.~\ref{sec:DIS}, the impact of this hadron on the final yield would then be captured by our expressions.

Following this first stage of string fragmentation, the composite states, and/or their decay products, can undergo further interactions with remnant particles of the bath, preheated or reheated plasma, and among themselves.
Such interactions may change the ultimate yield of the relic composite states.
Before taking these additional effects into account in Sec.~\ref{sec:DIS}, in the next sections we complete the modelling we proposed above, by describing the behaviour of the ejected quarks and deriving the Lorentz factor of the wall, $\gwp$.

\subsection{Ejected quarks and gluons and their energy budget}
\label{sec:ejected}
So far we dealt with what happens inside the bubble wall.
The process we described apparently does not conserve color charge: we started with a physical quark or gluon with a net color charge entering the bubble, and we ended up with a system of hadrons which is color neutral. Where has the color charge gone?

\paragraph{The necessity of ejecting a quark or gluon.}
To understand this, it is convenient to recall the physical modelling behind the process of string fragmentation that converts the initial fluxtube into hadrons, see e.g.~the original Lund paper~\cite{Andersson:1983ia}.
When the fluxtube length, in its center-of-mass frame, becomes of order $f^{-1}$, the string breaks at several points via the nucleation of quark-antiquark pairs from the vacuum. Now consider, in our cosmological picture, the quark-antiquark pair nucleated closest to the bubble wall. One of the two --- say the antiquark --- forms a hadron inside the wall. The only thing that can happen to the quark is for it to be ejected from the wall, because of the lack of charge partners inside the wall.
This process, somehow reminiscent of black hole evaporation, thus allows for charge to be conserved.
The momentum of the ejected quark, in the wall frame, has to be some order-one fraction of the confinement scale $f$, because that is the only energy scale in the process. For definiteness, in the following we will take this fraction to be a half.
This picture is visualized in Fig.~\ref{fig:string_breaking}, and it is analogous if $\TC_i$ is a gluon instead of a quark.
 
\medskip
\paragraph{Energy of the ejected quark or gluon.}
One then has one ejected quark (at least) or gluon per fluxtube, thus per quark or gluon that initially entered. Therefore, the number of techniquanta outside the bubble wall does not diminish upon expansion of the bubble.
This population of ejected techniquanta is energetically as important as that of hadrons inside the bubble.
Indeed the energy of an ejected quark or gluon (or quark pair), in the plasma frame, reads
\beq
E_\text{ej,p}\simeq \gwp f.
\label{eq:energy_ejected}
\eeq
This is of the same order as the total energy in the hadrons from the fragmentation of a single string,
\beq
E^\text{tot}_\text{A,p} = \frac{N_\psi^\text{string}(\ECM)}{2} E_\text{A,p} \simeq \gwp \frac{f}{2},
\label{eq:ECMp}
\eeq
obtained by multiplying $E_\text{A,p}$ of Eq.~(\ref{eq:EAp}) times half of the total number of hadrons produced per string (i.e.~we included only the energetic ones).
The population of ejected techniquanta cannot therefore be neglected in the description of the following evolution of this cosmological system.

\section{Bubble wall velocities}
\label{sec:wall_speed}

The wall boost in the plasma frame, $\gwp$, affects many key properties of our scenario, from the ejection of techniquanta to the number and energy of the hadrons produced by string fragmentation.
It is the purpose of this section to study the possible values it can take over the~PT.

\begin{figure}[t]
\begin{center}
\includegraphics[width=.65\textwidth]{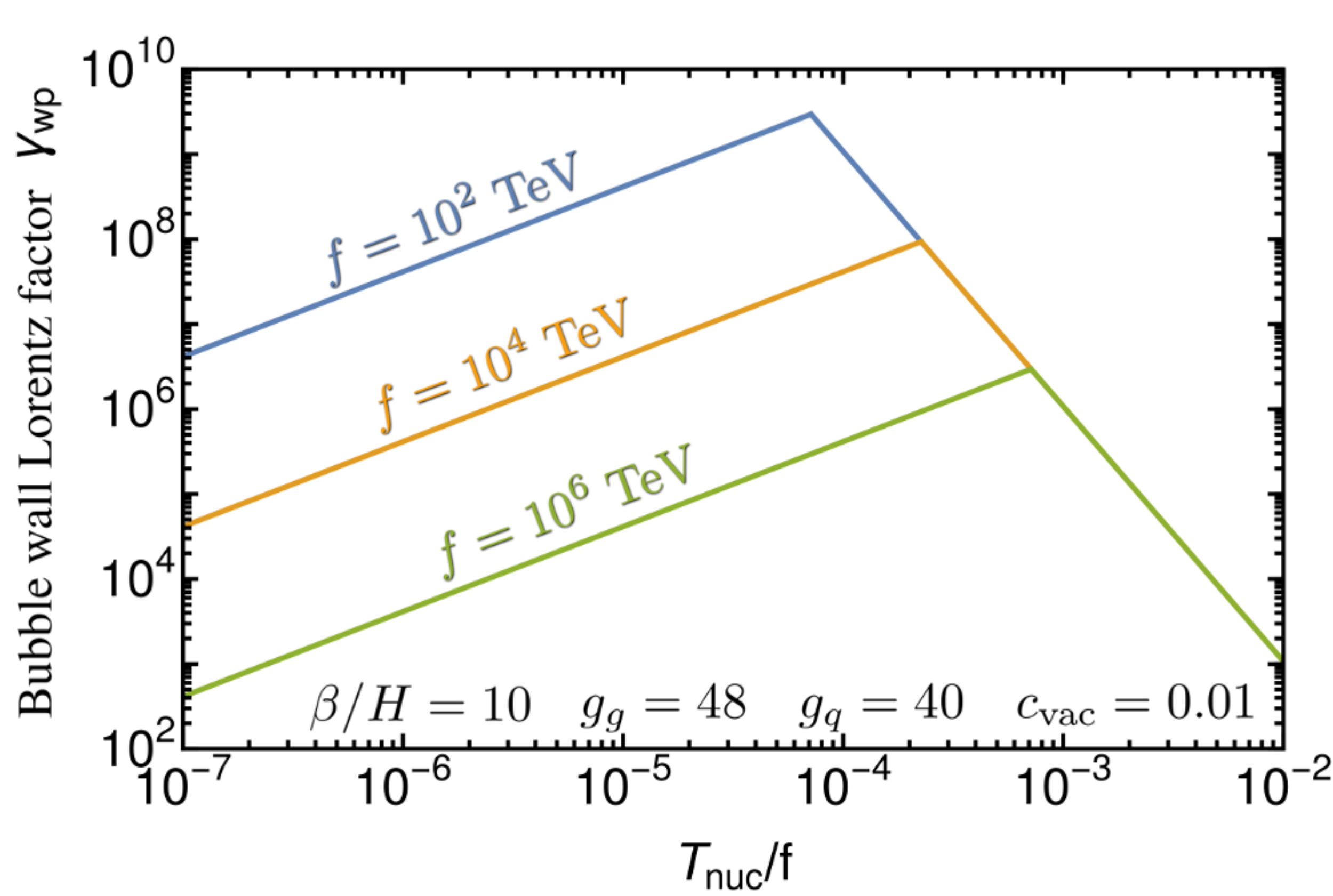}
\caption{
\label{fig:gwp} 
\it \small
The Lorentz factor of the wall at bubble percolation for various values of $f$ and amounts of supercooling, assuming $\beta/H=10$. For extreme supercooling (on the left side of the plot) $\gwp$ is in the runaway regime. In this regime, larger $f$ or smaller $\Tnuc$ leads to a smaller distance over which the bubble can accelerate. The former because of the smaller Hubble horizon and the latter due to the larger bubble size at nucleation. Therefore $\gwp$ decreases for more supercooling.
}
\end{center}
\end{figure}

\paragraph{Final results.}
As bubbles are nucleated and start to expand, $\gwp$ starts growing as well.
If nothing slows down the bubble-wall acceleration, then $\gwp$ keeps growing until its value at the time of bubble-wall collision, $\gwp^\text{runaway}$.  Sources of friction that could prevent this runaway regime are given by the equivalent, in this scenario, of the so-called leading order (LO) and next-to-leading order (NLO) contributions of~\cite{Bodeker:2009qy} and~\cite{Bodeker:2017cim} respectively.
We find it convenient to report right away our final result for the maximal possible value of $\gwp$,
\begin{equation}
\gwp^\text{max} \simeq \text{Min}\Big[
1.7 \,\frac{10}{\beta/H} \Big(\frac{0.01}{c_\text{vac}}\Big)^{\!\frac{1}{2}}\, \frac{\Tnuc}{f} \frac{\MPl}{f},~
1.0\times 10^{-3}\frac{c_\text{vac}}{0.01}\frac{80}{g_\TC} \Big(\frac{f}{\Tnuc}\Big)^{\!3}
\Big],
\label{eq:gwp_max}
\end{equation}
where the first entry is associated to $\gwp^\text{runaway}$, and the second to the boost as limited by the LO pressure, $\gwp^\text{LO}$. $\gwp^\text{LO}$ is always smaller than $\gwp^\text{NLO}$ in the parameter space of our interest, so that $\gwp^\text{NLO}$ does not enter Eq.~\eqref{eq:gwp_max}.
We learn that in the regime of very strong supercooling and/or of very large confinement scale $f$, which will be the most relevant one for the DM abundance, bubble walls run away. The behaviour of $\gwp$ is illustrated in Fig.~\ref{fig:gwp}.

\paragraph{The impact on GW.}
The behaviour of $\gwp$ also has important consequences for the gravitational wave signal from the phase transition~\cite{Caprini:2015zlo,Caprini:2019egz}. If $\gwp^\text{max}=\gwp^\text{runaway}$ then the vacuum energy is converted into kinetic energy of the bubble walls \cite{Ellis:2019oqb}. The gravitational wave (GW) spectrum sourced by scalar field gradient is traditionally computed in the envelope approximation \cite{Kosowsky:1992vn,Huber:2008hg, Jinno:2016vai}. However, the latest lattice results \cite{Cutting:2020nla, Lewicki:2020jiv} suggest an enhancement of the GW spectrum at low frequency due to the free propagation of remnants of bubble walls after the collision, the IR slope $\propto k^3$ becoming close to $\propto k^1$. This confirms the predictions from the analytical bulk flow model \cite{Jinno:2017fby, Konstandin:2017sat}. Note that  the IR-enhancement is stronger for thick-walled bubbles \cite{Cutting:2020nla}, which is the case relevant for nearly-conformal potential leading to strong supercooling, and thus for the PT considered here. (Instead, for thin-walled bubbles, after collision the scalar field can be trapped back in the false vacuum \cite{Konstandin:2011ds, Jinno:2019bxw,Lewicki:2019gmv}. Instead of propagating freely, the shells of energy-momentum tensor remain close to the collision point and dissipate via multiple bounces of the walls.)
Irrespectively of whether the IR slope at $ f \lesssim \beta$ is $\propto k^3$ or $\propto k^1$, at much lower frequency, $f \lesssim H$, the slope must converge to $k^3$ due to causality \cite{Durrer:2003ja,Caprini:2009fx, Cai:2019cdl}.
Oscillations of the condensate following the PT can provide an additional source of GW~\cite{Child:2012qg}. However, instead of $\beta^{-1}$ the time scale is set by the inverse scalar mass $\sim f^{-1}$ and the signal is Planck-suppressed $\propto \beta/f$ \cite{Cutting:2018tjt}.

If instead, $\gwp^\text{max} = \gwp^\text{NLO}$, the vacuum energy is converted into thermal and kinetic energy of the particles in the plasma already prior to the bubble wall collision. The contribution from sound waves or turbulence~\cite{Caprini:2015zlo,Caprini:2019egz}, however, in supercooled transitions is not yet clearly understood. Indeed, current hydrodynamical simulations, which aim to capture the contribution of the bulk motion of the plasma to the gravitational wave signal, do not yet extend into the regime in which the energy density in radiation is subdominant to the vacuum~\cite{Cutting:2019zws}. And analytical studies of shock-waves in the relativistic limit have just started \cite{Jinno:2019jhi}. In any case, we expect supercooled transitions to provide promising avenue for detection in future GW observatories.

\medskip
\noindent
We now proceed to a detailed derivation of Eq.~(\ref{eq:gwp_max}).

\paragraph{Linear growth.}
The energy gained upon formation of a bubble of radius $R$ is
$E_\text{bubble} = \frac{4}{3} \pi R^3 \Delta V_\text{vac}$,
where $\Delta V_\text{vac}$ is the difference between the vacuum energy density outside and inside the bubble.
The energy lost upon formation of a bubble of radius $R$ is $E_\text{wall} \simeq 4 \pi R^2 \gwp \sigma_w$,
where $\sigma_w$ is the surface energy density of the wall (surface tension) in the wall frame.
If a bubble nucleates and expands, its energy $E_\text{bubble}$ is transferred to the wall energy $E_\text{wall}$.
As soon as a nucleated bubble contains the region $\chi \simeq f$, neither $\Delta V_\text{vac}$ nor $\sigma_w$ change upon bubble expansion. Indeed both are a function of the bubble wall profile, which does not change in that regime (also see Fig.~\ref{fig:wall_profile}).
We thus recover the well-known property that $\gwp$ grows linearly in $R$,
\beq
\gwp =\frac{R}{R_0} \sim \Tnuc \,R\,,
\label{eq:gwp_growth}
\eeq
where $R_0$ is a normalisation of the order of the minimal radius needed for a bubble to nucleate, and where in the second relation we have used $R_0 \gtrsim L_w \sim \Tnuc^{-1}$ because we assumed the nucleated bubble to contain the region $\chi \simeq f$. A more precise treatment can be found, e.g.~in the recent~\cite{Ellis:2019oqb}, which confirms the parametric dependence of Eq.~(\ref{eq:gwp_growth}). 

\paragraph{At collision time.}
In a runaway regime, i.e.~for small enough retarding pressure on the bubble walls, $\gwp$ at collisions then reads
\beq
\gwp^\text{runaway}
\sim \Tnuc \, \beta^{-1}
\simeq 1.7 \, \frac{10}{\beta/H}\,\Big(\frac{0.01}{c_\text{vac}}\Big)^{\!\frac{1}{2}}\frac{\Tnuc}{f}\frac{\MPl}{f},
\label{eq:gwp_runaway}
\eeq
where $\beta^{-1}$ is the average radius of bubbles at collision, $H \simeq \Lambda_\text{vac}^2/(\sqrt{3} \MPl)$, and the value $\beta/H \simeq 10$ is a benchmark typical of supercooled phase transitions~\cite{Randall:2006py,Konstandin:2010cd,Konstandin:2011dr,Baldes:2018emh,Bruggisser:2018mrt,Megias:2018sxv,2ndpaper}, which we employ from now on.

The bubbles swallow most of the volume of the universe, and thus most techniquanta, when their radius is of the order of their average radius at collision $\beta^{-1}$. Therefore, in the regime of runaway bubble walls, the relevant $\gwp$ for all the physical processes of our interest (hadron formation from string fragmentation, quark ejection, etc.) will be some order one fraction of $\gwp^\text{runaway}$.
For simplicity, in the runaway regime we will then employ the simplifying relation $\gwp =\gwp^\text{runaway}$.
This will not only be a good-enough approximation for our purposes, but it will also allow to clearly grasp the parametric dependence of our novel findings.
Moreover, a more precise treatment, to be consistent, would need to be accompanied by a more precise solution for $\gwp$ than that of Eq.~(\ref{eq:gwp_growth}), i.e.~we would need to specify the potential driving the supercooled PT and solve for $\gwp$.
As the purpose of this paper is to point out effects which are independent of details of the specific potential, we leave a more precise treatment to future work.

\subsection{LO pressure}
\label{sec:PLO}

\paragraph{Origin.}
By LO pressure we mean the pressure from the partial conversion --- of the quark's momenta before entering the bubbles --- into hadron masses~\cite{Bodeker:2009qy}, plus that from the ejection of quarks.
We use the subscript LO in reference to~\cite{Bodeker:2009qy,Bodeker:2017cim}, because this pressure is of the form $\mathcal{P_{\rm LO}} \sim \Delta m^2 T^2$, where $\Delta m $ is the rest energy of the flux tube between the incoming techni-quanta and the wall.
However, in contrast to \cite{Bodeker:2009qy,Bodeker:2017cim}, here the pressure arises from non-perturbative effects.
 
\paragraph{Momentum transfer.}
The momentum exchanged with the wall, upon hadronization of a single entering quark plus the associated quark ejection, reads in the wall frame
\beq
\Delta p_\LO
= E_\text{in} - \sqrt{E_\text{in}^2 - \Delta m^2_\text{in}} + E_\text{ej}
\simeq f\,,
\label{eq:Deltap_LO}
\eeq
where $E_\text{in} \simeq 3\,\gwp \Tnuc$ is the energy of the incoming quark, $\Delta m^2_\text{in}$ is the fraction of that energy that is converted into `inertia' of the string, and $E_\text{ej} \simeq f/2$ is the energy of the ejected quark or gluon.
In the second equality, we have used $\Delta m^2_\text{in} \simeq \ECM^2 \simeq 3\,\gwp\, \Tnuc \, f$ from Eq.~(\ref{eq:ECMstring}) and $\gwp \gg f/\Tnuc$. Note that $\Delta p_\LO$ is independent of $p_{\rm in}$.

\paragraph{Pressure.}
In light of Sec.~\ref{sec:when_confinement}, we can safely consider a collision-less approach and neglect the interactions between neighboring quarks.
The associated pressure is given by
\beq
\mathcal{P}_\LO
= \sum_a g_a \int\frac{d^3 p_\text{in}}{(2 \pi)^3} \frac{1}{e^{|p_\text{in}|/\Tnuc}\pm 1}\,\Delta p_\LO,
\eeq
where $g_a$ is the number of internal degrees of freedom of a given species $a$ of the techniquanta.
Upon using Eq.~(\ref{eq:Deltap_LO}), we get
\beq
\mathcal{P}_\LO
\simeq  \frac{\zeta(3)}{\pi^2}\,g_\TC \,\gamma_{\rm wp}\,\Tnuc^3 f\,,
\label{eq:LOpressure}
\eeq
where we remind that $g_\TC =g_g + \frac{3g_q}{4}$.
This result can be understood intuitively from $\mathcal{P}_\LO \sim n_{\TC,w} \Delta p_\LO$, where $\gamma_{\rm wp}$ enters through $n_{\TC,w}$~\cite{Bodeker:2009qy}. Note that, in the absence of ejected particles, the pressure would have been a half of our result in Eq.~(\ref{eq:LOpressure}).

\paragraph{Terminal velocity.}
The resulting upper limit on $\gwp$ is obtained by imposing that the LO pressure equals that of the internal pressure from the difference in vacuum energies,
\beq
\mathcal{P}_\text{expand} = c_\text{vac} f^4\,,
\label{eq:pressure_expansion}
\eeq
and reads
\beq
\gwp^\text{LO} =  c_\text{vac} \frac{\pi^2}{\zeta(3)}\frac{1}{g_\TC} \Big(\frac{f}{\Tnuc}\Big)^{\!3}\,.
\label{eq:P_LO}
\eeq
We finally remark that $\mathcal{P}_\LO$ grows linearly in $\gwp$, unlike in `standard' PTs where it is independent of the boost. The reason lies in the fact that the effective mass $\Delta m^2_\text{in}$ grows with $\gwp$, whereas in `standard' PTs it is constant in $\gwp$. Our results then imply that, in confining phase transitions, the LO pressure is in principle enough to ensure the bubble walls do not runaway asymptotically. This is to be contrasted with non-confining PTs, where the asymptotic runaway is only prevented by the NLO pressure.\footnote{In our scenario, bubble walls can still run away until collision for some values of the parameters, and we anticipate they will. Unlike in non-confining PTs, the scaling of our LO pressure with $\gwp$ implies they could not runaway indefinitely if there were no collisions.}

\subsection{NLO pressure}
\label{sec:P_NLO}
\paragraph{Origin.}
The NLO pressure comes from the techniquanta radiating a soft gluon \cite{Bodeker:2017cim} which itself forms a string attached to the wall in the broken phase.  

\paragraph{Result.}
We derive it in detail in App.~\ref{app:NLO_pressure}. We find, cf. Eq.~\eqref{eq:NLOpressure_app}
\begin{equation}
\mathcal{P}_\NLO
\simeq  \big(g_g C_2[g]+ \frac{3}{4}g_q C_2[q]\big) \frac{8 \zeta(3)}{\pi} \frac{g_\text{conf}^2}{4\pi} \epsilon_\text{ps}\,\frac{\log\big(1+\frac{m_g^2}{k_*^2}\big)}{k_*/m_g}\,\gwp \Tnuc^3 m_g\,,
\label{eq:NLOpressure}
\end{equation}
where $C_2[g,q]$ are the second Casimirs of the representations of gluons and quarks under the confining group (if $SU(N)$, $C_2[g] = N$, $C_2[q] = (N^2-1)/2N$), $g_\text{conf}$ is the gauge coupling of the confining group, $\epsilon_\text{ps} \leq 1$ encodes the suppression from phase-space saturation of the emitted soft quanta $g$, important for large coupling $g_\text{conf}$, $m_g$ is an effective mass of the soft radiated gluons responsible for this pressure, and $k_*$ the IR cut-off on the momentum radiated in the direction parallel to the wall.
\paragraph{Vector boson mass.}
As we model the masses of our techniquanta as the inertia that their fluxtube would gain inside the bubble, these masses increase with increasing momentum of the techniquanta, in the wall frame. The NLO pressure is caused by emission of gluons `soft' with respect to the incoming quanta. Their would-be mass $m_g$ upon entering the wall cannot, therefore, be as large as that of the incoming quanta that emit them, $\Delta m_\text{in} \simeq \sqrt{3\,\gwp\, \Tnuc\, f}$.
At the same time, the effective gluon mass should at least allow for the formation of one hadron inside the wall, therefore we assume it to be of the order of the confinement scale, $m_g \sim f$. The fact that $m_g$ does not grow with $\gwp$ while $\Delta m_\text{in}$ does, is the reason why unlike in non-confining phase transitions, we find here that $\mathcal{P}_\NLO$ and $\mathcal{P}_\LO$ have the same scaling in $\gwp$ and in the amount of supercooling.

\paragraph{NLO pressure is sub-leading.}
By making the standard~\cite{Bodeker:2017cim} choice $k_* \simeq m_g$, and assuming $\epsilon_\text{ps} g_\text{conf}^2 <1 $, we then find that $\mathcal{P}_\NLO \ll \mathcal{P}_\LO$ in the entire parameter space of our interest. Thus, for simplicity, we do not report the NLO limit on $\gwp^{\rm max}$ in Eq.~(\ref{eq:gwp_max}).

Recently, Ref.~\cite{Hoeche:2020rsg} performed a resummation of the log-enhanced radiation that leads to the scaling $\mathcal{P}_\NLO \propto g_\text{conf}^2 \gwp^2 \Tnuc^4$. By using the analogue of that result for confining theories, we find that $\mathcal{P}_\NLO$ dominates over $\mathcal{P}_\LO$ in some region of parameter space, and therefore that the values of the parameters for which bubble walls run away slightly change.
Still, even by using that resummed result, we find that the region relevant for DM phenomenology corresponds to the region where bubble walls run away, so that the difference between the results of~\cite{Bodeker:2017cim} and~\cite{Hoeche:2020rsg} does not impact the DM abundance. As observed in~\cite{Vanvlasselaer:2020niz}, the pressure as determined in~\cite{Hoeche:2020rsg} does not tend to zero when the order parameter of the transition goes to zero, casting a shadow on that result. Therefore, both for this issue as well as for the limited impact on the DM abundance that we will discuss later, we content ourselves with a treatment analogous to~\cite{Bodeker:2017cim} in our paper.

\paragraph{Summary and runaway condition.}
At small supercooling (i.e. not too small $\Tnuc/f$) the bubble wall velocity reaches an equilibrium value set by the LO pressure.
At larger supercooling bubble walls collide before reaching their terminal LO velocity, and $\gwp$ is set by the runaway value Eq.~\eqref{eq:gwp_runaway}. By comparing Eq.~\eqref{eq:P_LO} with Eq.~\eqref{eq:gwp_runaway}, we find that bubble walls run away for
\beq
\frac{\Tnuc}{f}
\lesssim 1.2 \times 10^{-4}
\left(\frac{80}{g_\TC}\frac{\beta/H}{10}\,\frac{f}{\text{PeV}}\right)^{\!\frac{1}{4}}
\left(\frac{c_\text{vac}}{0.01}\right)^{\!\frac{3}{8}}
\label{eq:run-away_cond}
\eeq
The bubble wall Lorentz factor is plotted in Fig.~\ref{fig:gwp} against the amount of supercooling.

\subsection{Ping-pong regime}
\label{sec:pingpong}
\paragraph{Condition to enter.}
For even a single hadron to form inside the bubble, one needs $\ECM \geq m_\pi$, where $\pi$ is the lightest hadron of the new confining sector (e.g. a pseudo-goldstone boson). Via Eq.~(\ref{eq:ECMstring}), this implies
\beq
\gwp \gtrsim \gwp^\text{enter} = \frac{m_\pi^2}{3\,\Tnuc\,f}\,.
\label{eq:gamma_enter_def}
\eeq
\paragraph{Contribution to the pressure.}
For $\gwp \lesssim \gwp^\text{enter}$, which holds at least in the initial stages of the bubble expansion, the quarks and gluons are reflected and induce a pressure
\beq
\mathcal{P}_\text{refl} \sim n_{\TC,w} \times \Delta p_{\TC,w} \sim \Tnuc^3 \gwp \times \gwp \Tnuc \sim \gwp^2 \Tnuc^4\,.
\label{eq:pressure_reflectedquarks_init}
\eeq
This is to be compared with Eq.~(\ref{eq:pressure_expansion}), $\mathcal{P}_\text{expand} = c_\text{vac} f^4$, which implies the bubble wall could in principle be limited by this pressure to $\gwp \sim (f/\Tnuc)^2$. Nevertheless, as $(f/\Tnuc)^2 \gg \gwp^\text{enter}$, this pressure ceases to exist at an earlier stage of the expansion, namely once $\gwp = \gwp^\text{enter}$. Hence the maximum Lorentz factor remains encapsulated by Eq.~(\ref{eq:gwp_max}).
\paragraph{Ping-pong regime.}
In some extreme regions of parameter space, however, one could have $\gwp^\text{max} < \gwp^\text{enter}$, so that all techniquanta in the plasma are reflected at least once before entering a bubble. We leave a treatment of this `ping-pong' regime to future work.

\section{Amount of supercooling needed for our picture to be relevant}
\label{sec:our_picture_relevant}

\paragraph{Intuition about the limit of no supercooling.}
In the limit of no supercooling, one does not expect the fluxtubes to attach to the bubble wall, but rather to connect the closest charges that form a singlet and induce their confinement.
In other words, in the limit of no supercooling one expects the picture of confinement to be the one of `standard phase transitions'.
By continuity, there should exist a value of $\Tnuc$, smaller than $f$, such that the our picture ceases to be valid, and one instead recovers the more familiar confinement among closest color charges.
We now wish to determine it.
In order to do so, we note that the absence of ejected techniquanta is a necessary condition for the above to hold, therefore we now phrase the problem in terms of absence of ejected techniquanta.

\paragraph{Rate of detachment of $\bigstar$.}
We propose and analyse some effects that could lead to fluxtubes detaching from the bubble walls without ejecting particles.
To take place, these effects need to happen before the end-point of the fluxtube on the wall, $\bigstar$, ceases to exist, i.e.~when the string breaking inside the bubble has already taken place and a quark is ejected.
So we start by computing the rate $\Gamma_{\text{det}\bigstar}$ of detachment of $\bigstar$, the point where the fluxtubes is attached to the wall, from the wall itself. To estimate it, we again borrow the modelling of the classic paper on string fragmentation~\cite{Andersson:1983ia}.

The distances between the several points of breaking of a given string (that connects in our case $\TC_i$ and $\bigstar$) are space-like. In the frame of each point of breaking, that breaking is itself the first to happen, a time of order $N/f$ after the string formation (we adopt the scaling for strong sector gauge groups $SU(N)$~\cite{Casher:1978wy,Armoni:2003nz}).
This time therefore also applies to the outermost breaking point in our picture, i.e. that closest to the wall, whose frame approximately coincides with the wall frame. We remind the reader that the outermost breaking is the one that nucleates the quark or gluon that is eventually ejected. The rate we need can therefore be estimated as the inverse of the nucleation time of the outermost pair,
\beq
\Gamma_{\text{det}\bigstar\text{,w}} = \tau_{\text{det}\bigstar\text{,w}}^{-1} \simeq f/N\,.
\label{eq:rate_detachment}
\eeq

We now enumerate and model effects that could lead to fluxtubes detaching from the bubble walls without ejecting techniquanta, and compare their time scales with Eq.~(\ref{eq:rate_detachment}).
\begin{enumerate}
\item
\textbf{Flux lines overlap.} 
The faster a bubble-wall, the denser and thus the closer together in the wall frame are the quarks and gluons entering it. Eventually, they could get closer than the typical transverse size of a fluxtube $d_\text{tr} \simeq f^{-1}$~\cite{Bicudo:2017uyy}. When that happens, the fluxtubes between different color charges have a non-negligible overlap. We expect that in this situation it will not be clearly preferable energetically for these strings to attach directly to the wall. Thus there would be no ejected techniquanta.
This situation is of course realised also in the case of small supercooling $f/\Tnuc$, in addition to and independently of the case of fast bubble-walls.

We then obtain a rate of `string breaking by fluxtube overlap', $\Gamma_\text{overlap}$, as follows.
We define an effective associated cross section as the area of a circle on the wall, centered on any $\bigstar$ and with radius $d_\text{tr}$,
\beq
A_\text{overlap} = \pi d_\text{tr}^2 \simeq \pi f^{-2}\,.
\eeq
The associated rate then reads
\beq
\Gamma_\text{overlap} = A_\text{overlap} v \,n_{\TC,\text{w}} 
\simeq \frac{\gwp \zeta(3) g_\TC}{\pi} \frac{\Tnuc^3}{f^{2}}  \,,
\eeq
where $n_{\TC,\text{w}} = \gwp n_{\TC,\text{p}}$ is the density of techniquanta in the wall frame, $g_\TC = g_g + 3g_{q}/4$, and we have used that they are relativistic $v = 1$. The condition of no ejected techniquanta then reads
\beq
\Gamma_\text{overlap} > \Gamma_{\text{det}\bigstar\text{,w}}
\Rightarrow
\gwp \gtrsim \frac{2.6}{g_\TC N}\Big(\frac{f}{\Tnuc}\Big)^3\,.
\label{eq:noejquarks_detach}
\eeq

\item
The entire fluxtube connecting real color charges, so including its portion in the region $\chi \simeq \chi^* \ll f$ (see Fig.~\ref{fig:wall_diagram}), could enter the region $\chi =f$ before its portions in the region $\chi = f$ break and form hadrons, and eject particles. We see two ways this could happen.
\begin{enumerate}
\item[2.1] \textbf{Attractive interaction between neighboring flux lines.}  
The points $\bigstar$ are not static, because they move by the force exerted by the part of the string which is outside the wall, in the layer where $\langle \chi \rangle \simeq \chi^*$. Defining $y_\bigstar$ as the transverse distance, on the wall, between two $\bigstar$ points connected by a fluxtube, one has
\beq
\frac{d^2 y_\bigstar}{dt^2}
= \frac{F}{m_\bigstar}
\sim -\frac{d E_{q\bar{q}}/dy}{f}
\simeq -c_{q\bar{q}}\frac{{\chi^*}^2}{f}
\sim -c_{q\bar{q}} \frac{\Tnuc^2}{f}\,,
\eeq
where, consistently with our previous treatments, we have assigned to $\bigstar$ an inertia $m_{\bigstar} \sim f$.
If $y_\bigstar$ goes to zero in a time shorter than the breaking time $\tau_{\text{det}\bigstar\text{,w}} \sim N f^{-1}$, then the two fluxtubes connect and become fully contained in the region $\chi = f$ before they break and form hadrons, and thus there are no ejected techniquanta.
To determine this condition, we assume initially static points $\bigstar$, and thus we only need the initial distance between them $y_\bigstar (t=0) \simeq (\gwp n_{\TC,\text{p}})^{-1/3}$. We then obtain
\beq
y_\bigstar(t=\tau_{\text{det}\bigstar,\text{w}})
\simeq (\gwp n_{\TC,\text{p}})^{-1/3} - c_{q\bar{q}}\frac{\Tnuc^2}{f} \frac{\tau_{\text{det}\bigstar\text{,w}}^2}{2} \,.
\eeq
The resulting condition for no ejected quarks reads 
\beq
y_\bigstar(t=\tau_{\text{det}\bigstar\text{,w}}) < 0
\Rightarrow
\gwp \gtrsim \frac{ 6.6 \times 10^{-2}}{g_\TC  N^{6} }  \left( \frac{ 10 }{ c_{q\bar{q}} } \right)^{3}  \Big(\frac{f}{\Tnuc}\Big)^9\,.
\label{eq:noejquarks_classical}
\eeq

\item[2.2]
\textbf{Limit of no distortion of the flux lines.}  
When the string portion in the region $\chi \simeq \chi^* \ll f$ has a small enough length $d_\bigstar$, the possibility that it is pulled inside the region
$\chi = f$ could be energetically more convenient than the one of our picture, where it stays outside and instead energy goes in increasing the length of the strings that are perpendicular to the wall.
The energy price, for the string portion in the region $\chi \simeq \chi^* \ll f$ to enter the region $\chi = f$, reads in the wall frame
\beq
\Delta E_\text{pull-in,w}
\simeq c_{q\bar{q}} (f^2 - {\chi^*}^2)d_\bigstar
\simeq c_{q\bar{q}} f^2 d_\bigstar\,,
\label{eq:price_pulledin}
\eeq
where we stress that the length of the string portion $d_\bigstar$ is transverse to the bubble-wall velocity and therefore is not Lorentz contracted in the process of being pulled into the bubble. In the wall frame, it reads $d_\bigstar \simeq (\gwp n_{\TC,\text{p}})^{-1/3}$
The transition between $\chi \simeq \chi^* \ll f$ and $\chi = f$ is exponentially fast in the proper coordinate $s$ (see App.~\ref{app:wall_profile}), and happens over an interval (a distance, in the wall frame) $L_\text{f} \sim f^{-1}$.
The energy price of Eq.~(\ref{eq:price_pulledin}) should therefore be compared with the one to stretch two strings, inside the wall, by an amount $L_\text{f}$:
\beq
\Delta E_\text{stretch,w}
\simeq 2 c_{q\bar{q}} f^2 L_\text{f}/\gwc
\sim 2 c_{q\bar{q}} f \Big(\frac{f}{3\,\gwp \Tnuc}\Big)^{1/2},
\label{eq:Lf_introduce}
\eeq
where we have used that the string length in the expression for $E_{q\bar{q}}$, Eq.~(\ref{eq:stringenergy}), has to be evaluated in the string center-of-mass frame, and that $\gwc \simeq \sqrt{3\,\gwp \Tnuc/f}$ from Eq.~(\ref{eq:gwc}).
Therefore, it is energetically more convenient to pull the fluxtube inside the region $\chi \simeq f$, and so to have no ejected quarks, if
\beq
\Delta E_\text{pull-in} < \Delta E_\text{stretch}
\Rightarrow
\gwp \lesssim 0.035 \, g_\TC^2 \, \Big(f\,L_f\Big)^{\!6}\, \Big(\frac{\Tnuc}{f}\Big)^{\!3}\,.
\label{eq:noejquarks_pulled}
\eeq
Contrary to the previous two possibilities to have no ejected quarks, Eqs.~(\ref{eq:noejquarks_detach}) and (\ref{eq:noejquarks_classical}), the possibility in Eq.~(\ref{eq:noejquarks_pulled}) imposes an upper limit on $\gwp$. We anticipate that, in the regimes of supercooling interesting for our work $\Tnuc/f \ll 1$, Eq.~(\ref{eq:noejquarks_pulled}) cannot be satisfied consistently with $\gwp > 1$, so that it is not relevant for our work. 
\end{enumerate}

\end{enumerate}
%

\begin{figure}[t]
\begin{center}
\begin{adjustbox}{max width=1\linewidth,center}
\includegraphics[width=.5\textwidth]{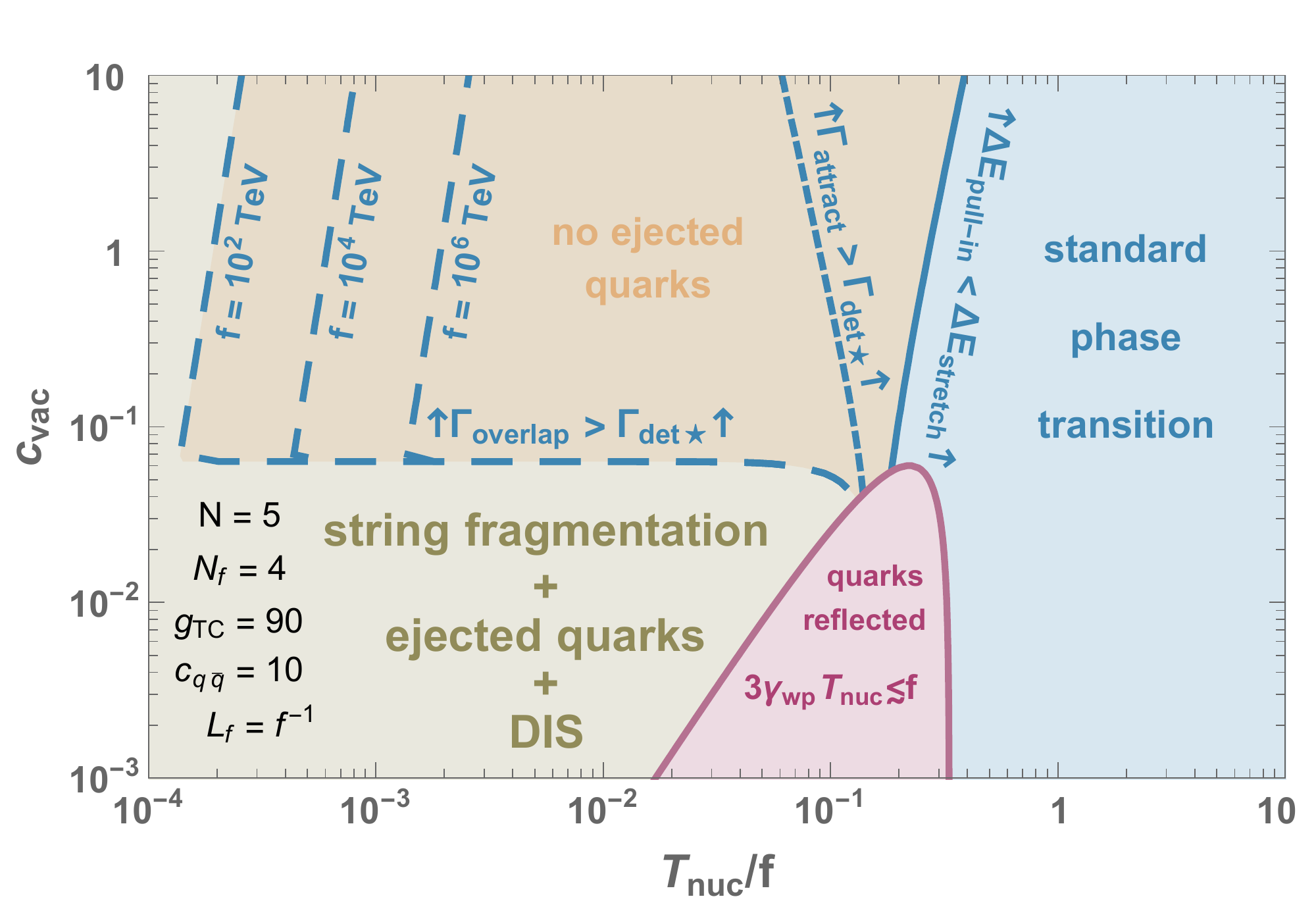}
\includegraphics[width=.5\textwidth]{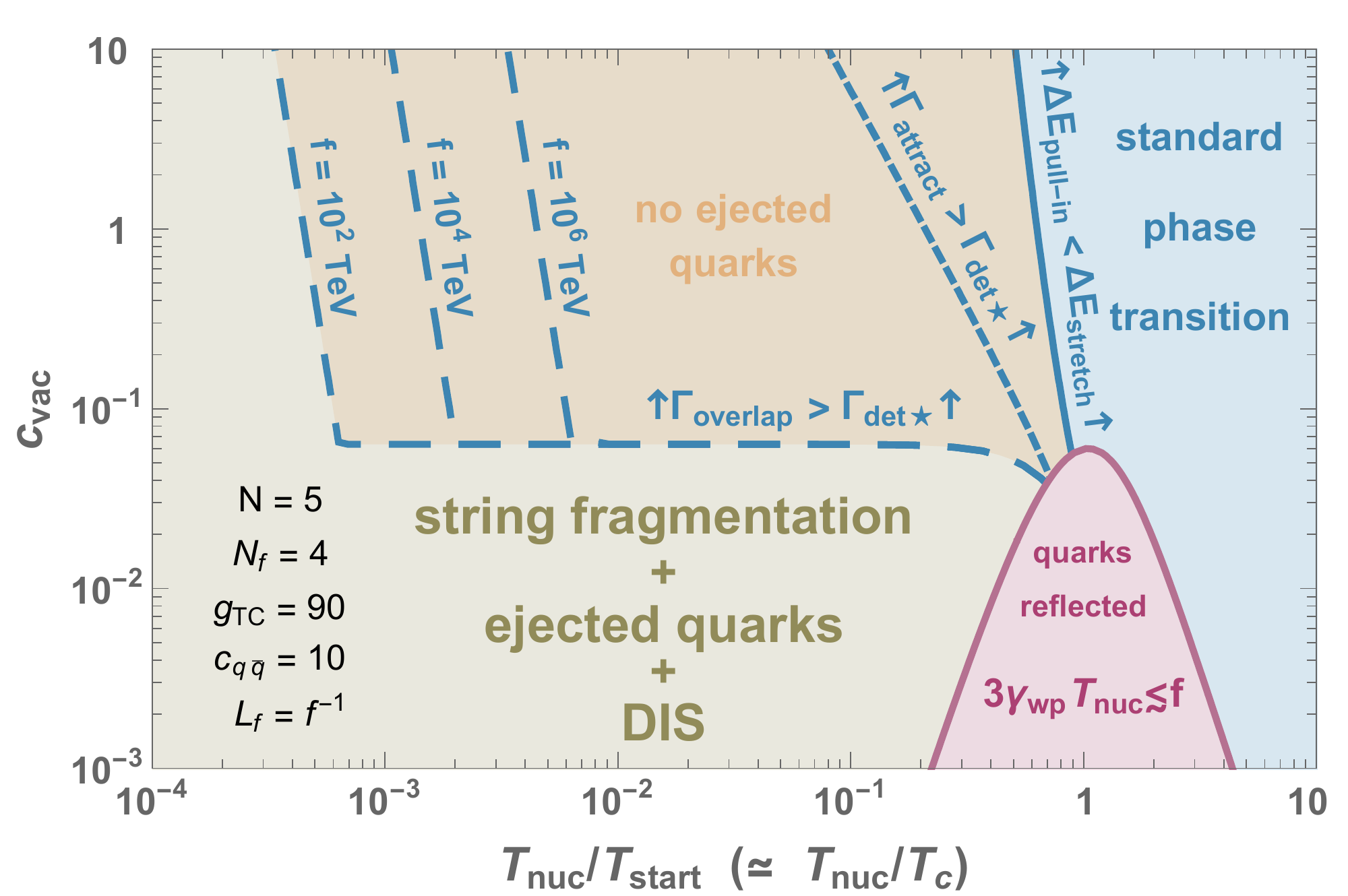}
\end{adjustbox}
\caption{
\label{fig:StdPictureRecovered} 
\it \small  \newline
\textbf{Blue Region}: The incoming techniquanta confine with their neighbours as in the standard picture of phase transitions that are not supercooled. \newline
\textbf{Olive Region}: All three inequalities, \eqref{eq:noejquarks_detach}, \eqref{eq:noejquarks_classical}, \eqref{eq:noejquarks_pulled}, are violated and the new effects pointed out in this study, i.e. string fragmentation, ejection of techniquanta and deep inelastic scattering, should be taken into account. 
\newline 
\textbf{Orange Region}: At least one but not all of the inequalities above hold, therefore there are no ejected techniquanta.
The dynamics taking place in this region remains to be investigated.
\newline
\textbf{Purple Region}: Quarks are too weakly energetic to enter the bubbles, see.~\ref{sec:pingpong}. \newline 
\textbf{Left of solid line}: Eq.~\eqref{eq:noejquarks_pulled}
is violated and it is energetically favourable for the flux lines to be distorted.  \newline 
\textbf{Left of dotted line}: Eq.~\eqref{eq:noejquarks_classical} is violated we can neglect the attractive interactions between neighboring flux lines.  \newline
\textbf{Left of dashed lines}: Eq.~\eqref{eq:noejquarks_detach} is violated and we can neglect the overlap of neighbouring flux lines.  \newline
The two plots only differ through their horizontal axis, see Sec.~\ref{sec:thermal_history} for the definitions of $c_{\rm vac}$ and $T_{\rm start}$, and App.~\ref{app:wall_profile} for that of $T_c$. 
To avoid the unphysical values $\gwp < 1$, we have added 1 to Eq.~\eqref{eq:gwp_max}.
}
\end{center}
\end{figure}

\paragraph{\bf Summary of required supercooling.}

In the regime where $\Tnuc \gtrsim f$, we expect that neither ejection of techniquanta nor string fragmentation should take place, and that the standard picture of quarks and gluons confining with their neighbors should be recovered (which we dub the `standard phase transition'). More precisely, if any of Eqs.~(\ref{eq:noejquarks_detach}), (\ref{eq:noejquarks_classical}) and (\ref{eq:noejquarks_pulled}) hold, we depart from our picture in at least one regard.
By demanding none of these inequalities hold, we expect the new effects of our study, namely flux line attached to the wall, string fragmentation, quark ejection and deep inelastic scattering, to take place. In the non-runaway regime, we require
		\begin{align}
	c_{\rm vac} & \lesssim  \frac{0.32}{N} \quad \text{and} \\
	 \frac{\Tnuc}{f} & \lesssim  \text{Min}\Big[  0.19 \left(\frac{5}{N}\right) \left( \frac{0.01}{c_{\rm vac}} \right)^{1/6} \left( \frac{10}{c_{q\bar{q}}} \right)^{1/2}, \,  \frac{ 0.12 }{ fL_{f} }  \left( \frac{c_{\rm vac}}{0.01} \right)^{1/6} \left( \frac{90}{g_{\rm TC}} \right)^{1/2}  \Big], \nonumber
	\end{align}
for our picture to hold. In the runaway regime, we instead require
\begin{align}  
 	\frac{\Tnuc}{f} &  \lesssim \nonumber \\
	\text{Min}\Big[ & 6.1 \times 10^{-5} \left( \frac{\beta/H}{10}\right)^{1/4}  \left( \frac{c_{\rm vac}}{0.01} \right)^{1/8} \left( \frac{90}{g_\TC} \right)^{1/4} \left( \frac{5}{N} \right)^{1/4} \left( \frac{ f }{ 10 \; \mathrm{TeV} } \right)^{1/4},   \nonumber \\
	& 6.4\times 10^{-3} \left( \frac{\beta/H}{10}\right)^{1/10} \left( \frac{c_{\rm vac}}{0.01} \right)^{1/20} \left( \frac{10}{c_{q\bar{q}}} \right)^{3/10} \left( \frac{90}{g_\TC} \right)^{1/10} \left( \frac{5}{N} \right)^{3/5} \left( \frac{ f }{ 10 \; \mathrm{TeV} } \right)^{1/10}   , \nonumber \\
	& \frac{ 1.2 \times 10^6 }{ (fL_{f})^{3} } \left( \frac{10}{\beta/H} \right)^{1/2}\left( \frac{0.01}{c_{\rm vac}} \right)^{1/4} \left( \frac{90}{g_\TC} \right) \left( \frac{ 10 \; \mathrm{TeV} }{ f } \right)^{1/2}     \Big].
\end{align}
for our picture to hold. Here we have used $\gwp$ in Eq.~\eqref{eq:gwp_max}. The conditions are visually summarised in Fig.~\ref{fig:StdPictureRecovered}.

 In light of this figure, we conclude that some new effects pointed out in our study are also relevant in confining phase transitions where $\Tnuc \sim T_{\rm start} \sim T_{c}$ (see App.~\ref{app:wall_profile} for the definition of the critical temperature $T_c$), e.g.
\cite{Schwaller:2015tja, Aoki:2017aws,Aoki:2019mlt, Bigazzi:2020avc,Ares:2020lbt,Huang:2020mso, Halverson:2020xpg}, provided $c_{\rm vac}$ is small enough.
A possible impact on the QCD phase transition, e.g. \cite{Buballa:2003qv,Fukushima:2010bq,Fukushima:2013rx,Herzog:2006ra,Schwarz:2009ii,Schettler:2010dp,Alho:2013hsa,Ahmadvand:2017xrw,Ahmadvand:2017tue,Chen:2017cyc}, remains to be investigated.

\paragraph{Averaged quantities only. }
We conclude this section by also stressing that all the conditions above refer to averaged quantities, and therefore do not take into account the leaks from tails of distributions. These leaks could for example imply that there are a few strings that hadronise without ejecting particles, even if all conditions Eqs.~(\ref{eq:noejquarks_detach}), (\ref{eq:noejquarks_classical}) and (\ref{eq:noejquarks_pulled}) are violated.
As these strings constitute a small minority of the total ones, these effects have a negligible impact on the phenomenology we discuss.
They could however be important in studying other situations of supercooled confinement.
Though certainly interesting, the exploration of these effects goes beyond the scope of this paper.

\section{Ejected quarks and gluons}
\label{sec:outside_bubble}

\subsection{Density of ejected techniquanta}
\label{sec:nej}

In the wall frame, since we have one ejected quark or gluon per each incoming one, we find
\beq
n_\text{ej,w} = n_{\TC,\text{w}}(r_\text{ej}) = \gwp(r_\text{ej}) n_{\TC,\text{p}}\,,
\eeq
where $n_{\TC,\text{p}} = g_\TC \zeta(3) \Tnuc^3/\pi^2$ is the density of the diluted bath in the plasma frame. The density of ejected techniquanta then depends on the time passed since bubble wall nucleation, or equivalently on the bubble radius at the time of ejection $r_\text{ej}$, via $\gwp(r)$ (see Sec.~\ref{sec:wall_speed}). 
In the plasma frame, and at a given distance $D$ from the center of the bubble, we then have\footnote{The factor $2$ arises when we boost the quark current $(\gwp\, n_{\TC,\text{p}} , \,  \gwp\, \vec{\beta} \,n_{\TC,\text{p}})$, with $\vec{\beta} = \vec{e}_r$, from the wall to the plasma frame.}
\beq
n_\text{ej,p}(D)
= 2\gwp^2(r_\text{ej}) \Big(\frac{r_\text{ej}}{D}\Big)^{\!2} n_{\TC,\text{p}}\,,
\label{eq:nejp}
\eeq
where we have included the surface dilution from the expansion between the radius at which a given quark has been ejected, $r_\text{ej}$, and the radius $D$ where we are evaluating $n_\text{ej,p}$. 
\paragraph{Radial dependence. }
It is convenient to express $n_\text{ej,p}$ as a function of the radial distance $x$ from the bubble wall in the plasma frame, where for definiteness $x= 0$ denotes the position of the wall and $x = L_\text{ej,p}$ the position of the techniquanta ejected first (which constitute the outermost layer). In order to do so, we determine the relation between the position $x$ of a quark and the radius $r_\text{ej}(x)$ when it has been ejected.
We assume that the bare mass of the quarks is small enough such that they move at the speed of light, like the gluons.
The wall at $x = 0$, instead, moves at a speed $v_\text{wall} \simeq 1-1/(2\gwp^2)$ (we have used the relativistic limit $\gwp \gg 1$), dependent on its radius. 
The coordinate $x$ of a given layer of ejected particles can then be found by integrating the difference between the world line of an ejected particle and that of the wall,
\beq
x
= \int_{t_\text{ej}}^{t_\mathsmaller{D}} \!\!\! dt(1-v_\text{wall}) 
\simeq \int_{t_\text{ej}}^{t_\mathsmaller{D}} \frac{1}{2 \gwp^2(t)}
\simeq \frac{1}{2 \Tnuc^2} \Big(\frac{1}{t_\text{ej}} - \frac{1}{t_\mathsmaller{D}}\Big)\,,
\label{eq:rej}
\eeq 
where we defined $t_\mathsmaller{D}$ and $t_\text{ej}$ as the times when the bubble radius is respectively $D$ and $r_\text{ej}$, and we used $\gwp(t) \simeq \Tnuc t$, cf. Eq~(\ref{eq:gwp_growth}), valid up to relative orders $1/\gwp^2 \ll 1$. 
It is convenient to rewrite Eq.~(\ref{eq:rej}) as
\beq
r_\text{ej} (x) \simeq \frac{D }{1 + 2\,\Tnuc^2\,D\, x}\,.
\label{eq:gwp_x_runaway}
\eeq
We finally obtain
\beq
n_\text{ej,p}(x)
= \frac{2\,\gwp^2 (x)}{\big(1+2\,\Tnuc^2 \,D \,x\big)^{\!2}} n_{\TC,\text{p}}
\simeq  \frac{2\,\Tnuc^2\,D^2}{ \big(1+2\,\Tnuc^2 \,D \,x\big)^{\!2} } n_{\TC,\text{p}}\,,
\label{eq:nejpx}
\eeq
where the last equality is valid as long as the bubbles run away, i.e. as long as Eq.~(\ref{eq:gwp_growth}) $\gwp \simeq \Tnuc \,r$ holds.

\begin{figure}[t]
\begin{center}
\includegraphics[width=.65\textwidth]{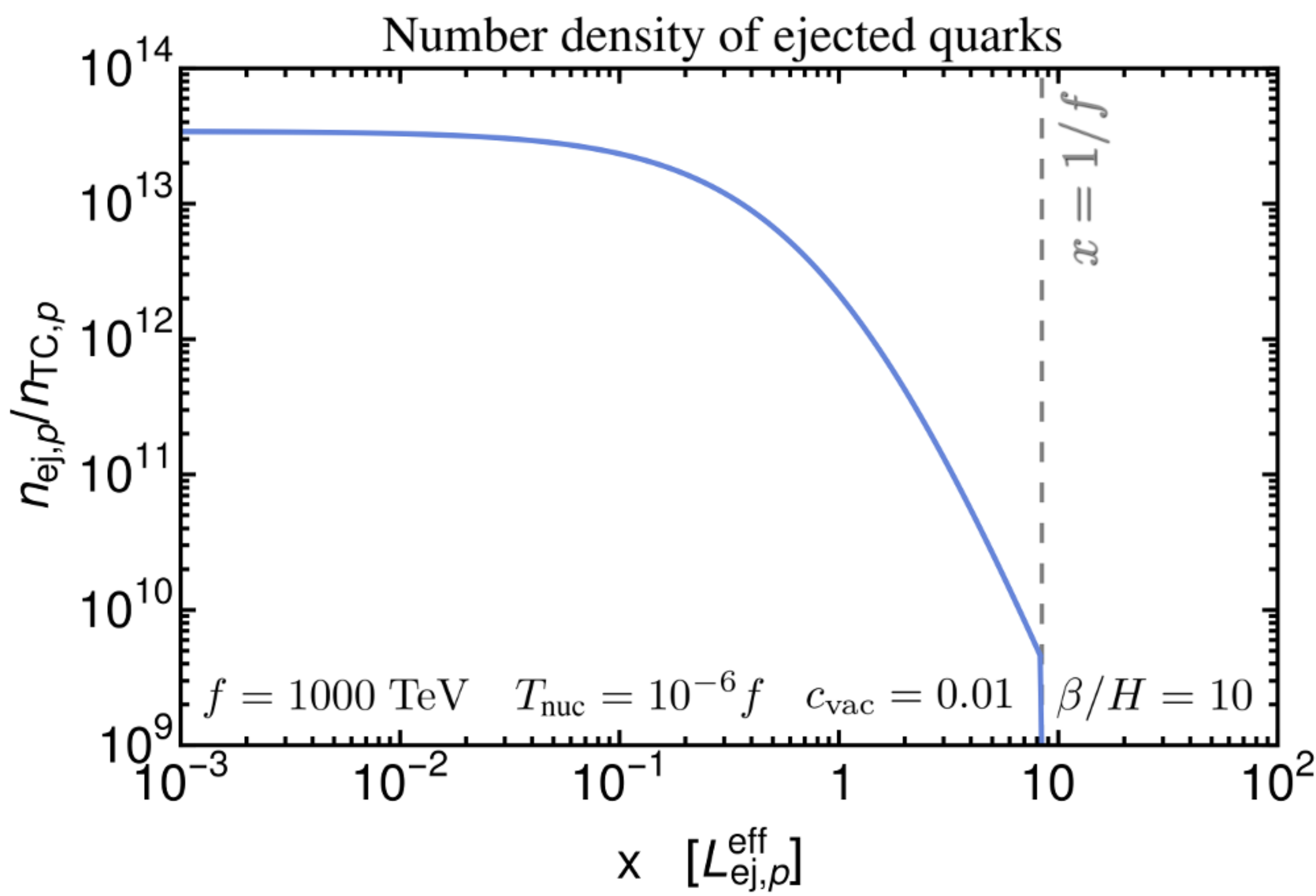}
\caption{
\label{fig:plasmaprofile} 
\it \small
The density of the ejected quarks in front of the bubble wall as a function of the distance $x$ in front of the bubble wall, Eq.~(\ref{eq:nejpx}), for an example parameter point. Here we have used the relation $\gwp \simeq D\Tnuc$. The distance to the outermost techniquanta $L_\text{ej,p} \approx 1/f$ is also shown. 
}
\end{center}
\end{figure}

\paragraph{Thickness of the layer of ejected techniquanta. }
Our result Eq.~\eqref{eq:nejpx} implies that the highest density, of ejected techniquanta, is located in the shell within a distance of the bubble wall 
\beq
L_\text{ej,p}^{\rm eff} \simeq \frac{1}{2\, \Tnuc^2\,D}\,.
\label{eq:Lej_eff}
\eeq
The density of ejected quarks $n_\text{ej,p}(x)$ extends to $x = L_\text{ej,p}$, i.e. to the outermost ejected layer, that we now show to be much larger than $L_\text{ej,p}^{\rm eff}$.
Indeed, $L_\text{ej,p}$ can be related to the time $t_\text{first}$ of ejection of the first techniquanta (corresponding to $\gwp \simeq m_\pi/\Tnuc$, Eq.~\eqref{eq:gamma_enter_def}).  Using $t_i \gg t_\text{first}$ and $t_\text{first} \sim m_\pi/\Tnuc^2$, we find
\beq
L_\text{ej,p} \simeq \frac{1}{t_{\rm first}\,2\,\Tnuc^2} \sim \frac{1}{f}\,,
\label{eq:Lej_velocity}
\eeq
where for simplicity we have assumed $m_\pi \simeq f$ as in QCD.
As long as $L_\text{ej,p} \gg L_\text{ej,p}^{\rm eff}$, as it holds for our estimate Eq.~\eqref{eq:Lej_velocity}, the value of $L_\text{ej,p}$ does not affect any of the results of this paper.
\footnote{One could easily envisage situations in which $m_\pi$ differs sizeably from $f$, e.g. because pions are much lighter or because of a possible dependence of the mass of the lightest resonances on the number of colours $N$. The exploration of if and how this possibility would affect our results (for example the conclusion that $L_\text{ej,p} \gg L_\text{ej,p}^{\rm eff}$), while certainly interesting, goes beyond the purposes of this paper.
}
The density profile of Eq.~(\ref{eq:nejpx}) is shown in Fig.~\ref{fig:plasmaprofile}.

\paragraph{Sanity check. }
As a check of our result Eq.~\eqref{eq:nejpx}, we verify that one has one ejected quark or gluon per each one that entered the bubble. Indeed, we compute
\beq
4 \pi D^2
\int_{0}^{L_\text{ej,p}}\!\!\!dx \,n_\text{ej,p}(x)
= \frac{4}{3} \pi D^3 n_{\TC,\text{p}}\,,
\label{eq:cons_number}
\eeq
where we have assumed $D \gg f/\Tnuc^2$, i.e. we have placed ourselves deep in the regime where hadrons can form inside bubbles (see Eq.~(\ref{eq:gamma_enter_def})).
Equation~(\ref{eq:cons_number}) guarantees that the number of ejected techniquanta in the layer of thickness $L_\text{ej,p}$ is equal to the total number of techniquanta that entered the bubble up to radius $D$.

\medskip
\paragraph{Interactions between ejected quarks. }
Let us finally comment why, we think, interactions among the ejected techniquanta cannot much alter their density.
The density of the particles in the incoming bath does not change out of their own interactions.
In the wall frame, both the density and the relative momentum of the ejected techniquanta are of the same order of those of the particles in the incoming bath. Therefore, we analogously expect that the density of the ejected techniquanta would also not change after ejection.
Since what will matter for the following treatment is the energy in the ejected techniquanta, rather than how this energy is spread among the various degrees of freedom, we content ourselves with this qualitative understanding and leave a more precise treatment to future work.

\subsection{Scatterings of ejected quarks and gluons before reaching other bubbles}
\label{sec:energy_transfer}

Before possibly reaching other expanding bubble-walls and their ejected techniquanta, ejected quarks and gluons could undergo scatterings with particles from the supercooled bath at temperature $\Tnuc$, and with techniquanta ejected from other bubbles. In this section we study the effects of these scatterings.

\paragraph{Ejected techniquanta are energetic. }
As soon as a bubble occupies an order one fraction of its volume at collision, the total energy in ejected particles is much larger than that in the supercooled bath outside the bubble.
Indeed, we have seen that for each quark or gluon in the supercooled bath that enters a bubble, there is at least an ejected one, and that the energy ejected per each incoming particle is much larger than the energy per each particle in the bath, $E_\text{ej,p}\simeq \gwp f \gg \Tnuc$, Eq.~(\ref{eq:energy_ejected}).
Assuming the degrees of freedom in quarks and gluons are not an extremely small fraction of those in the diluted medium, then the diluted medium outside the bubbles does not have enough energy to act as a bath for the ejected particles.
This implies that most ejected particles keep most of their energy upon passing through the supercooled bath.

\paragraph{Energy transfer between ejected techniquanta and diluted bath. }
By reversing the logic above, the ejected particles can deposit in the supercooled bath an energy much larger than its initial one.
Pushing this to the extreme, the ejected techniquanta could make the bath move away from the bubble wall, thus making our treatment so far valid only in the first stages of bubble expansion.
In order to assess this, we estimate
the rate of transferred energy between ejected techniquanta and particles from the bath outside the bubbles,
\beq
\Gamma_\text{ej-bath}
=  n_\text{ej} \int_{\Delta E_\text{min}}^{\Delta E_\text{max}} \!\! \!\!\!d \Delta E\,\frac{d\sigma v}{d \Delta E}\, \Delta E
\simeq n_\text{ej} \int_{-s}^{t_\IR} \!\! dt\frac{d\sigma}{d t}\, \sqrt{-t}\,,
\eeq
where $n_\text{ej}$ is the density of ejected techniquanta, $\Delta E$ is the energy transferred per single scattering, and where in the second equality we have taken the limit of relativistic particles $v \simeq 1$ and small energy transfer per single scattering $\Delta E$, so that the Mandaelstam variable $t$ can be expressed as $t \simeq -\Delta E^2$.
The quantity $d\sigma/dt$ depends on the specific model under consideration, in particular it depends both on whether the ejected particle is a quark or a gluon, and on the identity of the scatterer in the bath outside the bubbles. For definiteness, we model it as the cross section for fermion-fermion scattering mediated by a light vector with some effective coupling $\sqrt{4 \pi \alpha_\text{eff}}$,
\beq
\frac{d\sigma}{dt} = \frac{4\pi \alpha_\text{eff}^2}{t^2}\,.
\label{eq:dsigmadt}
\eeq
We then obtain
\beq
\Gamma_\text{ej-bath}
=  n_\text{ej} \frac{8 \pi\alpha_\text{eff}^2}{\sqrt{-t_\IR}}.
\label{eq:transfer_energy_rate}
\eeq
$\Gamma_\text{ej-bath}$ is of course not Lorentz invariant, it depends on the frame via the density of ejected techniquanta $n_\text{ej}$ determined in Sec.~\ref{sec:nej}.

\paragraph{Impact on diluted bath. }
The average energy transferred to a particle in the diluted bath at position $D$, when this particle goes across the layer of ejected techniquanta (so before it reaches the wall and initiates the processes described in Sec.~\ref{sec:inside_bubble}), then reads
\beq
Q_\text{ej-bath} \equiv
\int_{0}^{L_\text{ej,p}}\!\!\!dx\, \Gamma_\text{ej-bath,p} (x)\,,
\label{eq:Qeji}
\eeq
where we remind that the spatial coordinate $x$ is the distance between a given layer of ejected techniquanta and the wall at $x=0$. 
Upon use of Eqs.~(\ref{eq:transfer_energy_rate}) and \eqref{eq:nejpx}, we can then evaluate the average energy transferred to an incoming particle from the diluted bath, Eq.~(\ref{eq:Qeji}), as
\beq
Q_\text{ej-bath}
\simeq \frac{8 \pi\alpha_\text{eff}^2}{3}\frac{D \,n_{\TC,\text{p}}}{\sqrt{-t_\IR}}\,.
\label{eq:Qejbath}
\eeq
Note that the product $D \, n_{\TC,\text{p}}$ is Lorentz-invariant, so that $Q_\text{ej-bath}$ is indeed a Lorentz-invariant quantity.
To learn whether particles from the diluted bath are prevented from entering the wall, because of the interaction with the ejected techniquanta, we compare the energy they exchange with them upon passing their layer with their initial energy in the wall frame\footnote{Had we chosen another frame, we would have had to include the wall velocity in the condition.},
$E_{i,w}\simeq 3\,\gwp \Tnuc$,
\beq
\frac{Q_\text{ej-bath}}{E_{i,w}}
\simeq  \frac{8 \zeta(3)}{9\pi} \alpha_\text{eff}^2 \,g_\TC \frac{\Tnuc D}{\gwp} \frac{\Tnuc}{\sqrt{-t_\IR}}\,. 
\label{eq:QoverEi}
\eeq
The novel physical picture we described in Secs.~\ref{sec:inside_bubble} and~\ref{sec:outside_bubble} is valid as long as $Q_\text{ej-bath}/E_{i,w} \ll 1$.
As seen in Sec.~\ref{sec:wall_speed}, $\gwp$ initially grows linearly with the bubble radius, $\gwp \simeq \Tnuc\,D$, until the retarding pressure possibly becomes effective. It will turn out in Sec.~\ref{sec:DMabundance} that the runaway regime of linear growth is the one relevant for the phenomenology we will discuss. In that regime, the condition $Q_\text{ej-bath}/E_{i,w} \ll 1$ translates into
$\Tnuc/\sqrt{-t_\IR}~\ll~1$.
\paragraph{IR cut-off. }
The quantity $-t_\IR$ is the IR cutoff of the scattering, $-t_\IR \equiv m_V^2$, with $m_V$ some effective mass of the mediator responsible for the interactions that exchange momentum. 
In the absence of mass scales, which is the case for example for the SM photon and for the gluons, the effective mass $m_V$ is equal to the plasma mass of these particles in the thermal bath.
If the only bath was the diluted one, one would have $m_{V,\text{therm}}^2 \sim \alpha_\text{eff}\, n_{\TC,\text{p}}/\langle E_{\TC,\text{p}}\rangle \sim \Tnuc^2$ (see e.g.~\cite{Kapusta:2006pm}).
However, the process of our interest here happens in the much denser bath of ejected techniquanta, $n_\text{ej,p} \gg n_{\TC,\text{p}}$, so that we indeed expect $m_{V,\text{therm}}^2 \gg \Tnuc^2$, so that $Q_\text{ej-bath}/E_{i,w} \ll 1$ and our picture so far is valid.
More precisely, the screening mass for non-equilibrium systems scales as~\cite{Arnold:2002zm} ($f(p)$ is the non-equilibrium phase space distribution of the particles in the system)
\beq
m_{V,\text{therm}}^2
\simeq g_\TC \alpha_\text{eff} \int \frac{f(p)}{|p|}
\sim  \frac{n_\text{ej,p}}{\langle E_\text{ej,p}\rangle}
\sim \frac{\gwp}{f/\Tnuc} \Tnuc^2 \gg \Tnuc^2\,,
\label{eq:mV}
\eeq
where we have used $\langle E_\text{ej,p} \rangle\sim \gwp f$ and $n_\text{ej,p}  \sim  (D/L_\text{ej,p}) n_{\TC,\text{p}} \sim D^2 \Tnuc^5 \sim \gwp(D)^2 \Tnuc^3$.
Equations~\eqref{eq:QoverEi} and \eqref{eq:mV} teach us that, in the regions of parameter space where $\gwp \gg f/\Tnuc$, the energy received by each particle in the diluted bath, from scatterings with the ejected techniquanta, is much smaller than their energy in the wall frame $E_{i,w}\simeq 3\,\gwp \Tnuc$.\footnote{One could be worried that in the outer shell of size $L_{\rm ej,p} \sim  1/f$, the thermal mass $m_{V,\text{therm}}$ is much smaller than its value in the densest region in $x\simeq 0$, explicited in Eq.~\eqref{eq:mV}, such that $ \frac{Q_\text{ej-bath}}{E_{i,w}}$ becomes larger than $1$.
We can check that it is not the case by including the $x$-dependence of $m_{V,\text{therm}}$, Eq.~\eqref{eq:mV}, in the integral in Eq.~\eqref{eq:Qeji} 
\begin{equation}
\label{eq:fullint}
Q_\text{ej-bath, p} =
\int_{0}^{L_\text{ej,p}}\!\!\!dx\,\, n_\text{ej, p} \frac{8 \pi\alpha_\text{eff}^2}{\sqrt{n_\text{ej,p}/\langle E_\text{ej,p}\rangle}}\,,
\end{equation} 
where $n_\text{ej,p}$ is defined in Eq.~\eqref{eq:nejpx}, $L_\text{ej,p} \sim 1/f$ in Eq.~\eqref{eq:Lej_velocity}, $D \sim \gwp / \Tnuc$ in Eq.~\eqref{eq:gwp_growth}, and $E_\text{ej,p} \sim \langle E_\text{ej,p} \rangle\sim \gwp f$. 
 We compute the integral in Eq.~(\ref{eq:fullint}) and obtain
\begin{equation}
\frac{Q_\text{ej-bath}}{E_{i,w}} \simeq 
 \left\{
                \begin{array}{ll}
                 \frac{8\sqrt{2 \zeta(3)}}{9} \alpha_\text{eff}^2\, \sqrt{g_\TC \frac{f/\Tnuc}{\gwp}} \ll 1, \quad \text{if}~ \gwp \gtrsim f/\Tnuc \\
                  \frac{8\sqrt{2 \zeta(3)}}{3} \alpha_\text{eff}^2\, \sqrt{g_\TC \frac{\gwp}{f/\Tnuc}} \ll 1, \quad \text{if}~ \gwp \lesssim f/\Tnuc,
                \end{array}
              \right.
\end{equation}

This confirms that the energy of incoming particles, $E_{i,w} \simeq 3 \gwp \Tnuc$, is not affected by the shell of ejected~quarks.
  }
Since $E_{i,w}$ was the crucial input quantity for our treatment in Sec.~\ref{sec:inside_bubble}, the picture that emerged there is not affected by these scatterings.

\medskip

\paragraph{Energy transferred to techniquanta ejected from other bubbles. }
Finally, before ejected techniquanta can possibly enter another expanding bubble, they also have to pass through the layer of the techniquanta ejected from that other bubble. To investigate this, one can use the result derived above, Eq.~(\ref{eq:Qejbath}), with the specification that now $D$ is the maximal radius reached on average by expanding bubbles, because the shells of ejected quarks and gluons meet just before the bubble walls do.
We then find that the average energy transferred is much smaller than the energy of an ejected techniquanta in the plasma frame $\simeq \gwp f$,
\beq
\frac{Q_\text{ej-ej}}{\gwp f}
\simeq \frac{8 \zeta(3)}{3\pi} \alpha_\text{eff}^2 \,g_\TC \frac{\Tnuc}{f} \frac{\Tnuc D}{\gwp} \frac{\Tnuc}{\sqrt{-t_\IR}}
\ll 1\,.
\label{eq:QoverEej}
\eeq
Hence, for the purpose of determining the average energy of ejected quarks when they enter another bubble, one can safely ignore the interactions between the two shells.

\subsection{Ejected techniquanta enter other bubbles (and their pressure on them)}
\paragraph{Ejected techniquanta are squeezed. }
In the plasma frame, all ejected techniquarks are contained within a shell of length given by Eq.~(\ref{eq:Lej_velocity}) $L_\text{ej,p} \sim 1/f$, and most of them lie within a length given by Eq.~(\ref{eq:Lej_eff}) $L_\text{ej,p}^\text{eff} \sim 1/(\Tnuc^2 D) \ll 1/f$.
In the frame of the wall of the bubble they are about to enter, these lengths are further shrunk, so that ejected techniquarks are closer to each other than $1/f$ by several orders of magnitude.
Therefore we expect no phenomenon of string fragmentation when they enter other bubbles. So each ejected particle, upon entering another bubble, forms a hadron with one or more of its neighbours. This also implies there is no further ejection of other techniquanta.
Each of these hadrons carries an energy equal to that of the techniquanta that formed it, of order $\gwp f$ in the plasma frame.

\paragraph{Contribution to the retarding pressure. }
This conversion of ejected techniquanta into hadrons results in another source of pressure on the bubble walls, that acts for the relatively short time during which the bubble wall swallows the layer of ejected techniquanta.
In the frame of the bubble wall that they are entering, the energy of each ejected quark or gluon reads $E_\text{ej,w2} \simeq 2 \gwp^2 f$.
We then proceed analogously to what done in Sec.~\ref{sec:PLO}, and compute
\beq
\Delta p_\text{LO}^\text{ej}
= E_\text{ej,w2}-\sqrt{E_\text{ej,w2}^2 - \Delta m^2_\text{in}}
\simeq \frac{f}{4\,\gwp^2}\,,
\eeq
\beq
P_\text{LO}^\text{ej}
\simeq n_\text{ej,w2}\,\Delta p_\text{LO}^\text{ej}
\simeq  \frac{\zeta(3)}{\pi^2}\,g_\TC \,\gamma_{\rm wp}\Tnuc^3 f\,,
\eeq
where we have used $g_\TC = g_g + 3g_q/4$, $\Delta m^2_\text{in} \simeq f^2$, $n_\text{ej,w2} \simeq 2 \gwp n_\text{ej,p}$ and, for simplicity, the peak value $n_\text{ej,p} \simeq 2\,\gwp^2 n_{\TC,\text{p}}$ of Eq.~(\ref{eq:nejpx}).
The population of techniquanta ejected from other bubbles thus exert, on a given bubble wall, a pressure comparable to that exerted by the techniquanta incoming from the bath at LO, cf. Eq.~(\ref{eq:LOpressure}). 
Therefore, the pressure from ejected techniquanta does not alter the picture described so far --- a fortiori --- because it is exerted only just before bubble walls collide and not throughout their entire expansion.

\subsection{Ejected techniquanta heat the diluted SM bath}
\label{sec:SMbathheated_byejecta}
In Sec.~\ref{sec:energy_transfer} we found that the scatterings between ejected techniquanta and the diluted bath do not quantitatively change the picture of string fragmentation described in Sec.~\ref{sec:inside_bubble}.
These scatterings may however affect the properties of the particles, in the diluted bath, that do not confine.
These particles include all the SM ones that are not charged under the new confining group, so that for simplicity we denote them as `SM'.
By a derivation analogous to the one that lead us to Eq.~(\ref{eq:Qejbath}), we find that the average energy they exchange with the ejected quarks reads
\beq
Q_{\text{ej-}\SM}
\simeq \frac{8 \pi\alpha_\SM^2}{3}\frac{D \,n_{\TC,\text{p}}}{\sqrt{-t_\IR}}
\sim \alpha^2_\SM \,g_\TC \Big(\frac{\gwp}{f/\Tnuc}\Big)^{\!\frac{1}{2}} f\,,
\label{eq:QejSM}
\eeq
where we have used $\Tnuc D \simeq \gwp$ and $-t_\IR \sim \Tnuc^3 \gwp/f$, cf. Eq.~(\ref{eq:mV}).
We have denoted by $\alpha_\SM$ an effective coupling between SM particles and the techniquanta, which is model-dependent.

Now assume the techniquarks carry SM charges, e.g. as expected in composite Higgs models.
Then, in the wall frame, the fractional change of energy is of course similar to that derived in Eq.~\eqref{eq:QoverEi} for the incoming techniquanta. However the incoming techniquanta next undergo string fragmentation, and Eq.~\eqref{eq:QoverEi} does not affect that energy balance for $\gwp \gg f/\Tnuc$. In other words, string fragmentation renders this energy transfer irrelevant for the techniquanta, while the SM particles neutral under the confining group just proceed undisturbed so they keep track of it.
In particular, $Q_{\text{ej-}\SM}$, is much larger than the latter energy in the plasma frame $\sim \Tnuc$, and may even be slightly larger than the confinement scale $f$.\footnote{As already anticipated, in the regime of interest for DM phenomenology we will find that bubble walls run away, so that $\gwp^\text{max}$ is (much) smaller than $\sim 10^{-3} (f/\Tnuc)^3$, see Eq.~(\ref{eq:gwp_max}). 
Note also that, for Eq.~\eqref{eq:QejSM} only, $g_\TC = 3g_q/4$, i.e. the gluon contribution to heating the SM is negligible because they cannot carry SM charge.}

This need not be the case, however, as the new techniquanta may be very weakly interacting with the SM.
As they cannot interact too weakly, otherwise our assumption of instantaneous reheating would not hold, for simplicity we ignore this case in what follows and we assume that some techniquarks carry SM charges.

\section{Deep Inelastic Scattering in the Early Universe}
\label{sec:DIS}

The physical picture described so far results in a universe that, before (p)reheating from bubble wall collisions, contains three populations of particles.
\begin{itemize}
\item {\bf Population A.} Arises from hadronisation following string fragmentation. It consists of $N_\psi^\text{string}/2$ hadrons per quark or gluon in the initial bath, each on average with energy
\beq
E_A \simeq \frac{\gwp f}{N_\psi^\text{string}(\ECM)},
\label{eq:EA}
\eeq
in the plasma frame, and of roughly the same number of hadrons with much smaller energy. (The latter can be thought as coming from the half of the string closer to the center of the bubble wall.)
The physics resulting in this population is described in Sec.~\ref{sec:inside_bubble}, see Eq.~(\ref{eq:EAp}) for $E_A$ and Eq.~(\ref{eq:Npsi_string}) for  $N_\psi^\text{string}(\ECM = \sqrt{3\, \gwp \, \Tnuc\, f})$.
\item {\bf Population B.} Comes from the hadronisation of the ejected techniquanta. This population consists in $\sim$ one hadron per quark or gluon in the initial bath, each with energy
\beq
E_B \simeq \gwp f\,.
\label{eq:EB}
\eeq
So this population carries an energy of the same order of that of population A. Its physics is described in Sec.~\ref{sec:outside_bubble}, the energy $E_B$ is that of the initial quark or gluon, Eq.~(\ref{eq:energy_ejected}).
\item {\bf Population C.} Consists of the particles that do not feel the confinement force, that we denote `SM' for simplicity, each with a model-dependent energy given by Eq.~(\ref{eq:QejSM}), and whose total energy is much smaller than that in populations A and B.
\end{itemize}
The direction of motion of all these populations points, on average, out of the centers of bubble nucleation.

Hadrons from both populations A and B have large enough energies, in the plasma frame, that showers of the new confining sector are induced when they (or their decay products) scatter with the other particles in the universe and/or among themselves.
These deep inelastic scatterings (DIS):
\begin{itemize}
\item Increase the number density of composite states.
\item Decrease the momentum of each of these states with respect to the initial one $|\vec{p}_\psi|$.
\end{itemize}
Hence, such effects need to be taken into account to find the yield of any long-lived hadron.

The evolution of our physical system would require solving Boltzmann equations for the creation and dynamics of populations A, B and C in a universe in which preheating is occurring, and of the interactions of populations A, B and C among themselves and with the preheated particles produced from bubble wall collisions.
While certainly interesting, such a refined treatment goes beyond the purpose of this paper. 
In this Section, we aim rather at a simplified yet physical treatment, in order to obtain an order-of-magnitude prediction for the yield of long-lived hadrons.

\subsection{Scatterings before (p)reheating}
\label{sec:scatters_before_preheating}
We begin by considering the interactions among populations A, B and C.
\paragraph{Number densities of scatterers. }
Let us define $L_X$, with $X=A,B,C$, the effective thickness of the shells containing populations A, B, and C respectively.
For example, $L_\text{B,p} = L_\text{ej,p}^\text{eff}$ of Eq.~(\ref{eq:Lej_eff}).
We know that population A(B) consists on average of $N_\psi^\text{string}/2$ hadrons (one hadron) per each quark or gluon in the initial diluted bath, and that population C is the initial diluted SM population.
By conservation of the number of particles, we then obtain the number densities
\beq
n_A \simeq \frac{K^\text{string}}{2} \times \frac{D}{3 L_A} n_\TC,
\qquad
n_B \simeq \frac{D}{3 L_B} n_\TC,
\qquad
n_C \simeq \frac{D}{3 L_C} n_\SM,
\eeq
where $D$ is the average radius of a bubble at collision and we have used $L_X \ll D$.

\medskip
\paragraph{Energy transferred between scatterers. }
We now determine the average momentum, transferred to a particle from population $X$, upon going across a shell of population $Y$. 
In order to do so, we use our result Eq.~(\ref{eq:transfer_energy_rate}) for the rate of transferred energy and compute
\beq
Q_{\text{Y} \to \text{X}}
\simeq \Gamma_{\text{Y} \to \text{X}} L_Y
\simeq n_\text{Y} L_\text{Y} \frac{8\pi \alpha_\text{X-Y}^2}{\sqrt{-t_\IR}}
\simeq \frac{8\zeta(3)}{3\pi} \alpha_\text{X-Y}^2 g_\text{Y} \frac{\Tnuc}{\sqrt{-t_\IR}} \,\gwp \Tnuc\,,
\label{Q_YX}
\eeq
where $\alpha_\text{X-Y}$ is the effective interaction strength of the scatterings of interest,
$g_\text{Y}$ the number of degrees of freedom in density of population Y (where we include a factor of $N_\psi^\text{string}/2$ for Y = A), and we have used the relation $\Tnuc D = \gwp$ valid in the runaway regime.
We conclude that:
\begin{itemize}
\item {\bf Populations A and B.} The energies of the hadrons of population A and B in the plasma frame, respectively $\gwp f/N_\psi^\text{string}$ and $\gwp f$, are both much larger than the energy they can exchange with any of the other baths among A,B,C, by a factor that scales parametrically as $f/\Tnuc$ or larger (because for all populations we have $-t_\IR \sim n/\langle E\rangle > \Tnuc^2$, see the discussion in Sec.~\ref{sec:energy_transfer}).
Therefore these elastic scatterings are not effective in reducing the energy of the hadrons of either population A or population B.
\item {\bf Population C.} On the contrary, $Q_{\text{A,B} \to \text{C}}$ can be of the same order of the energy of each particle in population C, Eq.~(\ref{eq:QejSM}), which therefore are significantly slowed down by these interactions.
Importantly for our treatment, this does not alter the fact that population C was energetically subdominant with respect to populations A and B.
\end{itemize}

\paragraph{No significant DIS between populations A, B and C.}
Finally, we determine whether any of the scatterings among particles in populations A,B,C could result in significant hadron production, via deep inelastic scattering. A single scattering event potentially results in a shower of the new confining sector if the exchanged momentum is larger than the confinement scale, $t^2 > f^2$.
This condition is allowed by kinematics, because the center-of-mass energy of the scatterings between any of the populations above is much larger than $f$.
A significant amount of DIS happens if the DIS scattering rate $\Gamma_{\text{Y} \to \text{X}}^\text{DIS}$ of a particle from population $X$, upon going across a shell of population $Y$, is much larger than the inverse of the length of the shell $Y$.
We then compute
\beq
\Gamma_{\text{Y} \to \text{X}}^\text{DIS} L_\text{Y}
\simeq n_\text{Y} \sigma_\text{X-Y} v\, L_\text{Y}
\simeq \frac{4 \zeta(3)}{3 \pi} \alpha_\text{X-Y}^2 g_\text{Y} \frac{\gwp}{(f/\Tnuc)^2}\,,
\label{eq:DIS_XY}
\eeq
where again we have used the runaway relation $\Tnuc D = \gwp$ and, for definiteness, we have assumed the scattering cross section has the form of Eq.~(\ref{eq:dsigmadt}).
Therefore, no significant DIS happens in the regions where $\gwp \ll (f/\Tnuc)^2$.
This condition will turn out to be always satisfied in the parameter space of our interest, so we can ignore the DIS among populations A, B and C in what follows.

\subsection{Scatterings with the (p)reheated bath}
\label{sec:DIS_preheated}
By \textit{preheating}, we intend the stage between the time when bubble walls collide and start to produce particles (e.g. from the resulting profile of the condensate), and the \textit{reheating} time when these particles have thermalised into a bath.
We now discuss the scatterings of populations A and B with the particles produced at preheating, that we have assumed to be efficient. 
The contribution of population C to the final yield of hadrons is subdominant with respect to the one of populations A and B because, as seen in Secs.~\ref{sec:SMbathheated_byejecta} and \ref{sec:scatters_before_preheating}, the total energy in population C is much smaller than that in populations A and B.

\paragraph{Energy of the (p)reheated bath. }
The preheated particles are produced with energies, in the plasma frame, of the order of the mass of the scalar condensate,\footnote{In the picture we have in mind, non-perturbative effects such as Bose enhancement or parametric resonance (see e.g.~\cite{Kofman:1997yn}) are not relevant: the first because the SM particles are interacting, thus they exchange momentum and do not occupy the same phase space cells; the second because the variation of their masses from the dilaton's oscillations is smaller than their mass at the minimum. Note that, unlike what occurs in many inflationary scenarios, we expect only a small hierarchy $ T_{\rm RH} \lesssim \langle E_\text{prh} \rangle $.}
\beq
\langle E_\text{prh} \rangle \simeq m_\chi < f\,.
\eeq
Their total energy scales as
\beq
E_\text{prh}^\text{tot} \sim f^4 V\,,
\eeq
with V the volume of a large enough region of the universe.
For comparison, the total energy in populations A and B scales as
\beq
E_\text{A,B}^\text{tot} \sim \gwp f \Tnuc^3 V\,,
\eeq
which is much smaller than $f^4 V$ because $\gwp \ll (f/\Tnuc)^3$, Eq.~(\ref{eq:gwp_max}).
So the preheated particles can act as a thermal bath for all the other populations A, B and C, because the energy of A, B, and C is subdominant in the energy budget of the universe.

\paragraph{Inelastic versus elastic scattering. }
Scatterings of hadrons (or their decay products) with the preheated bath will, therefore, eventually slow down and thermalise populations A and B.
However, these scatterings can also exchange energies much larger than $f$,  thus inducing deep inelastic scatterings. Indeed their center-of-mass energy squared reads
\beq
s_\text{A,B} \simeq 2\, m_\chi E_\text{A,B}\,,
\label{eq:s_DIS}
\eeq
where $E_\text{A} \simeq \gwp f /N_\psi^\text{string}(\ECM)$ and $ E_\text{B} \simeq \gwp f$.
Eq.~(\ref{eq:s_DIS}) is the result of our simplifying assumption to neglect masses and to average to zero scattering angles with particles in a bath: define $p_E= E(1,\,\hat{E})$, $p_\text{prh} = m_\chi(1,\,\hat{m})$, then $s = (p_E+p_\text{prh})^2 \simeq 2 E \,m_\chi (1- \hat{E} \cdot \hat{m}) \simeq 2 E\, m_\chi$.
We now determine if those center of mass energies are entirely available for particle production via DIS, or if instead they are reduced by several low-momentum-exchange interactions.
In order to do so, we evaluate the rate of energy loss of a particle from population A or B, $\Gamma^\text{loss}_\text{A,B}$, as the ratio between the rate of energy it exchanges with the preheated bath, that we evaluate analogously to Eq.~(\ref{eq:transfer_energy_rate}), and its initial energy $E_\text{A,B}$.
We then compare this quantity with the rate for a deep inelastic scattering to happen with the full energy available $s_\text{A,B}^{1/2}$,
\beq
\frac{\Gamma^\text{loss}_\text{A,B}}{\Gamma^\text{DIS}_\text{A,B}}
\simeq \frac{n_\text{prh} 8 \pi\alpha_\text{eff}^2/(E_{A,B} \sqrt{-t_\IR})}{n_\text{prh} 4 \pi\alpha_\text{eff}^2/s_{A,B}}
\simeq \frac{m_\chi}{\sqrt{-t_\IR}}
\sim \frac{1}{\sqrt{c_\text{vac}}}\frac{m_\chi^2}{f^2}\,.
\label{eq:loss_over_DIS}
\eeq
In the last equality, we have again used the screening mass for non-equilibrium systems~\cite{Arnold:2002zm}
\beq
-t_\IR
\sim \frac{n_\text{prh}}{\langle E_\text{prh} \rangle}
\sim c_\text{vac} \frac{f^4}{m_\chi^2}\,,
\eeq
where we have used that by conservation of energy $n_\text{prh} \sim \rho_\RH/\langle E_\text{prh} \rangle$, and where we have expressed the energy density of the reheated bath $\rho_\RH$ using the results of Sec.~\ref{sec:thermal_history}.

We conclude that, if
\beq
\Big(\frac{m_\chi}{f}\Big)^2 \ll c_\text{vac}^{1/2}\,,
\label{eq:condition_full_DIS}
\eeq
the full center-of-mass energies $s_\text{A,B}$ are available for deep inelastic scattering, i.e. populations A and B do not lose a significant amount of their energy via interactions with the preheated bath.
For simplicity, in what follows we assume this model-dependent property to hold.

\subsection{Enhancement of hadron abundance via DIS}
\label{sec:DIS_hadron_abundance}

\paragraph{The picture: a cascade of DIS. }
The number of composite states arising from a hard scattering depends on how the strings fragment, so on the same physics that set the abundance of the composite states when the techniquanta cross the bubble walls, discussed in Sec.~\ref{sec:fragmentation_multiplicity_energy}.
Each scattering, depending on its center-of-mass energy, produces a number $N_\psi^\text{string}$ of hadrons $\psi$, that we model in the same was as in Eq.~(\ref{eq:Npsi_string}).
Given the large initial energies $s_\text{A,B}$, the daughter hadrons typically still have enough energy to themselves induce further deep inelastic scatterings with the particles in the preheated bath, and hence additional hadron production.
Analogously, SM particles produced in such DIS typically have large enough energies to also initiate showers of the new confining force with their subsequent scatterings.
This process iterates until the average energy of scatterings drops below the confinement scale.

\paragraph{Number of hadrons produced per scattering. }
For reasons given in Sec.~\ref{sec:fragmentation_multiplicity_energy}, together with simplicity, we assume that the available energy $\sqrt{s}$ at each scattering splits equally among all the outcoming particles.  We then write the average of this number as
\beq
N^\DIS(s) =  N_\psi^\text{string}(\sqrt{s}/2)\,,
\label{eq:Npsi_NSM}
\eeq
where the factor of 2 in the argument of $N_\psi^\text{string}$ arises because Eq.~(\ref{eq:Npsi_string}), which defines $N_\psi^\text{string}$, assumes that $\sqrt{s}$ is the center of mass energy of the scattering of two particles neutral under the new confining force.
If a hadron is included among the two scatterers, then QCD studies find that the final number of hadrons can be obtained by just halving the energy in the center of mass frame~\cite{Rosin:2006av}, also see footnote~\ref{foot:Npsi_ee_vs_pp}.\footnote{
Note that if a hadron instead decays to two SM particles before it scatters, which is model-dependent, then $\sqrt{s}/2$ is again the good argument for the function $N_\psi^\text{string}$, because then one has two particles each with half the initial energy, but both neutral under the new confining force. In this case, however, Eq.~\eqref{eq:Npsi_NSM} becomes $N^\DIS(s) = 2 N_\psi^\text{string}(\sqrt{s}/2)$.
When iterating the treatment to many scatterings, we find that this extra factor of~2 does not impact the final abundance of hadrons, which can be understood by thinking that the same initial energy is spread faster to zero.}

\paragraph{Energies of produced hadrons. }
Explicitly, we assume $E'_\text{com} = \sqrt{s}/N^\DIS$, where $E'_\text{com}$ is the energy of any outgoing particle (SM and/or composite) in the center-of-mass frame of the scattering.
To iterate to many scatterings, we write $E'_\text{com}$ in the plasma frame as
$E' = \gamma'\,E_{\rm com}'(1 -   \hat{v}'\cdot \hat{v})$, where $\gamma'$ and $\hat{v}'$ are the associated Lorentz boost and its direction, and $\hat{v}$ is the direction of motion of the outgoing particle in the center-of-mass frame of the scattering.
By averaging $\hat{v}'\cdot \hat{v}$ to zero for simplicity, we obtain
\beq
E' = \gamma'\,E_{\rm com}'\,.
\eeq
We then determine $\gamma'$ by observing that the energy of each particle, in the center-of-mass frame of the scattering, is both $E_{\rm com} = \sqrt{s}/2$ and $E_{\rm com} = \gamma' E_\text{prh} (1 + \hat{v}' \cdot \hat{E}_{\rm com})$, where $E_\text{prh}$ is the energy in the plasma frame of the particles in the preheated bath. By averaging $\hat{v}' \cdot \hat{E}_{\rm com}$ to zero for simplicity, we obtain the Lorentz boost
\beq
\gamma' \simeq \frac{\sqrt{s}}{2 \langle E_\text{prh}\rangle}\,.
\eeq
Using Eq.~(\ref{eq:s_DIS}) for $s$ we finally obtain
\beq
E'_\text{A,B} \simeq \frac{1}{N^\DIS} E_\text{A,B}\,.
\eeq
(If we did not average over angles, we would have obtained $E'_\text{A,B} = (E_\text{A,B}/N^\DIS)(1-\hat{v}'\cdot \hat{v})(1-\hat{E} \cdot \hat{m})/(1+\hat{v}' \cdot \hat{E}_{\rm com})$).
So, after a hard scattering the energy of each outgoing particle in the plasma frame is roughly the initial energy divided by a factor $N^\DIS$. The subsequent $s$ is then reduced by the same factor, ensuring a convergence of $N^\DIS(s)$ to unity, via Eq.~(\ref{eq:Npsi_string}), after only a few iterations.
This also teaches us that the average energy of the particles, produced this way, quickly decreases to values lower than about $m_*$.

\paragraph{Number of hadrons produced by a chain of DIS. }
Let us now estimate the yield of final hadrons by following the above arguments. Assuming interactions are fast enough, also those following the first one happen with preheated particles of the same average energy $\langle E_\text{prh}\rangle$.
Now define the number of states (both composite and not) $N_{k}$ produced at the $k^\text{th}$ interaction. This can be expressed as
\beq
N_k(s)
\simeq N^\DIS
\Big(\frac{s}{
N_{k-1}
\times N_{k-2}
\times \cdots
\times N_{1}}\Big),
\eeq
where we remind the reader that the function $N^\DIS$ is obtained from Eqs.~(\ref{eq:ncharged_s}) and (\ref{eq:Npsi_NSM}).
Starting from a single resonance produced from the fragmentation of strings between quanta inside the bubble, after this chain of scattering processes one obtains a total number of resonances given by the product $\prod_k N_{k}(s)$.
We find numerically that this product can be expressed as
\beq
K^\DIS_\text{A,B}
 \simeq \frac{s_\text{A,B}}{m_*^2}\,.
 \label{eq:KDIS_AB}
\eeq
In other words, the iterative process we described converts the initial available energy into the rest mass of hadrons $m_*$.
Since our aim here is not to achieve a more precise treatment, we refrain from refining the assumption that the momenta are distributed evenly among the particles coming out of a scattering process. 
In the same spirit of building a physically-clear picture without drowning in model-dependent details, we do not cover here the possibility that every scattering produces, in addition to the composite states, a comparable or larger amount of SM particles. (That would result in $N^\DIS > N_\psi^\text{string}$ and in a faster degrowth of the available scattering energy to $m_*$ at each step.) In addition to simplicity, this can be justified by observing that, in the limit of large number of degrees of freedom in the dark sector, our assumption that they carry SM charges will make their production dominant with respect to the one of SM particles.

\paragraph{Additional comments.}
We conclude our derivations with two comments concerning its validity.
\begin{itemize}
\item
If the full center-of-mass energies are not available for DIS, i.e. if Eq.~(\ref{eq:condition_full_DIS}) does not hold, then one could use the same result  $K^\DIS_\text{A,B}$ of Eq.~(\ref{eq:KDIS_AB}), upon substituting $s_\text{A,B} = 2E_\text{A,B} m_\chi$ with the largest energy for which $\Gamma^\text{loss}_\text{A,B} \ll \Gamma^\text{DIS}_\text{A,B}$, that can be derived via Eq.~(\ref{eq:loss_over_DIS}).

\item We have ignored the production of heavy particles from the collisions of bubble walls~\cite{Hawking:1982ga, Watkins:1991zt, Konstandin:2011dr, Braden:2014cra}.
This is justified as it has been shown that it only occurs when the minima of the potential are nearly degenerate and seperated by a sizable barrier~\cite{Falkowski:2012fb, Katz:2016adq}, which is not the case for the close-to-conformal potentials we have in mind. 
Hence we expect only particles lighter than the scalar condensate to be produced during reheating following the wall collision.
\end{itemize}

\subsection{DIS summary}
\label{sec:DIS_summary}

The yield of hadrons, resulting from the processes of deep inelastic scattering described above, receives contributions from:
\begin{itemize}
\item {\bf Population A.} That is, the hadrons produced from string fragmentation as described in Sec.~\ref{sec:inside_bubble}.
Their contribution reads
\beq
Y^{\SC+\rm string+\DIS }_\text{A}
\simeq \frac{1}{2}  K^\DIS_\text{A}  N_\psi^\text{string}(\ECM) D^\SC Y_\TC^{\rm eq}
\simeq \frac{\gwp f m_\chi}{m_*^2}\,D^\SC Y_\TC^{\rm eq}\,,
\label{eq:KDIS_popA}
\eeq
where we have used $K^\DIS_\text{A} = s_\text{A}/m_*^2$, cf. Eq.~\eqref{eq:KDIS_AB}, with $s_\text{A} \simeq 2 m_\chi E_\text{A} \simeq 2 m_\chi \gwp f/N_\psi^\text{string}(\ECM)$, cf. Eqs.~\eqref{eq:s_DIS} and \eqref{eq:EA}.
Note that the above expression captures also the regime where each string fragmentation produces on average one hadron, because the energy of that single hadron is  roughly $\gwp f/2$, see the related discussion in Sec.~\ref{sec:string_summary}.

\item {\bf Population B.} That is, the hadrons produced out of the techniquanta ejected from the bubbles, described in Sec.~\ref{sec:outside_bubble}.
Their contribution reads
\beq
Y^{\SC+\rm string+\DIS }_\text{B}
\simeq K^\DIS_\text{B} D^\SC Y_\TC^{\rm eq}
\simeq 2\,\frac{\gwp f m_\chi}{m_*^2}\,D^\SC Y_\TC^{\rm eq}\,,
\eeq
where we have used $K^\DIS_\text{B} = s_\text{B}/m_*^2$, cf. Eq.~\eqref{eq:KDIS_AB}, with $s_\text{B} \simeq 2\,m_\chi E_\text{B} \simeq 2\,m_\chi \gwp f$, cf. Eqs.~\eqref{eq:s_DIS} and \eqref{eq:EB}.
\end{itemize}

Thus, the combined contribution to the total hadron yield is given by
\beq
	\label{eq:DISrelic}
	Y^{\SC+\rm string+\DIS }
	\simeq K^\DIS D^\SC \,Y_\TC^{\rm eq}
	\simeq  3\,\frac{\gwp\, f\, m_\chi}{m_*^2}\,D^\SC\, Y_\TC^{\rm eq}\,,
\eeq
where we have defined
\beq
K^\DIS = \frac{1}{2}K^\DIS_\text{A} N_\psi^\text{string}(\ECM)  + K^\DIS_\text{B}\,.
\label{eq:KDIS_definition}
\eeq
Note finally that, in the regime of runaway bubble-walls, one obtains the parametric scaling $Y^{\SC+\rm string+\DIS} \propto (\Tnuc/f)^3 \gwp$. Which is much larger than the simple supercooling dilution, $\sim (\Tnuc/f)^3$, in the regions of parameter space where our analysis holds, namely for $\gwp > f/\Tnuc$.

\section{Supercooled Composite Dark Matter}
\label{sec:DMabundance}

\subsection{Initial condition for thermal evolution}

Finally, all unstable resonances decay either to SM or to the long-lived or stable hadrons, which we take to form DM.
To obtain the yield of any such hadron $i$ at the onset of reheating, one should use the expression
	\beq
	\label{eq:DISrelic_i}
	Y_i^{\SC+\rm string+\DIS } =  \mathrm{BR}_i\,K^\DIS  D^\SC Y_\TC^{\rm eq}\,,
	\eeq
where $K^\DIS$, $D^\SC$ and  Y$_\TC^{\rm eq}$ are defined, respectively, in Eqs.~\eqref{eq:KDIS_definition}, \eqref{eq:DSC} and~\eqref{eq:Yeqi}.
$\mathrm{BR}_i$ is a pseudo-branching ratio, of the energy available to the confining techniquanta, into $\psi_i$ particles.
	Estimates of $\mathrm{BR}_i$ for the cases where $\psi_i$ is a meson and a baryon are given in App.~\ref{app:brstring}, which show a broad range of underlying-model dependent values are possible, albeit with a large uncertainty.
	For example, in a QCD-like theory where $\psi_i$ is a baryon with mass $\sim 4\pi f$ and the pions have mass $\sim f$, one obtains values $\mathrm{BR}_i \sim 10^{-6}$, while larger values $\mathrm{BR}_i$ are obtained for baryon-pion mass rations closer to one, or if $\psi_i$ is a meson. Hence, we will take $\mathrm{BR}_i$ to be a free parameter.

		\begin{figure}[t]
\centering
\begin{adjustbox}{max width=1.\linewidth,center}
\includegraphics[width= 0.5\textwidth]{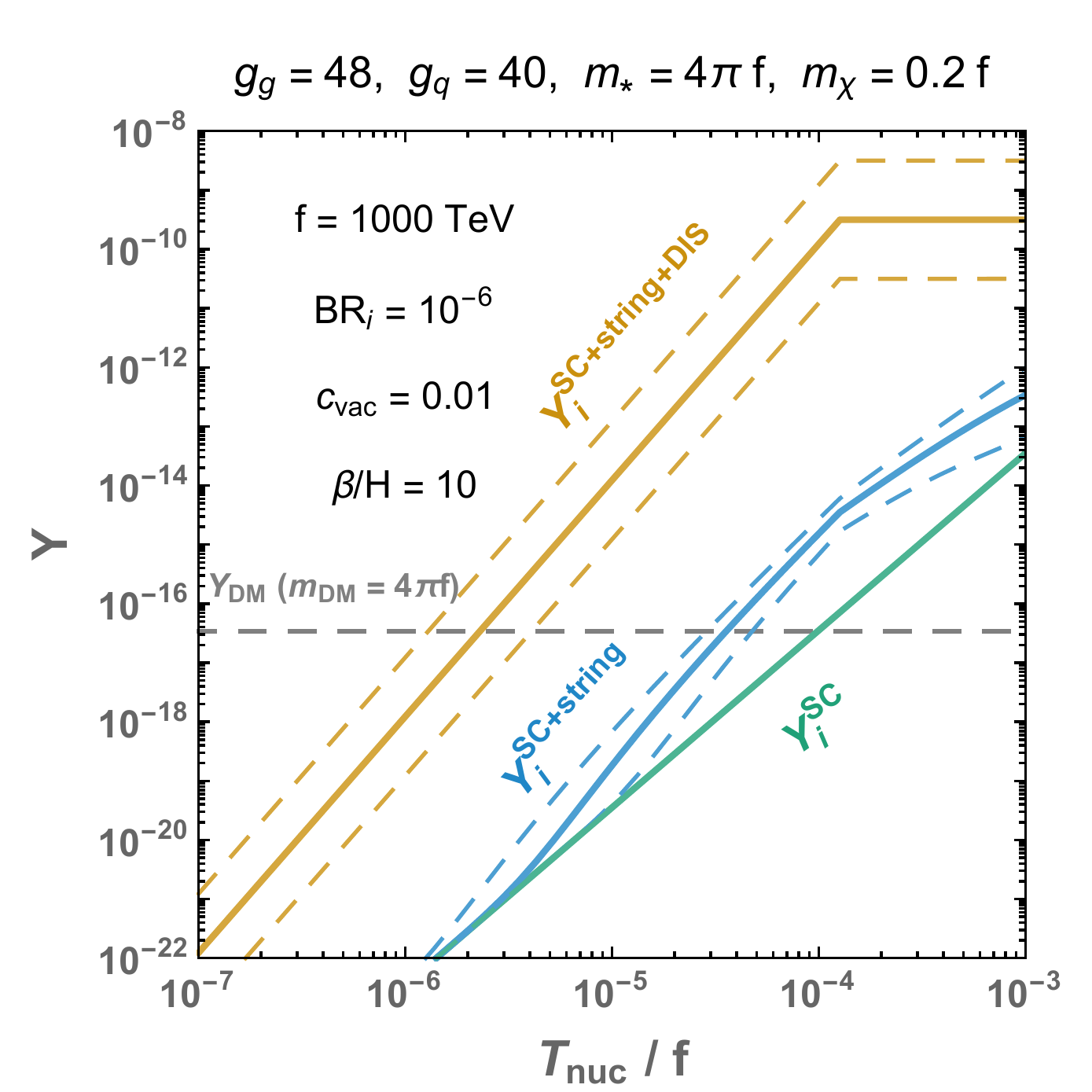}
\includegraphics[width= 0.5\textwidth]{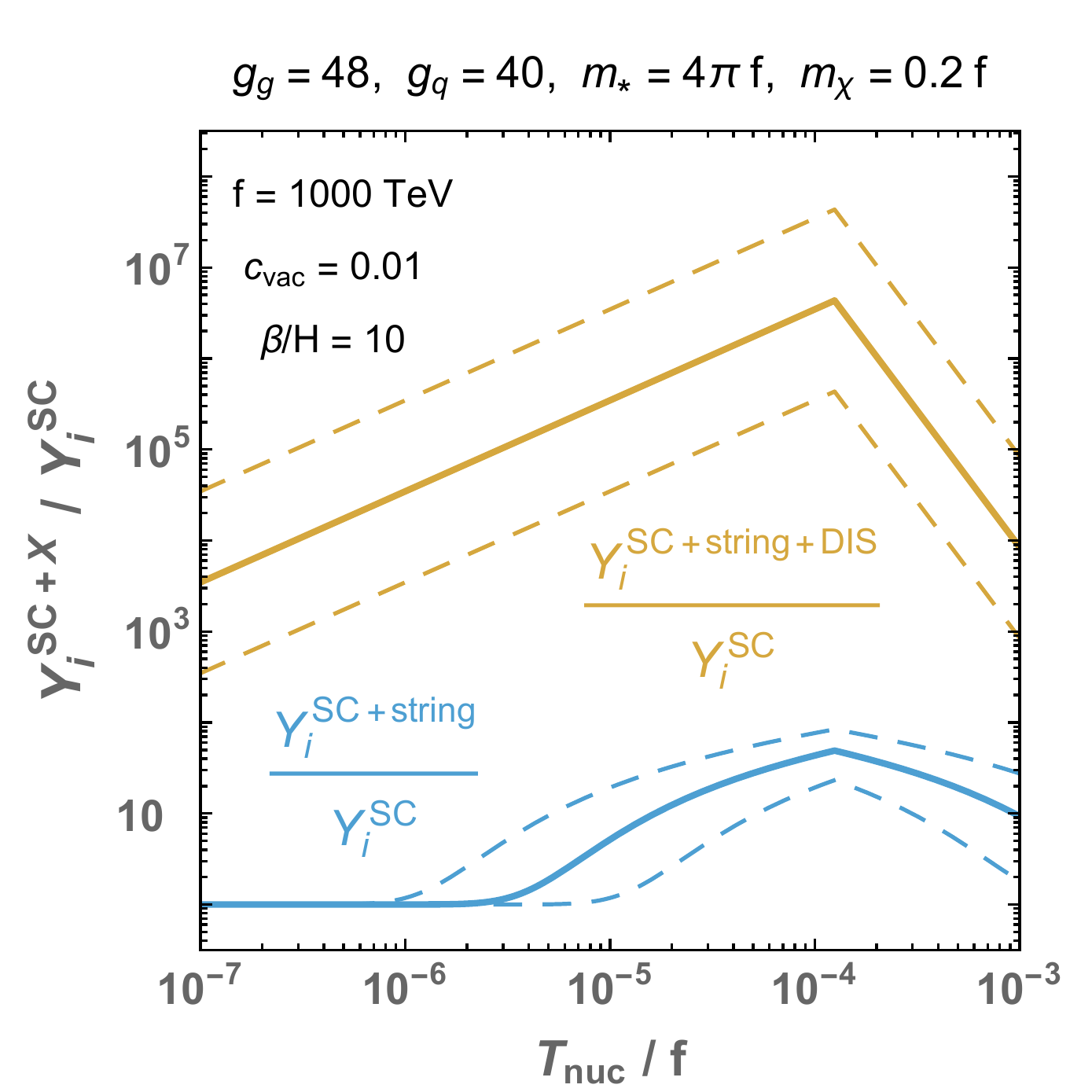}
\end{adjustbox}
\caption{\it \small Left: the yields following supercooling Eq.~(\ref{eq:SC_relic_i}), string fragmentation Eq.~\eqref{eq:SC_and_string_relic_i} and deep inelastic scattering Eq.~\eqref{eq:DISrelic_i}. The yield matching the observed relic abundance of DM for $\mDM = 4\pi\,f$ is also shown. The dashed lines show the effect of varying $\gwp$ by an order of magnitude in either direction around Eq.~(\ref{eq:gwp_max}). This illustrates the sensitivity of the yield to our determination of $\gwp$. Right: ratios of the same yields. The peak corresponds to the maximum $\gwp$. }
\label{fig:contributions} 
\end{figure}

	 For completeness, the supercooling plus string  and supercooling yields read
	\begin{eqnarray}
	Y_i^{\SC+\rm string} &=&  \mathrm{BR}_i\,K^{\text{string}}  D^\SC Y_\TC^{\rm eq}, \label{eq:SC_and_string_relic_i}\\
	Y_i^\SC &=&  \mathrm{BR}_i\, D^\SC \frac{\frac{3}{4}g_q}{g_\TC} Y_\TC^{\rm eq}\,, \label{eq:SC_relic_i}
	\end{eqnarray}
	where $K^{\text{string}}$ is defined in Eq.~\eqref{eq:Kstring}. We have included a factor $\frac{3}{4}g_q/g_\TC$ in $Y_i^\SC$ to account for the fact that, in the case of no string fragmentation nor DIS, gluons do not contribute to the final abundance of heavy composite states of quarks. It would be absent if one was interested in light composite states of quarks. The yield of the various contributions is shown in Fig.~\ref{fig:contributions}.

It will turn out that the measured DM abundance is achieved in the regime of runaway bubble walls. In that regime, the resulting expression for the DM yield has a simple parametric form that eventually results in the DM abundance being independent of the DM mass, if it is to match onto observation $Y_{\rm DM}  \simeq 0.43 \, \mathrm{eV}/\MDM$~\cite{Aghanim:2018eyx}, which we find convenient to report here.
By using~Eqs.~(\ref{eq:Yeqi}), (\ref{eq:DSC}), (\ref{eq:gwp_runaway}), and (\ref{eq:DISrelic}),  with $g_{Rf} = g_\SM = 106.75$, we find
\beq
Y_{i,\text{runaway}}^{\SC+\rm string+\DIS } 
\simeq 0.43\,\frac{\text{eV}}{m_*}
\times
\frac{\mathrm{BR}_i}{10^{-6}}
\, \frac{g_\TC}{120}
\,\left(\frac{0.01}{c_\text{vac}}\right)^{\!\frac{5}{4}}
\,\frac{m_\chi/f}{0.2}
\,\frac{4\pi}{g_*}
\,\left(\frac{\Tnuc/f}{10^{-5.7}}\right)^{\!4}\,.
\label{eq:DISrelic_runaway}
\eeq

\subsection{Thermal contribution}
\label{sec:sub-thermal-abundance}

To complete our discussion, we must still determine the effects on the yield of any DM interactions with the thermal bath after supercooling, DIS, and reheating.
The importance of thermal effects following reheating was already pointed out in~\cite{Hambye:2018qjv} (therein dubbed the subthermal contribution). Following the phase transition and particle production through DIS, the SM bath and the DM have returned to kinetic equilibrium. The scattering energy is now insufficient to break the resonances, but these may still annihilate into SM particles or be produced in the inverse process. Thus, just after the reheating, the DM abundance evolves according to the well known Boltzmann equation~\cite{Kolb:1990vq}
	\begin{align}
	\label{eq:first_case}
	\frac{dY_{\mathsmaller{\rm DM}}}{dx} = -   \sqrt{ \frac{ 8\pi^2 \gSM }{ 45 }} \frac{ M_{\rm pl} \MDM \, \left<\sigma v_{\mathsmaller{\rm rel}} \right>}{x^2} \, \left( Y_{\mathsmaller{\rm DM}}^2 -  Y_{\mathsmaller{\rm DM}}^{\mathrm{eq} \, 2} \right),
	\end{align}
where we use $x\equiv\MDM/T$ as the time variable, and $M_{\rm pl}$ is the reduced Planck mass. For simplicity we only consider velocity independent cross sections here. As an intitial condition we take the relic abundance at the reheat temperature, $Y_{\mathsmaller{\rm DM}}(T_{\rm RH}) = Y_{\mathsmaller{\rm DM}}^{\SC+\rm string+\DIS }$, estimated following string fragmentation and DIS enhancement in Eq.~\eqref{eq:DISrelic}. 
For our plots we solve the Boltzmann equation numerically. If the cross section and reheating temperatures are sufficiently large the system will be driven back into equilibrium. The relic density is then largely set by freezeout dymanics, albeit with somewhat different initial conditions. On the other hand, if the cross section and reheat temperatures are small enough, the relic density is set by dilution, string fragmentation and DIS, with only negligible thermal corrections following reheating.
Using the dilution mechanism of the PT, of course, we can avoid the usual unitarity constraint on the maximum thermal relic DM mass~\cite{Griest:1989wd} (see e.g.~\cite{vonHarling:2014kha,Baldes:2017gzw} for recent appraisals).

\begin{figure}[t!]
\begin{center}
\begin{adjustbox}{max width=1.0\linewidth,center}
\includegraphics[width=.5\textwidth]{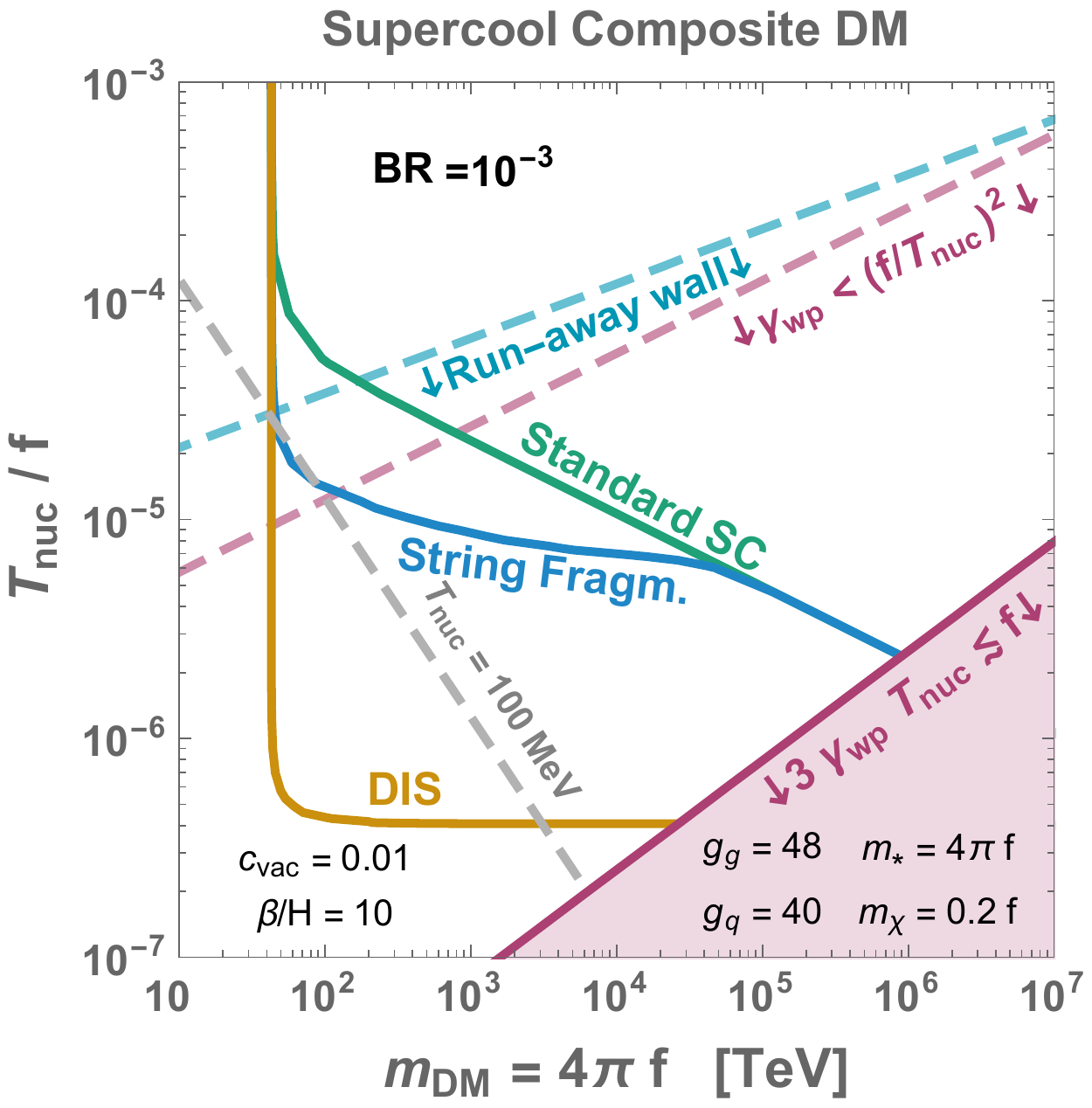} \quad
\includegraphics[width=.5\textwidth]{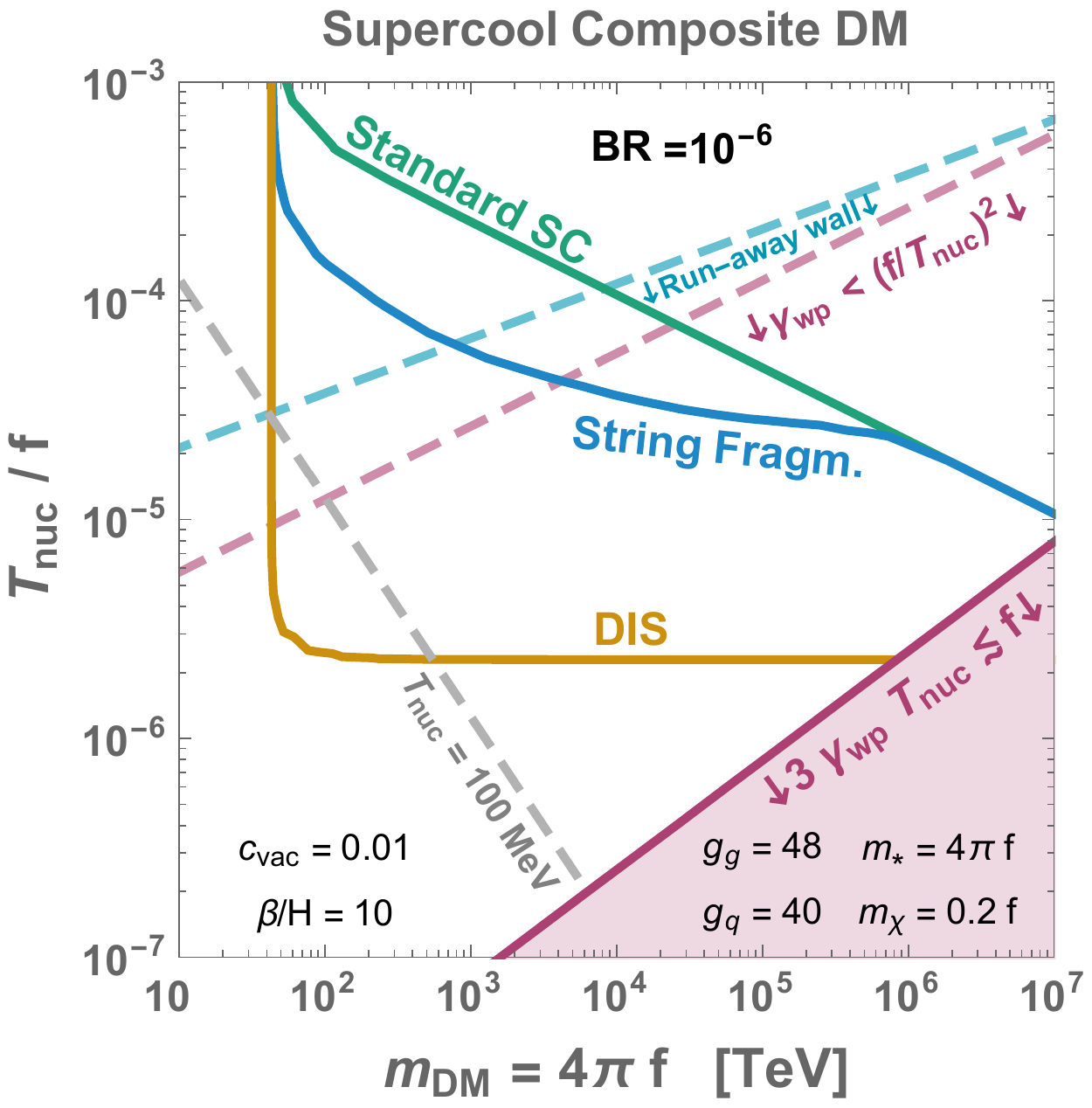}
\end{adjustbox}
\caption{
\label{fig:compositeDM_generic} 
\it \small
Solid lines: supercooling $\Tnuc/f$ and DM mass $\MDM$ required to obtain the observed DM abundance. The parameters chosen imply a reheating temperature $T_\RH \simeq 0.13~f$, see Sec.~\ref{sec:thermal_history}.
All lines include the thermal contribution discussed in Section~\ref{sec:sub-thermal-abundance}.
The line with initial condition $Y^{\rm \SC+string+\DIS}_i$ of Eq.~(\ref{eq:DISrelic_i}) corresponds to the yellow contour.
For comparison, we show in green (blue) the contour that one would obtain by skipping directly from the supercooling (supercooling plus string fragmentation) step to the thermal corrections, respectively Eqs.~\eqref{eq:SC_relic_i} and~\eqref{eq:SC_and_string_relic_i}.
All contours converge at some $\MDM$ where thermal effects following reheating become dominant, of the order of $\MDM \approx 100 \, \mathrm{TeV}$ because we fixed $\langle \sigma v_{\mathsmaller{\rm rel}}\rangle =4\pi/\MDM^2$. Below this mass, the relic density is necessarily suppressed compared to the observed DM density, due to efficient DM annihilation after reheating. 
In the purple region $3\,\gwp\,\Tnuc \lesssim f$ the quarks are reflected by the first wall they encounter, but may enter the bubbles in following stages of their evolution, and the DM abundance lines ignore possible modifications arising from this `ping-pong' effect.
They also ignore that, for values of $\gwp$ only slightly larger than $f/\Tnuc$ and depending on other model-dependent parameters, the energetics of our treatment may be more complicated, see Eqs.~(\ref{eq:QoverEi}) and~(\ref{eq:mV}).
The dashed gray line delimits the area $\Tnuc < O(100) \, \mathrm{MeV}$ where the supercooled phase transition could happen because of QCD dynamics.
The dashed light blue line indicates the regimes where bubble walls run away, cf. Eq.~\eqref{eq:run-away_cond}. The dashed purple line indicates the regime where $\gwp < (f/\Tnuc)^2$, and the fact it lies above the horizontal part of the DIS line confirms that our treatment has been consistent when ignoring the DIS of Eq.~(\ref{eq:DIS_XY}).
}
\end{center}
\end{figure}

\subsection{Dark matter relic abundance}

We now combine all our results together and determine the amount of supercooling required to match the observed relic abundance $Y_{\rm DM}  \simeq 0.43 \, \mathrm{eV}/\MDM$. Examples are shown in Fig.~\ref{fig:compositeDM_generic} for some representative choices of the parameters. From these figures we can draw a number of conclusions. 

\begin{enumerate}
\item[i)] If we assume $\left<\sigma v_{\mathsmaller{\rm rel}} \right> \propto 1/\MDM^{2}$, thermal effects will necessarily dominate if the DM is light enough. This occurs because $T_{\rm RH}$ cannot realistically be arbitrarily suppressed below $f$, for sensible choices of $g_{Ri}$ and $g_{Rf}$. This regime corresponds to the point in which the contours turn vertical in Fig.~\ref{fig:compositeDM_generic}. At which value of $\MDM$ this occurs depends on the precise choice for $\left<\sigma v_{\mathsmaller{\rm rel}} \right> $.
For definiteness, in Fig.~\ref{fig:compositeDM_generic} we choose $\langle \sigma v_{\mathsmaller{\rm rel}} \rangle = 4\pi/\MDM^2$ as typical of baryon scatterings in a strongly coupled sector.
Thermal effects can of course be further suppressed if we depart from the efficient reheating assumption made here~\cite{Hambye:2018qjv}.

\item[ii)] String fragmentation and DIS lead to large corrections to the composite DM relic density, compared to the naive supercooling dilution. This implies a mismatch between the relic abundances of primordial elementary and composite relics alluded to before. Whether the composite or elementary relic would have the greater abundance depends on the details of confinement (for elementary relics BR$_i=1$). If the composite relic is say, a light meson which is produced abundantly, the multiplicative DIS process can be highly efficient in populating these states following the PT. This implies we require much more supercooling to match onto the observed DM relic abundance. On the other hand, if the composite relic is some heavy state, perhaps a baryon, it could be produced in a highly suppressed rate both in string fragmentation and DIS. In this latter case, the required amount of supercooling to match onto the DM relic density is also reduced. The two cases are illustrated with two different assumptions for the branching ratios in Fig.~\ref{fig:compositeDM_generic}.\footnote{
We checked that the DIS line is unaffected if we use the treatment of the NLO pressure of~\cite{Hoeche:2020rsg}, instead of the one of~\cite{Bodeker:2017cim} that we have employed in this paper, cf. Sec.~\ref{sec:P_NLO} and App.~\ref{app:NLO_pressure}. On the other hand, the run-away wall dashed line, and hence the string fragmentation line, could be affected by this choice. For simplicity as well as in light of the criticism of~\cite{Hoeche:2020rsg} appeared in~\cite{Vanvlasselaer:2020niz,Azatov:2021ifm}, we employed the results of~\cite{Bodeker:2017cim} in this paper.}

\item[iii)] In some cases, we find $T_{\rm nuc} \lesssim 100$ MeV, as delineated in Fig.~\ref{fig:compositeDM_generic}. Thus QCD effects could assist in completing the PT~\cite{Witten:1980ez,Iso:2017uuu,vonHarling:2017yew,Hambye:2018qjv}. On the other hand, if QCD effects help the transition to occur, they can also suppress the eventual gravitational wave signature~\cite{Baldes:2018emh} (simply because the QCD effects increase the tunneling probability and thus will act to shorten the timescale of the PT). The details will depend on the physics entering the effective potential of the scalar $\chi$ and need to be studied in a model dependent way.

\end{enumerate}

Together with the gravitational wave signal from the PT, there may also be model dependent collider, direct, and indirect detection signatures associated with the DM from the strongly coupled sector. We will investigate these further, together with their interplay with the novel string fragmentation and DIS effect, in a concrete realisation of such a confining sector in a companion paper~\cite{2ndpaper}.

\section{Discussion and Outlook}
\label{sec:outlook} 
The possible existence of a new confining sector of Nature is motivated by several independent problems of the Standard Model of particle physics and by cosmology.
This encourages the identification of predictions of confining sectors, that are independent of the specific problem they solve.
One such prediction is the possibility that the finite temperature phase transition in the early universe, between the deconfined and confined phase, is supercooled.
This possibility has received a lot of attention in recent years, see e.g.~\cite{Iso:2017uuu,vonHarling:2017yew,Hambye:2018qjv,Baldes:2018emh,Bruggisser:2018mrt, Baratella:2018pxi, Agashe:2019lhy, DelleRose:2019pgi, Ellis:2019oqb,vonHarling:2019gme}.

In this paper, we have pointed out and modelled a novel dynamical picture taking place in every supercooled confining phase transition, that (to our knowledge) had been missed in the literature.
This novel picture stems from the observation that, when fundamental techniquanta of the confining sector are swept into expanding bubbles of the new confining phase, the distance between them is large with respect to the confinement scale.
Therefore the energy of the fluxtubes connecting techniquanta is so large that string breaking produces many hadrons per fluxtube, with large momenta in the plasma (CMB) frame, in a sense analogously to QCD hadrons produced in electron-positron collisions at colliders.
These hadrons and their decay products subsequently undergo scatterings with other particles in the universe, with center-of-mass energies much larger than both the confinement scale and the temperature that the universe reaches after reheating.
The dynamics just described is partly pictured in Figs.~\ref{fig:wall_diagram}~and~\ref{fig:string_breaking}.

The processes of string fragmentation and `deep inelastic scatterings in the sky', synthetised above, have a plethora of implications.
A key quantity to study them is the pressure on the bubble walls induced by this novel dynamics, which we have determined in Sec.~\ref{sec:wall_speed}, see Eq.~(\ref{eq:gwp_max}) and Fig.~\ref{fig:gwp} for the resulting bubble-wall velocities.
An interesting aspect of our findings is that the so-called `leading-order' pressure is proportional to the boost factor of the bubble wall, unlike in the case of non-confining supercooled PTs~\cite{Bodeker:2009qy,Bodeker:2017cim}.

We then quantified the values of supercooling below which one recovers the `standard phase transition', where confinement happens between nearest charges.
By relying on the modelling we proposed in Sec.~\ref{sec:our_picture_relevant} we found, interestingly,
that the PT does not proceed in the `standard' way already for minor supercooling, i.e. if bubbles are nucleated and expand just after vacuum energy starts to dominate.
Our proposed dynamics should not only be employed in the large supercooling region, but also in the minor supercooling one depending on the value of another model-dependent parameter, see Fig.~\ref{fig:StdPictureRecovered}.
The regimes in between these regions (one being the `ping-pong' regime of Sec.~\ref{sec:pingpong}) will be studied in future work, to not charge this paper with too much content.

Next, we have focussed on the implications of our dynamical picture for the abundance of long-lived or stable particles that are composite states of the new confining sector. They are summarised in the Synopsis, Sec.~\ref{sec:synopsis}, and a quantitatively accurate expression of the final yield of a given composite particle is given in Eq.~(\ref{eq:DISrelic_runaway}), for concreteness in the regime where bubble walls run away.
Compared to the simple dilution of relics induced by supercooling of non-confinement transitions, these processes enhance their abundance by parametrically large factors. Therefore they have to be taken into account whenever a property of the universe, e.g.~the DM and/or the baryon abundance, depends on the final yield of hadrons. As an example, their dramatic impact on the abundance of supercooled composite DM can be seen in Fig.~\ref{fig:compositeDM_generic}.

Concerning DM in particular, this study constitutes a novel production mechanism of DM with mass beyond the unitarity bound~\cite{Griest:1989wd}. It would be interesting and timely to study its experimental signals, given the new wave of telescopes that is starting to take data of high-energy neutrinos and gamma rays (e.g.~KM3NeT, LHAASO, CTA) and given their potential in testing heavy DM, e.g.~see~\cite{Cirelli:2018iax}.
One such study will appear in a forthcoming publication~\cite{2ndpaper}.

During the course of carrying out this study we have made a number of simplifications, for the purpose of obtaining a general and clear enough picture of the physics involved.
For example, the various populations of particles created by this novel dynamics, such as the ejected techniquanta and the hadrons that follow the bubble walls, could be better described by Boltzmann equations, by the use of simulations etc., rather than with our simple treatment that focused on their average properties.

%

Finally, this dynamics opens broader and exciting avenues of investigation, that we think deserve exploration.
For example, it would be interesting to study  its interplay with recent interesting ideas regarding phase transitions~\cite{Creminelli:2001th,Randall:2006py,Konstandin:2011ds,Konstandin:2011dr,
Falkowski:2012fb,Ipek:2018lhm,Bai:2018dxf,Baker:2019ndr,Chway:2019kft,Bloch:2019bvc,DelleRose:2019pgi,Kitajima:2020kig}, or its impact on the production of gravitational waves in supercooled confining phase transitions.
As for the latter, our study of the bubble wall Lorentz factor in~Sec.~\ref{sec:wall_speed} constitutes a necessary first step.


\medskip

\section*{Acknowledgements}

We thank G\'eraldine Servant for several precious discussions at many stages of this work, and for a careful reading of the manuscript.
We thank Benedict von Harling, Oleksii Matsedonskyi, Philip Soerensen, Ryusuke Jinno, Thomas Konstandin and Gilad Perez for useful discussion, and Maximilian Dichtl for spotting typos in a previous version of this work.
FS thanks Olaf Behnke, Claudio Bonati, Roberto Contino, Yohei Ema, Michael Geller, Kyohei Mukaida, Davide Pagani, Frank Tackmann and Daniele Teresi for additional useful discussion.

\paragraph*{Funding information}
The work of Y.G. and F.S. is partly supported by the Deutsche Forschungsgemeinschaft under Germany’s Excellence Strategy - EXC 2121 Quantum Universe - 390833306 and by a PIER Seed Project funding (Project ID PIF-2017-72). F.S. is also supported in part by a grant ``Tremplin nouveaux entrants et nouvelles entrantes de la FSI''. I.B. is postdoctoral researcher of the F.R.S.-FNRS with the project ``\emph{Exploring new facets of DM}." 
\medskip

\appendix

\section{Wall profile of the expanding bubbles}
\label{app:wall_profile}
\FloatBarrier

\begin{figure}[t]
\centering
\begin{adjustbox}{max width=1.2\linewidth,center}
\raisebox{0cm}{\makebox{\includegraphics[ width=0.5\textwidth, scale=1]{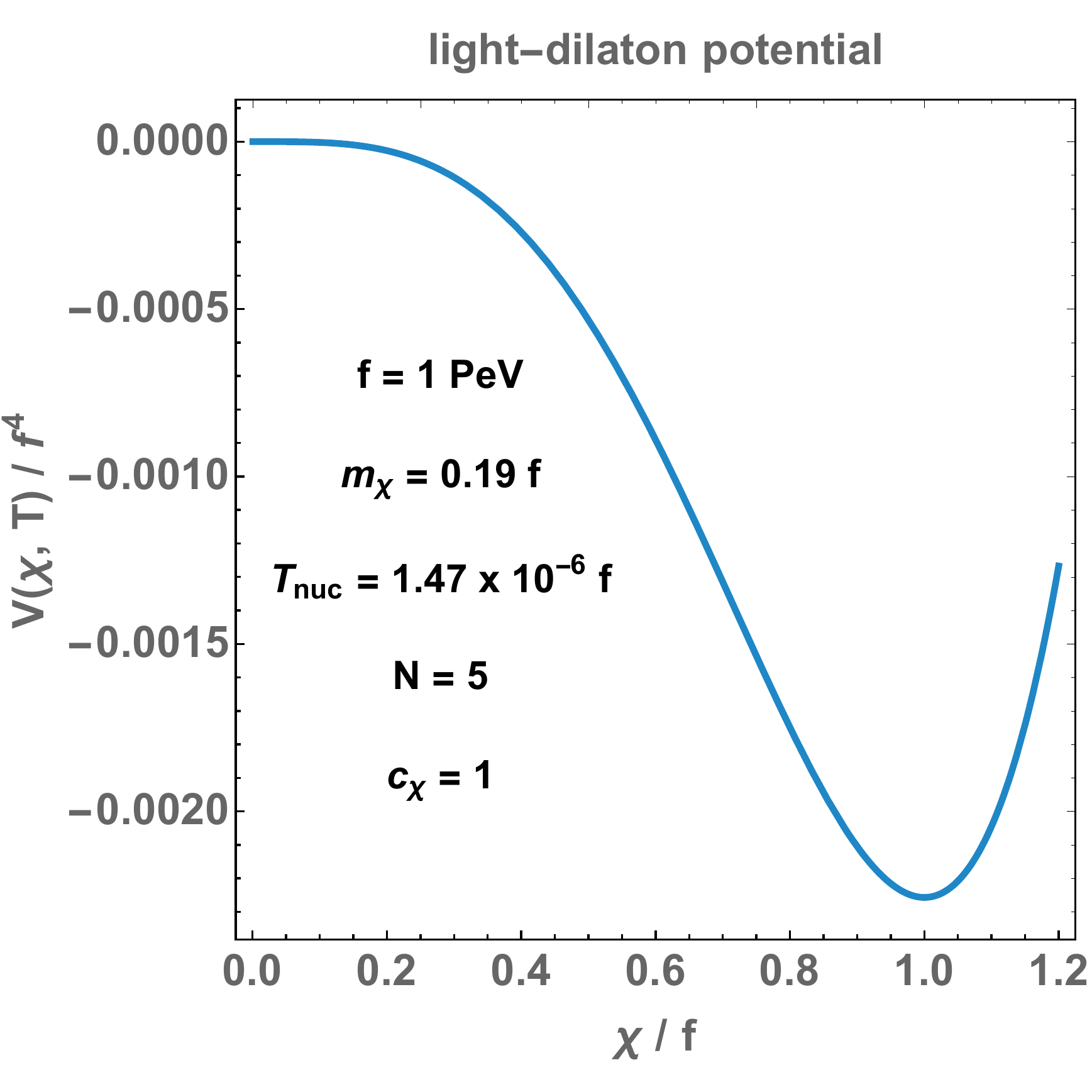}}}
\raisebox{0cm}{\makebox{\includegraphics[ width=0.5\textwidth, scale=1]{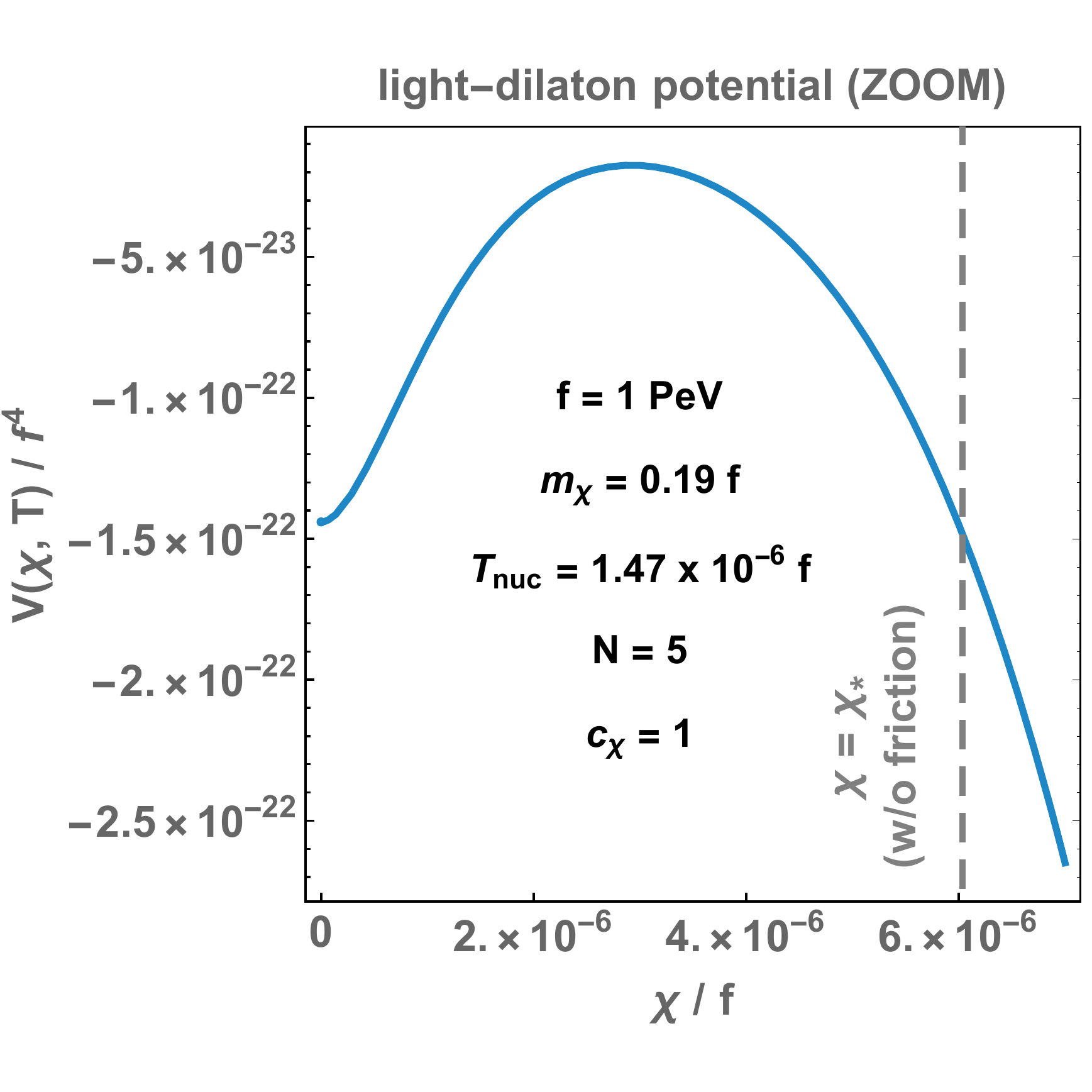}}}
\end{adjustbox}
\caption{\it \small \textbf{Left:} \label{fig:lightdilatonpot} light-dilaton potential with temperature corrections in Eq.~\eqref{eq:total_dilaton_potential}. \textbf{Right:} Zoom on the thermal barrier. The tunneling point $\chi_*$, in the case where the friction term in the Euclidean equation of motion is neglected,  is also shown.}
\end{figure}

\paragraph{The light-dilaton potential.}

In this section we suppose that confinement occurs from the condensation of a nearly-conformal strongly-interacting sector, when an approximate scale-invariance gets spontaneously broken. If the source of explicit breaking is small, the spontaneous breaking of scale invariance generates a pseudo Nambu-Goldstone boson, the dilaton which we parameterize as \cite{Goldberger:2008zz}
\begin{equation}
\dilaton(x) = \dilatonvev e^{\frac{\sigma(x)}{\dilatonvev}},
\end{equation}
where $f$ is the confining scale and where $\sigma(x)$ transforms non-linearly $\sigma(x)  \rightarrow \sigma{(\lambda x)} + \log{\lambda}$ under the scale transformation $x  \rightarrow \lambda x$. 
Its potential is given by \cite{Bruggisser:2018mrt}
 \begin{equation}
 \label{eq:zero_temp_pot_dil}
V_{\dilaton}^{\mathsmaller{\rm T=0}}(\dilaton)  =c_{\dilaton}\, g_{\dilaton}^2 \, \dilaton^4 \, \left[1- \frac{1}{1+\gamma_{\epsilon}/4}\left(\frac{\dilaton}{\dilatonvev}\right)^{\gamma_{\epsilon}} \right],
\end{equation}
with 
\begin{equation}
\label{eq:gammaepsilonVSmdilaton}
\gamma_{\epsilon} \simeq -\frac{1}{4 }\frac{m_{\dilaton}^2}{c_{\dilaton}\,g_{\dilaton}^2~f^2} <1,
\end{equation}
where $m_{\dilaton}$ is the dilaton mass, and $c_{\dilaton}$ is a constant of order $1$, which we fix to $c_{\chi} =1$.
The dilaton coupling constant $g_{\dilaton}$ is chosen to reproduce the glueball normalization
\begin{equation}
g_{\dilaton} \simeq \frac{4\pi}{N},
\end{equation}
with $N$ being the rank of the confining gauge group.
 The validity of the EFT relies on the smallness of the parameter $|\gamma_{\epsilon}| \ll 1$ (here taken negative) which controls the size of the explicit breaking of scale invariance, and thus of the dilaton mass.
 
Note that in the limit where $|\gamma_{\epsilon}| \ll 1$, the dilaton potential at zero-temperature reduces to the Coleman-Weinberg potential \cite{Coleman:1973jx}, i.e. 
\begin{equation}
V_{\dilaton}^{\mathsmaller{\rm T=0}}(\dilaton)   \overset{|\gamma_{\epsilon}|\ll 1}{=}  -\gamma_{\epsilon} \, c_{\dilaton}\,g_{\dilaton}^2\, \dilaton^4 \, \textrm{log} \left( \frac{\dilaton}{\dilatonvev} \right).
\label{eq:dilaton_potential_taylor_cw}
\end{equation}

\paragraph{Thermal corrections.}
To model thermal effects, we follow \cite{Randall:2006py,Bruggisser:2018mrt}, and consider the finite-temperature corrections generated by the particles charged under the confining force (the CFT bosons)
 \begin{equation}
  V_T( \dilaton,\, T) = \sum_{i \in \rm\mathsmaller{CFT}~ bosons} \frac{n T^4}{2 \pi^2} J_B\left( \frac{m_{i}^2}{T^2} \right), \qquad \textrm{with} \quad m_{i} \simeq g_{\dilaton }\,\chi.
 \label{eq:finite_temp_dilaton}
\end{equation}
The total number of CFT bosons $n$ is fixed to\footnote{The effective number of gluons in the deconfined phase $45N^2/4$ being different from $2(N^2-1)$ is a property valid at thermal equilibrium. It results from the peculiar strongly-coupled dynamics of the CFT. However, due to the large wall Lorentz factor, the CFT gas entering the wall can be considered as collisionless, cf. Sec.~\ref{sec:when_confinement}. This is why in the main text we consider the number of gluons entering the wall as $g_g = 2(N^2-1)$.}
\begin{equation}
\sum_{\mathsmaller{\rm CFT}~\rm bosons} n =\frac{45 N^2}{4} \equiv  \tilde{g}_{g},
\label{eq:number_dof_CFT}
\end{equation}
in order to recover the free energy of $\mathcal{N}=4$ $SU(N)$ large $N$ super-YM dual to an AdS-Schwarzschild space-time \cite{Creminelli:2001th}
\begin{equation}
\label{eq:free_energy_CFT}
V_T( 0,\, T) \simeq -b\, N^2\, T^4, \qquad \textrm{with} \quad b= \frac{\pi^2}{8}.
\end{equation}
By writing Eq.~\eqref{eq:free_energy_CFT}, we have neglected the contribution from the fermions present in the plasma.
For simplicity, we suppose that the dilaton degree of freedom $\chi$ still exists in the deconfined phase, such that the total potential for the dilaton is
\begin{equation}
V_{\rm tot}(\dilaton,\, T) =  V_{\dilaton}(\dilaton) + V_T( \dilaton,\, T) ,
\label{eq:total_dilaton_potential}
\end{equation}
where $V_{\dilaton}(\dilaton)$ and $ V_T(\dilaton, \, T) $ are given by Eq.~\eqref{eq:zero_temp_pot_dil} and Eq.~\eqref{eq:finite_temp_dilaton}. We plot the potential in Fig.~\ref{fig:lightdilatonpot}.
The supercooling stage starts when the energy density becomes vacuum-dominated
\begin{equation}
\frac{\pi^2}{30}g_{\rm Ri} T_{\rm start}^4 \simeq c_{\rm vac} f^4 \quad \implies \quad T_{\rm start}\simeq \left(\frac{30 c_{\rm vac}}{g_{\rm Ri} \pi^2}  \right)^{\! 1/4} f, \label{eq:Tstart_app}
\end{equation}
with $c_{\rm vac}=\dfrac{m_{\sigma}^2}{16f^2}$ and $g_{\rm Ri}= g_{\rm SM}+ g_{\rm TC}$ where (see Eq.~\eqref{eq:number_dof_CFT})
\begin{equation}
g_{\rm TC} = g_{\rm q} + \tilde{g}_{g} \simeq \frac{45N^2}{4}. \label{eq:gTC_app}
\end{equation}

\paragraph{Space-like region: the bounce profile.}
We solve the tunneling temperature by solving the equation 
\begin{equation}
\label{eq:nucTempInstApprox}
\Gamma(\Tnuc) \simeq H(T_{\rm nuc})^4.
\end{equation}
with \cite{Coleman:1977py, Callan:1977pt}
\begin{equation}
\Gamma(\Tnuc) = R_{0} ^{-4} \left( \frac{S_4}{2\pi} \right)^2 {\rm exp} \left(-S_4 \right) ,
\end{equation}
where $R_{0} \sim 1/\Tnuc$ is the bubble radius at nucleation and $S_4$ is the $O_4$-bounce action 
\begin{equation}
S_{4}=2\pi^2\int dr~r^3~\left[\frac{1}{2}\phi^{'}(r)^2+V\left(\phi(r)\right)\right],
\end{equation}
which we compute from solving the Euclidean equation of motion ($d=4$)
\begin{equation}
\phi''(s) + \frac{d-1}{s}\phi'(s) = \frac{dV}{d\phi},
\label{eq:bounce_eom}
\end{equation}
with boundary conditions
\begin{equation}
\phi'(0)=0, \qquad \textrm{and} \qquad  \lim_{r \to \infty} \phi(r) = 0.
\label{eq:BC_bounce}
\end{equation}
$s = \sqrt{\vec{r^2}+t_E^2} = \sqrt{\vec{r^2}-t^2}$ is the space-like light-cone coordinate and $t_E = i\,t$ is the Euclidean time.

\begin{figure}[t]
\centering
\begin{adjustbox}{max width=1.2\linewidth,center}
\raisebox{0cm}{\makebox{\includegraphics[ width=0.5\textwidth, scale=1]{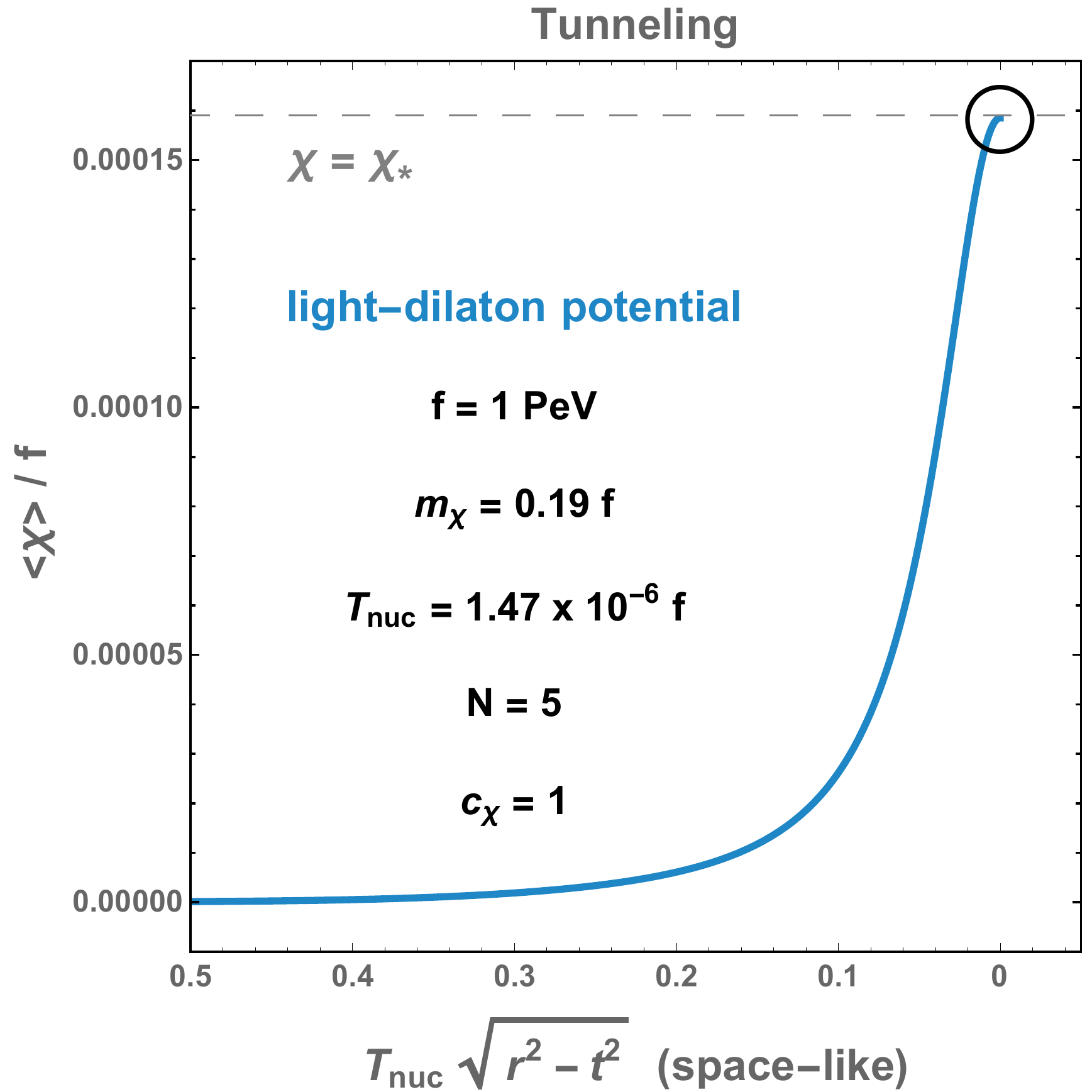}}}
\raisebox{0cm}{\makebox{\includegraphics[ width=0.5\textwidth, scale=1]{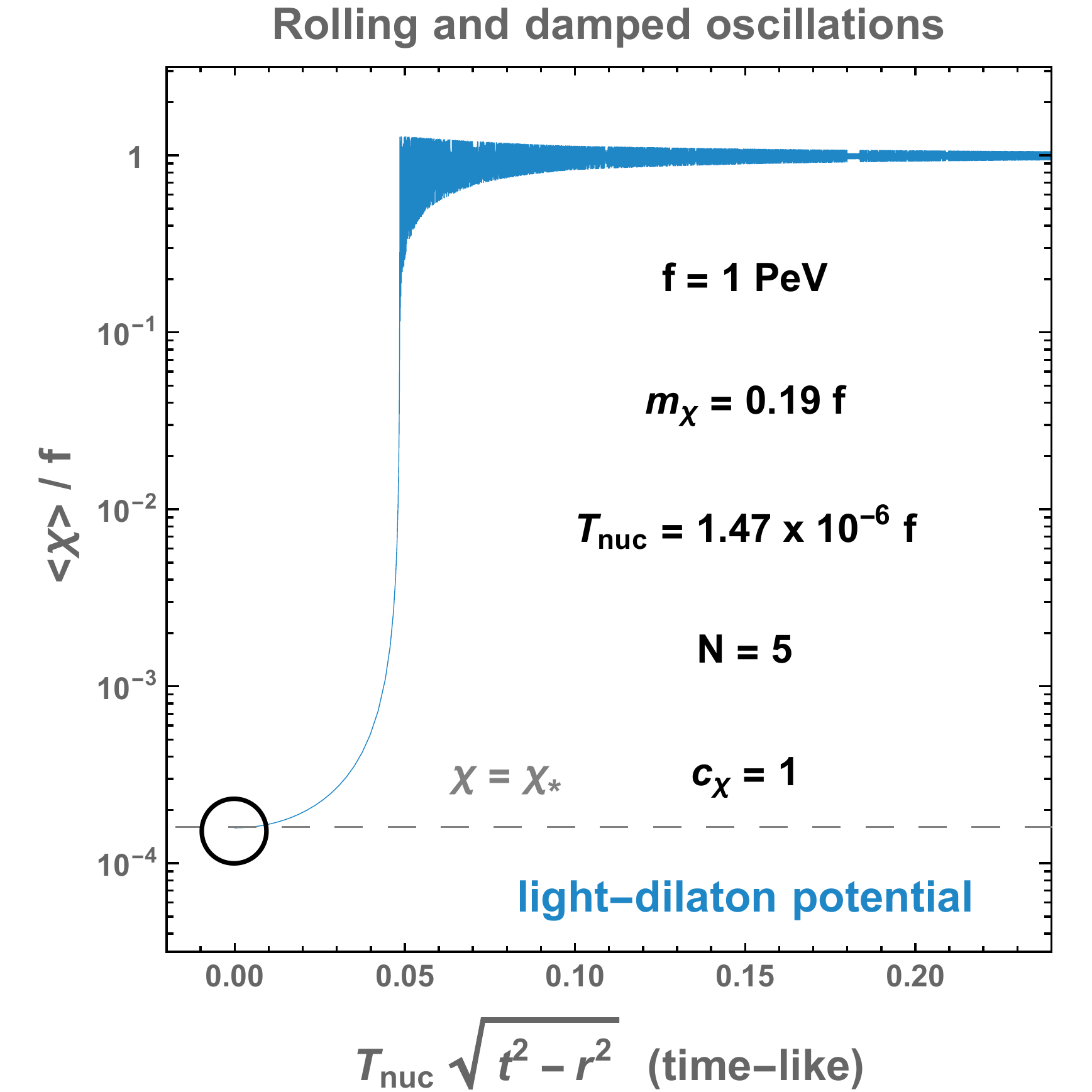}}}
\end{adjustbox}
\caption{\it \small \label{fig:scalarTimeLike} \textbf{Left:} Bounce profile at nucleation. It interpolates between the false vacuum $\left<\dilaton\right> = 0$ outside the bubble and the release point $\left<\dilaton\right>_{*}$ at the center of the bubble. \textbf{Right:} Evolution of the scalar field after tunneling. First, the scalar field rolls along the shallow part of the nearly-conformal potential and then realized oscillations with period $\sim \dilatonvev^{-1}$ with a damping time $\lesssim T_{\mathsmaller{\rm nuc}}^{-1}\gg\dilatonvev^{-1}$. Taking into account the decay of the scalar field would reduce the damping time after the first oscillation. The full bubble wall profile can be obtained after connecting the two figures through the two black circles.}
\end{figure}

We plot the bounce profile in the left-hand panel of Fig.~\ref{fig:scalarTimeLike} for given parameters relevant for the study. 
The value at the center of the bubble --- the tunneling point $\dilaton_*$ --- can be estimated analytically by energy conservation between $\chi=\chi_*$ and the false vacuum in $\chi =0$ if we neglect the friction term in the equation of motion in Eq.~\eqref{eq:bounce_eom},
\begin{equation}
V_{\rm tot}(\dilaton_*) \simeq V_{\rm tot}(0), \qquad \rightarrow \qquad \frac{\chi_{*}}{f} \simeq \frac{1}{\sqrt{2}\,{\rm log}^{1/4}(f/\chi_{*} )}\frac{T}{T_c}.
\end{equation}
Here (coincidence numeric) $T_c$ is the critical temperature, defined when the two minima of the free energy are equal
\begin{equation}
\label{eq:Tc_def}
c_{\rm vac} f^4 + V_T(f,\,T_c) - V_T(0,\,T_c) \equiv 0,
\end{equation} 
Note that for confining phase transition with $m_i(f) \gtrsim f$, the quantity $V_T(f,\,T_c)$ in Eq.~\eqref{eq:Tc_def} vanishes\footnote{
We recall that the thermal functions in Eq.~\eqref{eq:finite_temp_dilaton} verify the property $\lim_{x\to\infty} J_{\rm B/F}(x) = 0$. From using Eq.~\eqref{eq:Tstart_app}, we observe that
\begin{equation}
4.0~ \frac{m_i(f)}{f} \left(\frac{0.1}{c_{\rm vac}} \right)^{1/4}\left( \frac{g_{\rm Ri}}{80} \right)^{1/4} \gg 1 \qquad \implies \qquad m_i(f)/T_{\rm start} \gg 1. \label{eq:TC_Tstart_condition}
\end{equation}
The connection between $T_c$ and $T_{\rm start}$ in Eq.~\eqref{eq:TC_Tstart} applies for all phase transitions satisfying Eq.~\eqref{eq:TC_Tstart_condition}.} and $T_c$ is related to the temperature at which supercooling starts $T_{\rm start}$ in Eq.~\eqref{eq:Tstart_app} through
\begin{equation}
T_{\rm c} \simeq 3^{1/4} \left( \frac{g_{\rm Ri}}{g_\TC} \right)^{1/4} T_{\rm start}. \label{eq:TC_Tstart}
\end{equation}
$g_{\rm Ri}$ is the total number of relativistic d.o.f in the symmetric phase while $g_\TC$ only counts those which are involved in the phase transition ($g_\TC <g_{\rm Ri}$).
In the scenario studied in this appendix, upon assuming $g_{\rm Ri}\simeq g_{\rm TC}$ with $g_{\rm TC}$ given in Eq.~\eqref{eq:gTC_app}, we get
\begin{equation}
 T_c = \left(\frac{m_{\dilaton}^2 \,f^2}{16\,b\,N^2}\right)^{1/4} = \left( \frac{\left|\gamma_{\epsilon}\right|\,c_{\dilaton}\,g_{\dilaton}^2}{4\,b\,N^2} \right)^{1/4}~f. \label{eq:Tc_app}
\end{equation}

The tunneling point $\chi_*$ in absence of friction is shown in Fig.~\ref{fig:lightdilatonpot}, while the tunneling point from numerically solving the bounce equation is visible in Fig.~\ref{fig:scalarTimeLike}.
Plugging the numbers chosen for making the plots, we find $\chi_{*}/f \simeq 6.0 \times 10^{-6}$ for the analytical value and $\chi_{*}/f \simeq 1.6 \times 10^{-4}$ for the numerical value. This difference was expected since the analytical estimate neglects the friction term in Eq.~\eqref{eq:bounce_eom}.

\paragraph{Time-like region: rolling and damped oscillations. }

As soon as the bubble expands, the scalar field starts to roll toward the true vacuum $\left<\dilaton\right> = f$ and realize damped oscillations. The field dynamic is captured by the Klein-Gordon equation for an inhomogeneous field
	\begin{equation}
	\square \phi -\frac{\partial V}{\partial \phi} = 0.
	\end{equation}
We first use the $SO(3)$ symmetry to reduce the 3 Cartesian coordinates to the radial $r$ coordinate
	\begin{equation}
	\frac{\partial^2 \phi}{\partial r^2} +\frac{2}{r}\frac{\partial \phi}{\partial r} - \frac{\partial^2 \phi}{\partial t^2} -\frac{\partial V}{\partial \phi} = 0.
	\end{equation}
We used the Minkowski metric since we can neglect the universe expansion during the time of bubble propagation.
We then use the $SO(3,1)$ symmetry which reduces $r$ and $t$ to the time-like light-cone coordinate $s = \sqrt{t^2-r^2}$ only \cite{Jinno:2019bxw}
	\begin{equation}
	\label{eq:scalar-time-like}
	\frac{\partial^2 \phi}{\partial s^2} +\frac{3}{s}\frac{\partial \phi}{\partial s} +\frac{\partial V}{\partial \phi} = 0.
	\end{equation}
Note the opposite sign in front of the potential $V$ between the space-like (or Euclidean) equation of motion in Eq.~\eqref{eq:bounce_eom} and the time-like (or Minkowskian) equation of motion in Eq.~\eqref{eq:scalar-time-like}.
Here the damping is purely geometrical, reminiscent of the $SO(3,1)$ symmetry and we do not consider the damping due to the dilaton decay or due to the interaction with the plasma (see e.g. \cite{Dorsch:2018pat}). In right panel of Fig.~\ref{fig:scalarTimeLike}, we display the scalar field profile obtained after integration of the time-like equation in Eq.~\eqref{eq:scalar-time-like}, using the initial condition $\chi(s=0) = \chi_*$ given by the bounce solution in Eq.~\eqref{eq:bounce_eom}. 

\paragraph{The full bubble wall profile. }
The full bubble wall profile is obtained after matching the profile in the space-like region, left panel of Fig.~\ref{fig:scalarTimeLike}, with the profile in the time-like region, right panel of Fig.~\ref{fig:scalarTimeLike}.
One can see that the first confining scale, that the incoming techniquanta are subject to upon entering the wall, is the exit scale
\beq
\chi_* \gtrsim \Tnuc\,.
\eeq
Our explicit computation also shows that the length of the section of the bubble wall where $\langle \chi \rangle = \chi_*$, in the wall frame, satisfies
\beq
L_\text{w} \lesssim \Tnuc^{-1}\,,
\eeq
as we assumed in Eq.~(\ref{eq:Lw}) in the main text.
Then, $\langle \chi \rangle$ transits to its zero-temperature value $f$ over a length, in the wall frame, of order $f^{-1}$.

\section{NLO pressure on the bubble walls}
\label{app:NLO_pressure}

\paragraph{Transition splitting. }

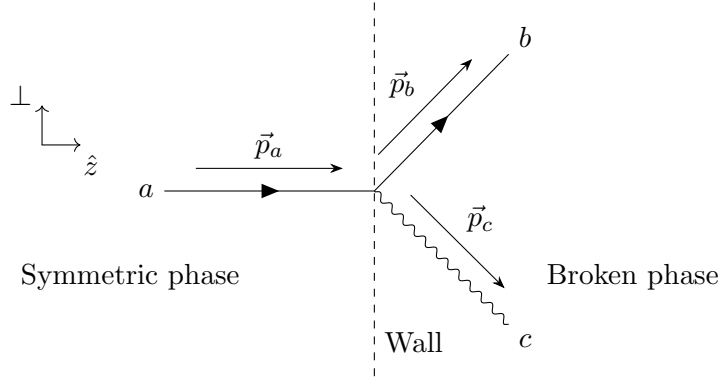
\begin{figure}[t]
\centering
\begin{adjustbox}{max width=1\linewidth,scale=1,center}
\begin{tikzpicture}

\begin{feynman}
\vertex (a) {\(a\)};
\vertex [right=3 cm of a] (b);
\vertex [above right=2.5cm of b] (c) {\(b\)};
\vertex [below right=2.5cm of b] (d)  {\(c\)};

\diagram* {
 (a) -- [fermion, momentum=\( \vec{p}_{a}\)] (b),
 (b) -- [fermion, momentum=\(\vec{p}_b\)] (c),
 (b) -- [boson, momentum=\(\vec{p}_c\)] (d),
};
\end{feynman}

\node [left=1cm of a] (a0) {};
\node [above=1cm of a0] (e) {};
\node [below=0.5cm of e, minimum width=0 pt,inner sep=0pt] (f) {};
\node [right=0.5 of f] (g) {};
\draw [->,shorten <= -0.8pt] (f) -- (e) node[left] {$\perp$};
\draw [->,shorten <= -0.3pt] (f) -- (g) node[below] {$\hat{z}$};

\node [above=2.5cm of b] (b1) {};
\node [below=2.5cm of b] (b2) {};
\draw [dashed] (b1) -- (b2) node[pos=0.9,right] {Wall};

\node [below=1cm of b] (b3) {};
\node [left=1.5cm of b3] (ps) {Symmetric phase};
\node [right=2cm of b3] (ph) {Broken phase};

\end{tikzpicture}
\end{adjustbox}
\caption{\it \small \label{fig:transition_splitting} NLO contribution to the retarding pressure: while entering inside the bubble wall, an incoming particle `$a$' radiates a vector boson `$c$' which gets a mass in the confined phase.}
\end{figure}

In Sec.~\ref{sec:PLO}, we have presented the retarding pressure due to the change in inertia of the system incoming-quark + gluon-flux-attached-to-the-wall when entering inside the confined phase, as well as the retarding pressure due to the ejected quark. In this section we introduce a possible correction which arises in presence of a finite gauge coupling constant. The correction term, which is called NLO pressure, arises from the possibility for the incoming particle to radiate a soft boson which gets a mass in the broken phase \cite{Bodeker:2017cim}
\begin{equation}
\mathcal{P}_\NLO =\sum_{a} \nu_a \int \frac{d^3 p_a}{(2\pi)^3}f_a(p_a)\,\frac{p_{a,\,z}}{p_{a,\,0}} \times \sum_{bc} \int dP_{a\to bc} \times (p_{a,\,s}^z - p_{b,\,h}^z - p_{c,\,h}^z),
\end{equation}
where $h,s$ stands for the `Higgs' and the symmetric phases. $p_a$ and $p_b$ are the momenta of the incoming particle before and after the splitting while $p_c$ is the momentum of the radiated boson\footnote{Note that our notation for `a', `b' and `c' is different from \cite{Bodeker:2017cim} where the roles of `b' and `c' are interchanged.}, see Fig.~\ref{fig:transition_splitting}. We summed over all the species $a$ likely to participate in the process, $\nu_a$ being their number of degrees of freedom. 
The differential splitting probability is given by
\begin{equation}
\int dP_{a\to bc} \equiv \int  \frac{d^3 p_c}{(2\pi)^3 2p_c^0}  \frac{d^3 p_b}{(2\pi)^3 2p_b^0} \left< \phi | \mathcal{T} |p_c,p_b\right> \left< p_c,p_b | \mathcal{T} |\phi \right>,
\end{equation}
with the transition element
\begin{align}
 \left< p_c,p_b | \mathcal{T} |p_a \right> &= \int d^4x  \left< p_c,p_b | \mathcal{H}_{\rm int} |p_a \right>,\\
 &= (2\pi)^3 \,\delta^{(2)} (\vec{p}_{a,\perp} - \vec{p}_{b,\perp}-\vec{p}_{c,\perp}) \, \delta (E_a - E_b -E_c) \,\mathcal{M},
\end{align}
where
\begin{equation}
\label{eq:M_matrix}
\mathcal{M} \equiv \int dz ~ \chi_{c}^*(z) \chi_{b}^*(z) V(z)\chi_{a}(z).
\end{equation}
We obtain \cite{Bodeker:2017cim}
\begin{multline}
\mathcal{P}_\NLO =\sum_{a,bc} \nu_a \int \frac{d^3 p_a}{(2\pi)^3 2E_a}f_a(p_a) ~\frac{d^3 p_c}{(2\pi)^3 2E_c} ~ \frac{d^3 p_b}{(2\pi)^3 2E_b}~ [1\pm f_{c}][1\pm f_{b}] ~(p_{a,\,s}^z - p_{b,\,h}^z - p_{c,\,h}^z)\\
\times  (2\pi)^3 \,\delta^{(2)}  (\vec{p}_{a,\,\perp} - \vec{p}_{b,\,\perp} -\vec{p}_{c,\,\perp}) \, \delta (E_a - E_b - E_c) ~|\mathcal{M}|^2.
\end{multline}
Now we assume $p_{a}^{z} \simeq E_{a}$, $p_b^z \simeq E_b \simeq E_a$ and $p_c^z \simeq E_c -\dfrac{m_c^2(z)+k_{\perp}^2}{2E_c}$ where $k_{\perp}$ is the transverse momentum of the emitted boson, from which we get
\begin{equation}
p_{a,\,s}^z - p_{b,\,h}^z - p_{c,\,h}^z \simeq \frac{m_c^2(z)+k_{\perp}^2}{2E_c},
\end{equation}
and
\begin{equation}
\mathcal{P}_\NLO =\sum_{a,\,bc} \nu_a \int \frac{d^3 p_a}{(2\pi)^3 (2E_a)^2}f_a(p_a) \frac{d^2 k_\perp}{(2\pi)^2 }  \frac{d E_c}{(2\pi) 2E_c} [1\pm f_c][1\pm f_{b}] \frac{m_c^2(z)+k_{\perp}^2}{2E_c}~|\mathcal{M}|^2.
\label{eq:P1to2_interm}
\end{equation}

\paragraph{WKB approximation. }
Next, we make use of the WKB approximation, 
\begin{equation}
\chi_{c}(z) \simeq {\rm exp } \left(i \int_0^z p_c^z(z') dz' \right) \simeq e^{i E_c z} \exp \left(  - \frac{i}{2E_c} \int_0^z(m_c^2(z')+k^2_\perp)~ dz' \right),
\end{equation}
which allows to write the product of wave functions in terms of a phase-dependent quantity $A$,
\begin{equation}
\chi_a(z)\chi_b^*(z)\chi_c^*(z) = {\rm  exp} \left( \frac{i}{2E_a} \int_0^z ~A(z')~dz' \right)\,,
\label{eq:M_matrix_1p5}
\end{equation}
with 
\begin{equation}
-A = m_a^2 -\frac{m_b^2 + k_\perp^2}{1-x}- \frac{m_c^2 + k_\perp^2}{x}  \simeq \dfrac{k^2_{\perp}+m_c^2}{x}.
\end{equation}
We have introduced the variable $x \equiv E_c /E_a$ and assumed $x \ll 1$ in the last equality.
We can now split the integral over $z$ across the wall in Eq.~\eqref{eq:M_matrix} into a contribution from the broken phase and a contribution from the symmetric phase. We assume that the vertices $V$ and WKB phases $A$ on each side of the wall are $z$-independent and we denote them by ($V_h$, $A_h$) and ($V_s$, $A_s$), such that we obtain
\begin{equation}
\mathcal{M} \simeq V_s \int_{-\infty}^0 dz\,{\rm exp} \left[ iz\frac{A_s}{2E_a} \right] + V_h \int_{0}^{\infty} dz\,{\rm exp} \left[ iz\frac{A_h}{2E_a} \right] = 2iE_a \left(  \frac{V_h}{A_h} - \frac{V_s}{A_s} \right).
\label{eq:M_matrix_2}
\end{equation}

\paragraph{Radiation of a soft transverse boson. }
It can be shown \cite{Bodeker:2017cim} that the most important process contributing to the pressure at large $E_a$ is likely to be $X(p_a) \to V_T(p_c)~ X(p_b)$ where $V_T$ is a transverse vector boson. The corresponding vertex function is phase-independent, $V_h=V_s$, and equal to
\begin{equation}
|V^2| = 4\,g^2\,C_2[R]\,\frac{1}{x^2}\,k_\perp^2,
\end{equation}
where $g$ is the gauge coupling constant and $C_2[R]$ is the second casimir of the representation $R$ of the incoming particle with respect to the gauge group.
Therefore, Eq.\eqref{eq:M_matrix_2} becomes 
\begin{equation}
|\mathcal{M}|^2 \simeq 16\,g^2 \,C_2[R] \,(E_{a})^2 \,\frac{m_V^4}{k_\perp^2\,(k_\perp^2+m_V^2)^2}.
\end{equation}
where we have replaced $m_V \equiv m_c$.
The $k_\perp$ integral becomes
\begin{equation}
\int \frac{d^2 k_\perp}{(2\pi)^2} ~\frac{1}{k_\perp^2(k_\perp^2+m_V^2)} = \frac{\log\left( 1+\frac{m_V^2}{k_*^2} \right)}{4\,\pi \,m_V^2},
\end{equation}
where $k_*$ is the IR cut-off on $k_{\perp}$. It is expected to be of order of the vector mass $k_* \sim m_V$.

\paragraph{Final NLO pressure. }
Finally, injecting the last two equations into Eq.~\eqref{eq:P1to2_interm} yields
\begin{equation}
\mathcal{P}_\NLO =\sum_{a,\,bc} \nu_a \int \frac{d^3 p_a}{(2\pi)^3 }f_a(p_a)\frac{d E_c}{(2\pi) E_{c}^2} [1\pm f_c][1\pm f_b] ~ g^2 \,C_2[R] \, m_V^2\,\frac{\log\left( 1+\frac{m_V^2}{k_*^2} \right)}{4\pi}.
\end{equation}
The Pauli blocking or Bose enhancing factor $1 \pm f_{b}$ is of order $1$, while $1\pm f_c$ sums to $1$ when considering both absorption and emission processes.
Hence, the result simplifies to
\begin{equation}
\mathcal{P}_\NLO =\sum_{a} \nu_a\,b_a  \,C_2[R]\,  \frac{8 \zeta(3)}{\pi} \frac{g^2}{4\pi}\, \epsilon_\text{ps}\, \frac{\log\left( 1+\frac{m_V^2}{k_*^2} \right)}{k_*/m_V}\, \gwp \Tnuc^3 \,m_V\,,
\label{eq:PLO_intermediateStep}
\end{equation}
where $b_a = 1~(3/4)$ for bosons (fermions) and $\alpha \equiv g^2/4\pi$. The Lorentz factor $\gamma_{\rm wp}$ between the wall and the plasma comes from $d^3p_a$. We have introduced $\epsilon_\text{ps} \leq 1$ to encode the suppression from phase-space saturation of the emitted soft techni-gluon,which is important for large coupling $g$, and which we justify in the next paragraph.

\paragraph{Phase-space saturation. }

At order $g^4$, the emitted gauge boson can interact among each other. These processes are weighted by $g^2\,f(k)$ with respect to NLO case studied in the last section. The occupancy function $f(k)$ can be estimated to be of order
\begin{equation}
f(k) \sim \frac{n}{\Delta k} \sim \frac{g^2\,\gwp \, \,\Tnuc^3}{m_V^3},
\end{equation}
where $\Delta k \sim m_V^3$ is the available phase space and $n\sim \mathcal{P}_{1\to 2} /(p_a^z-p_b^z-p_c^z) $ with $(p_a^z-p_b^z-p_c^z) \sim m_V$. Hence, we can not consider the individual transition splitting processes as independent from each other as soon as
\begin{equation}
\gwp ~\gtrsim~ \frac{m_V^3 }{g^4\Tnuc^3}.
\label{eq:parameter_space_saturation}
\end{equation}
At such large $\gwp$, we expect the NLO pressure to change behavior.
See \cite{Bodeker:2017cim} for more details and particularly about some hints of $\mathcal{P}_\NLO$ going from $\propto \gwp$ to $\gwp^{4/7}$.
For simplicity, we just encode this effect into the coefficient $\epsilon_\text{ps} \leq 1$ in Eq.~(\ref{eq:PLO_intermediateStep}).

\paragraph{Case of a $SU(N)$ confining sector. }
In the scenario we are interested, the deconfined phase contains $g_q$ techni-quark and $g_g$ techni-gluons, and the NLO pressure would be induced by the possibility for these techni-quanta, to radiate a soft techni-gluon acquiring a mass $m_V = m_g$ in the confined phase. Hence, Eq.~\eqref{eq:PLO_intermediateStep} becomes
\beq
\mathcal{P}_\NLO
\simeq  \big(g_g C_2[g]+ \frac{3}{4}g_q C_2[q]\big) \frac{8 \zeta(3)}{\pi} \frac{g_\text{conf}^2}{4\pi}\, \epsilon_\text{ps}\,\frac{\log\big(1+\frac{m_g^2}{k_*^2}\big)}{k_*/m_g}\,\gwp \Tnuc^3 \, m_g\,.
\label{eq:NLOpressure_app}
\eeq
where $g_\text{conf}$ is the gauge coupling of the confining group, and where $C_2[g] = N$, $C_2[q] = (N^2-1)/2N$ if the confining gauge group is $SU(N)$. Note that in the parameter space which we consider ($c_{\rm vac}=0.01$, $g_{\rm{\mathsmaller{ TC}}} = 78$) the LO pressure in Eq.~\eqref{eq:P_LO} prevents the condition in Eq.~\eqref{eq:parameter_space_saturation} to be satisfied such that we expect $\epsilon_\text{ps}$ to be close to unity.

\section{Example estimates of the string to DM branching ratio}
\label{app:brstring}

In Sec.~\ref{sec:string_breaking}, we have discussed that, after supercooling, the quarks enter inside the confined phase, with a typical seperation $\sim \Tnuc^{-1}$, much larger than the confining scale $f$, such that a highly energetic fluxtube forms.
We have shown that this string, which is unstable under quark-anti-quark pair nucleation, breaks into $K_{\rm string}$ pieces.
The dynamics of strings is then also relevant in the processes of deep inelastic scatterings of section~\ref{sec:DIS}.
In this section, we estimate the branching ratio of a string to a given hadron $i$, introduced in~Eq.~\eqref{eq:DISrelic_i}, in two different cases.
First, when $i$ is a light meson, in which case we expect the yield of $i$ to be independent of its mass and given by a combinatoric factor implying the number of flavors. Second, when $i$ is a heavy baryon in which case one expects the yield to be Boltzmann suppressed.

\subsection*{Light meson -- Combinatorics}

In the limit of large string energy, $\ECM \gg f$, one expects the fragmentation of the string to be democratic with respect to the different bound-states if they are light enough. In that case, the string-to-i branching ratio is given by a combinatoric factor depending on the number of flavors $N_{f}$ and the number of quark constituents (either $2$ for meson and $N_\TC$ for baryons). In the particular case of a light meson $q_1 \bar{q}_2$, one obtains
	\begin{equation}
	\mathrm{Br(\text{string} \rightarrow} i)  = \begin{dcases} 1/N_f^2, \qquad ~\text{if } q_1 = q_2, \\
						2/N_f^2 ,  \qquad  ~\text{if } q_1 \neq q_2.
	\end{dcases}
	\label{eq:combinatorics}
	\end{equation}

\subsection*{Heavy baryon -- Boltzmann suppression}

For this example a useful model for us will be the thermal one~\cite{Chliapnikov:1994qc,Becattini:1995if,Pei:1996kq,Chliapnikov:1999qi}, which was able to fit LEP data of particle yields up to a $10\%$ error~\cite{Andronic:2008ev}, even with an initial state far from thermal equilibrium. In this model, the yield of heavy mesonic or baryonic resonances is suppressed by a Boltzmann factor~\cite{Chliapnikov:1994qc,Becattini:1995if,Pei:1996kq,Chliapnikov:1999qi}, in which the strong scale plays the usual role of temperature. The yield of heavy resonances can be modelled by
	\begin{equation}
	\langle N_{i} \rangle \sim A_i  \frac{ (2J_i + 1) }{ \mathrm{Exp}\left[ M_{i}/B_i \right]} , 
	\end{equation}
where $M_i$ and $J_i$ are the mass and spin of the state $i$ respectively. Here $A_i$ is an overall normalisation, which will depend on whether the particle is a pseudoscalar meson, vector meson, or baryon etc. In QCD it was found to differ by $\lesssim 10$ between vector mesons, tensor mesons, and baryons~\cite{Chliapnikov:1999qi}.
For these particles $B_{i}$ was found to be a common factor between the groups, $B_{i} \equiv B \sim 150$ MeV~\cite{Chliapnikov:1999qi}.
Note the pseudoscalar mesons in QCD, however, which are lighter, follow a softer spectrum.

Following the above discussion, we shall construct a toy model for the baryonic particle yield from our string fragmentation. In order to retain some simplicity in our model we will consider all particles to share a common $B_i = m_* = g_* f$. In our toy model we consider $SU(N_{c})$ theories, with techniquarks in the fundamental representation, in which baryons will contain $N_{c}$ quarks. Mesons on the other hand will contain a quark-antiquark pair independent of $N_c$. In order to take into account the reduced probability of creating a baryon as opposed to a meson it is therefore suitable to include an additional suppression in the prefactor $A_{i} $ for baryons~\cite{Mitridate:2017oky}
	\begin{equation}
	p_{\mathcal{B}i} = \begin{dcases} \frac{1}{1+2^{N_{c}-1}/N_c}, \qquad ~\text{if $i$ is a baryon}, \\
						1 , \qquad \qquad \qquad \qquad \text{if $i$ is a meson}.
	\end{dcases}
	\end{equation}
Other than this we take a common $A_{i} = p_{\mathcal{B}i}A$.  
Applying energy conservation, we thus find the average number of  the composite state $i$ produced per string breaking to be 
	\begin{equation}
	\langle N_{i} \rangle
	\simeq  \, \frac{p_{\mathcal{B}i}(2J_i + 1) }{ \mathrm{Exp}\left[ M_{i}/m_\pi \right] } \, \left( \sum_k  \frac{p_{\mathcal{B}j} (2J_k + 1) }{ \mathrm{Exp}\left[ M_{k}/m_\pi \right] } \frac{M_k}{m_\pi} \right)^{-1} \langle N_\psi \rangle
	\equiv   \mathrm{BR}_i \,\langle N_\psi \rangle,
	\end{equation}
where the sum runs over all the states in the spectrum, and we remind that $\pi$ denotes the lightest composite state(s). In this case it is clearly possible to have a highly suppressed $\mathrm{BR}_i$, e.g. $\mathrm{BR}_i \simeq 10^{-6}$ for $m_i = m_* \simeq 4 \pi f$, $m_\pi \simeq f$, $N_c=10$, $N_\pi = 3$.

\medskip
\small

\bibliographystyle{JHEP}
\bibliography{String_Fragmentation}
\end{document}